\documentclass[
nofootinbib,
floatfix,
superscriptaddress,
preprintnumbers,
amsmath,amssymb,
twocolumn]{revtex4-2}

\usepackage{anysize}

\usepackage{amsfonts}
\usepackage{amssymb}
\usepackage{graphics}
\usepackage{epsfig}
\usepackage{graphicx}
\usepackage{epsfig}
\usepackage{amssymb}
\usepackage{amsfonts}
\usepackage{amsmath}
\usepackage{xcolor}
\usepackage{float}
\usepackage{enumerate}
\usepackage{slashed}
\usepackage{hyperref}
\hypersetup{colorlinks=true,linkcolor=blue,citecolor=blue}

\usepackage{bbold}

\usepackage{braket}

\usepackage{tikz}
\usetikzlibrary{decorations.pathmorphing}
\usepackage{pgfplots}
\usepgfplotslibrary{patchplots}

\numberwithin{equation}{section}

\newcommand {\beq} {\begin{equation}}
\newcommand {\eeq} {\end{equation}}
\newcommand{\bea}{\begin{eqnarray}}
\newcommand{\eea}{\end{eqnarray}}
\newcommand{\bit}{\begin{itemize}}
\newcommand{\eit}{\end{itemize}}
\def\nl{\nonumber \\}

\def\a{\alpha}

\def\b{\beta}

\def\l{\lambda}

\def\s{\sigma}

\def\p{\partial}

\def\le{\left(}
\def\ri{\right)}

\def\beq{\begin{equation}}
\def\eeq{\end{equation}}

{\left\lbrace\begin{array}{@{}l@{}}}%
{\end{array}\right.}

\def\1{ \mathbb{1}}

\begin{document}

\preprint{OU-HET-1171}

\title{Volume complexity of dS bubbles }
\author{Roberto  Auzzi}
\email{ roberto.auzzi@unicatt.it}
\affiliation{Dipartimento di Matematica e Fisica,  Universit\`a Cattolica del Sacro Cuore, Via della Garzetta 48, 25133 Brescia, Italy, }
\affiliation{ INFN Sezione di Perugia,  Via A. Pascoli, 06123 Perugia, Italy, }
\author{Giuseppe Nardelli}
\email{giuseppe.nardelli@unicatt.it}
\affiliation{Dipartimento di Matematica e Fisica,  Universit\`a Cattolica del Sacro Cuore, Via della Garzetta 48, 25133 Brescia, Italy,  }
\affiliation{TIFPA - INFN, c/o Dipartimento di Fisica, Universit\`a di Trento,  38123 Povo (TN), Italy }
\author{Gabriel Pedde Ungureanu}
\email{gpeddeun@sissa.it}
\affiliation{SISSA, Via Bonomea 265, Trieste, Italy }
 \author{ Nicol\`o Zenoni}
\email{nicolo@hetmail.phys.sci.osaka-u.ac.jp}
\affiliation{Department of Physics, Osaka University, Toyonaka, Osaka 560-0043, Japan }

\begin{abstract}

In the framework of the static patch approach to de Sitter holography
introduced in \cite{Susskind:2021esx}, 
 the growth of holographic complexity has a hyperfast 
behaviour, which leads to a divergence in a finite time. This is very different
from the AdS spacetime, where  instead the complexity rate 
 asymptotically reaches a constant value. We study holographic 
 volume complexity in a class of asymptotically AdS geometries
 which include de Sitter bubbles in their interior.
With the exception of the static bubble case,
the complexity obtained from the volume of the smooth extremal surfaces
which are anchored just to the AdS boundary has a similar behaviour to the AdS case,
because it  asymptotically grows linearly with time.
The static bubble configuration has a zero complexity rate
and corresponds to a  discontinuous behaviour,
which resembles a first order phase transition.
 If instead we consider extremal
surfaces which are anchored at both the AdS boundary and  the de Sitter stretched horizon,
we find that complexity growth is hyperfast, as in the de Sitter case.
 
\end{abstract}

\maketitle


\section{Introduction}

The AdS/CFT correspondence \cite{Maldacena:1997re} provides an interesting 
theoretical laboratory to address many important open questions
in quantum gravity. However, our observed  universe is rather different 
from  an AdS background. It is then a crucial problem to
find a quantum gravity formulation for a cosmological setting.
It is interesting to investigate generalizations of holography
for de Sitter (dS) spacetime. This is a challenging problem,
because in  dS  there is no natural notion
of timelike boundary contrary to asymptotically AdS spacetime.

In order to provide a holographic description of  dS,
the dS/CFT correspondence \cite{Strominger:2001pn,Bousso:2001mw,Balasubramanian:2002zh} proposes 
a duality between quantum gravity in dS$_{D}$ spacetime
and a $(D-1)$-dimensional CFT living on a spacelike boundary
at the future spacelike infinity in dS.
Examples of explicit dS/CFT correspondence have been 
proposed for higher spin gravity in four dimensions \cite{Anninos:2011ui}
and for 3-dimensional Einstein gravity \cite{Hikida:2021ese,Hikida:2022ltr}.
The boundary theory in these cases is not unitary and rather exotic
compared to the textbook examples of CFTs,
because it describes  dS  from the perspective of a
 metaobserver who lives at the future infinity.

A finite entropy can be associated with the area $A$ of 
the dS cosmological horizon \cite{Gibbons:1977mu}
surrounding a static patch observer.
Following \cite{Bousso:2000nf,Bousso:2000md,Chandrasekaran:2022cip},
if we consider a dS spacetime which includes particles and black holes,
we can define a generalized entropy $S_{\rm gen}$
which includes the cosmological horizon entropy and the
ordinary entropy $S_{\rm out}$ of the matter which can be seen
by the observer at the center of the static patch
\beq
S_{\rm gen}=\frac{A}{4 G} + S_{\rm out} \, ,
\eeq
where we denote by $G$ the Newton constant.
It has been argued that the maximum possible value of 
$S_{\rm gen}$ is saturated by the empty dS spacetime
\cite{Maeda:1997fh,Bousso:2000nf,Bousso:2000md}.
The presence of a bound for $S_{\rm gen}$
motivates another class of approaches 
to holography for dS space,
see for example \cite{Banks:2003cg,Goheer:2002vf,Parikh:2004wh,Banks:2006rx,Anninos:2011af,Susskind:2021omt},
in which gravity in dS is conjectured to be dual to a quantum mechanical
system with a finite number of degrees of freedom.

Quantum information provides a useful conceptual framework to 
implement several entries of the dictionary of holographic dualities,
and might give precious insights on how to formulate
  holography for the dS spacetime \cite{Sanches:2016sxy,Nomura:2017fyh,Nomura:2019qps}.
   An interesting generalization
 of the Ryu-Takayanagi  \cite{Ryu:2006bv,Hubeny:2007xt} entanglement entropy  formula 
 has been proposed in the context of static patch horizon holography in dS
   \cite{Susskind:2021esx,Shaghoulian:2021cef,Shaghoulian:2022fop}.
 In this proposal, the AdS boundary is replaced in dS by 
 a stretched horizon which is taken just inside the cosmological horizon
 that surrounds a static patch observer.
The Bekenstein-Hawking entropy associated to
 the cosmological horizon \cite{Gibbons:1977mu} is then interpreted as
 the entanglement entropy between the left and right static patches.
   Further recent works include
\cite{Susskind:2021dfc,Susskind:2022dfz,Lin:2022nss,Murdia:2022giv,Susskind:2022bia,Rahman:2022jsf,Bhattacharjee:2022ave}.

Computational complexity is another concept in quantum information theory
which may play an important role in holography, 
see \cite{Susskind:2014moa,Susskind:2018pmk,Chapman:2021jbh} for reviews.
Indeed, entanglement entropy saturates too fast 
 to describe the growth of the Einstein-Rosen bridge
inside a black hole horizon \cite{Hartman:2013qma} in terms of the boundary CFT.
On the other hand, quantum complexity saturates in a much larger timescale compared to
the thermalization one, so it has the correct behavior
\cite{Susskind:2014rva,Susskind:2014moa}  to overcome the 
limitations of entanglement entropy. 
In theoretical computer science, complexity \cite{Aaronson:2016vto} measures
how hard it is to build a generic target state from a simple reference one,
applying a set of elementary gates. For quantum systems 
with a finite number of degrees of freedom, a continuous notion of complexity 
was introduced by Nielsen \cite{Nielsen1}
in terms of the length of geodesics
in the space of unitary operators.
There is a large amount of arbitrariness in the definition of 
complexity, due to the choice of reference  state
and of the computational cost
of elementary  gates.
Despite these ambiguities, quantum complexity is expected to
exhibit several  robust and universal properties \cite{Brown:2021rmz},
provided that the computational costs of elementary gates 
is chosen in such a way that the complexity metric has negative curvature
\cite{Nielsen-Dowling,Brown:2016wib,Brown:2019whu,Auzzi:2020idm,Brown:2021euk,Basteiro:2021ene}.
It is still an open problem to generalize Nielsen's approach
to complexity in quantum field theory.
Many advances have been made in defining complexity
in free field theory \cite{Jefferson:2017sdb,Chapman:2017rqy}.
The definition of complexity in CFT is still a work in progress,
see  \cite{Caputa:2017urj,Caputa:2018kdj,Erdmenger:2020sup,Flory:2020eot,Chagnet:2021uvi}.

 Three main proposals have been extensively studied as a holographic dual
of quantum computational complexity:
\begin{itemize}
\item Complexity=volume (CV)  \cite{Stanford:2014jda}, in which
 complexity is proportional to the volume of the maximal 
codimension-one spatial surface anchored at a boundary time slice.
\item Complexity=action (CA)   \cite{Brown:2015bva,Brown:2015lvg}, in which
complexity is proportional  to the gravitational action evaluated on
 a codimension-zero bulk region, called the Wheeler DeWitt patch.
 \item Complexity=volume $2.0$, in which  complexity is proportional to
 the spacetime volume of the Wheeler DeWitt patch \cite{Couch:2016exn}.
\end{itemize}
Further generalizations  have been investigated in \cite{Belin:2021bga,Belin:2022xmt,Omidi:2022whq}.
All these holographic proposals reproduce the expected behavior of quantum complexity
for an AdS black hole: in the regime of bulk classical gravity,
which should be appropriate for times which are less than exponential
in the entropy of the system \cite{Susskind:2015toa}, complexity asymptotically grows linearly with time.

As studied in \cite{Susskind:2021esx,Jorstad:2022mls},
the definition of holographic complexity can be extended to  static
patch horizon holography in dS \cite{Susskind:2021esx,Shaghoulian:2021cef,Shaghoulian:2022fop}.
Much differently compared to AdS,
holographic complexity
in dS exhibits a hyperfast growth, \emph{i.e.}
the complexity growth  is so fast that it diverges at a finite critical time.
In \cite{Susskind:2021esx}, Susskind proposed   
the hyperfast growth to be a signature that the Hamiltonian of the 
dual of dS is not of the usual $k$-local type. 
Here $k$-local means that 
the Hamiltonian is the sum of terms that simultaneously act
at most on $k$  degrees of freedom, where $k$ is 
parametrically of order unity in the limit
of a large number of degrees of freedom. 

It is tempting to relate the hyperfast growth of complexity in dS
to the exponential growth of spacetime.
In order to further investigate the reason of a different time
 dependence of holographic complexity compared to AdS,  
  it is interesting to contemplate intermediate situations.
At the crossroad between AdS and dS holography, 
we can consider gravity backgrounds with an
asymptotically AdS$_{D}$ boundary which include dS$_{D}$ regions in their interior.
Examples  in $D=2$ include the centaur
geometry  \cite{Anninos:2017hhn,Anninos:2018svg}, which  can be built
in dilaton-gravity theories\footnote{ Related studies of dS bubbles in dilaton gravity 
include \cite{Biasi:2022ktq}.}.
For centaurs, the dS part of the spacetime is not hidden behind an AdS black hole horizon.
Holographic volume complexity in these backgrounds 
was recently studied in \cite{Chapman:2021eyy}.
In this case, the evolution of complexity is qualitatively different compared
to both AdS and dS cases, because complexity is a decreasing function of time.

In this paper, we study volume complexity 
in higher-dimensional examples of asymptotically  AdS spacetimes
with dS bubbles in their interior
 \cite{Freivogel:2005qh,Fu:2019oyc}.
 This kind of geometries can be realized, for instance,
in an Einstein-scalar theory
where the potential has various minima separated by a domain wall.
The $D \geq 3$ case significantly differs from the $D=2$ one, since
in higher dimensions the dS portion of the geometry is always screened by an AdS horizon at late time.
In order to simplify the model, it is useful to consider 
the limit in which the thickness of the domain wall surrounding the dS interior is small compared to the other physical scales,
 as studied  in \cite{Blau:1986cw,Farhi:1986ty,Farhi:1989yr}
  for a flat external space
  and in  \cite{Freivogel:2005qh,Fu:2019oyc} for an AdS external region.
For simplicity, we specialize to $D=3$  bulk spacetime dimensions
 and we consider spherically symmetric geometries
 consisting  of a dS region and an asymptotically AdS spacetime
 separated by a domain wall with negligible thickness.
We set the AdS scale to one, and we parameterize the 
dS cosmological constant by  $\l$ and the domain wall tension by  $\kappa$.
Outside the bubble, the solution is a BTZ black hole \cite{Banados:1992wn}
 with mass proportional to $\mu$.
While the parameters $\l$ and $\kappa$ specify the theory, $\mu$
is a property of the state.

\begin{figure}
\includegraphics[scale=0.3]{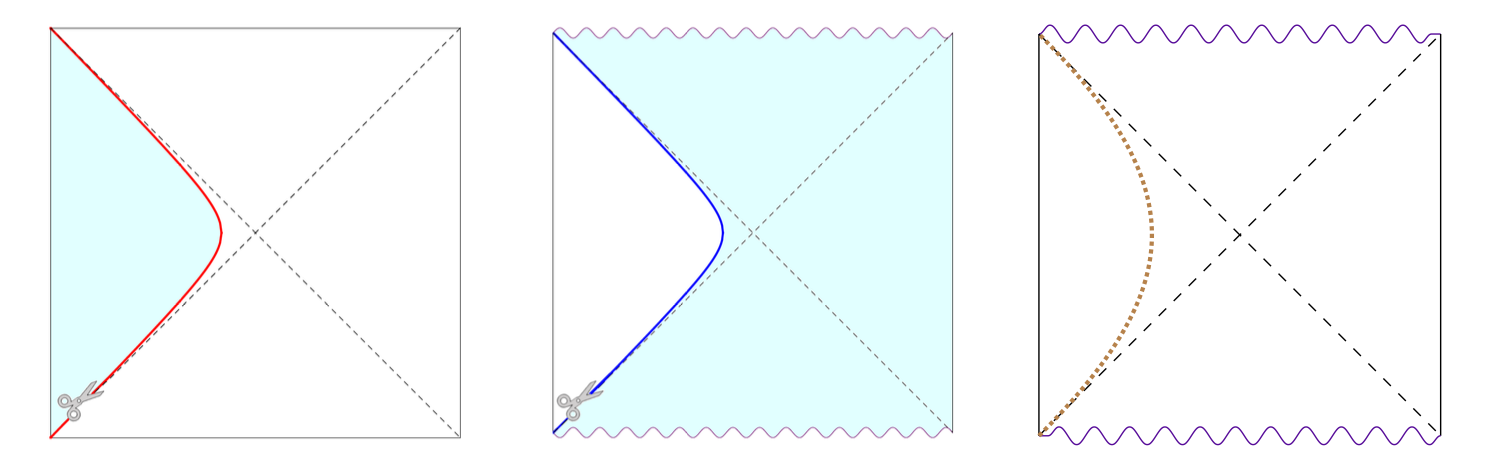}
\caption{Penrose diagrams for a static bubble.
We show in red and in blue the bubble trajectory
 in dS (left) and in AdS (center), respectively.
 Our geometry (on the right) is obtained by joining the dS portion 
 to the left of the red line with the AdS region 
 to the right of the blue line.
 }
\label{penrose-static-bubble}
\end{figure}

We focus on backgrounds which are invariant under time-reversal symmetry $t \to -t$.
For the critical value $\mu=\mu_0$, where
\beq
\mu_0=\frac{\sqrt{(\kappa^2+\l-1)^2 +4 \l}-(\kappa^2+\l-1)}{2 \l} \, ,
\label{mu-zero}
\eeq
 the only time-reversal invariant solution is the static bubble, see figure \ref{penrose-static-bubble}.
For $\mu>\mu_0$, the bubble starts from zero radius and expands 
to infinite size, without entering the external true vacuum region.
From the viewpoint of an external observer, the bubble remains behind a black hole horizon.
However, no such bubble solution exists which enjoys time-reversal invariance.
So, in this paper we focus on the regime $\mu \leq \mu_0$.

\begin{figure}
\includegraphics[scale=0.35]{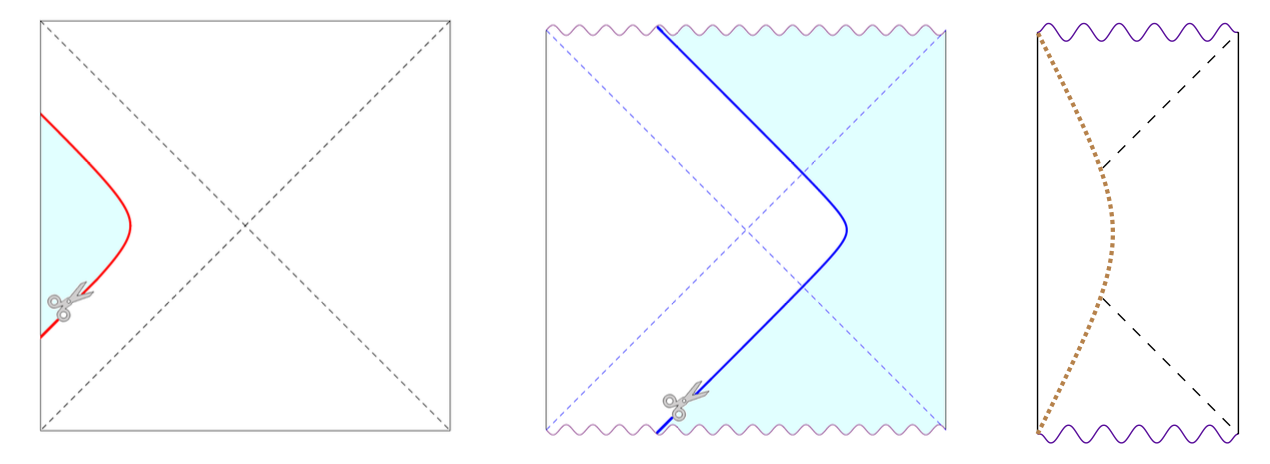}
\caption{Penrose diagrams for a very small bubble $0 < \mu < \mu_s$.}
\label{penrose-collapsing-bubble-1}
\end{figure}

\begin{figure}
\includegraphics[scale=0.32]{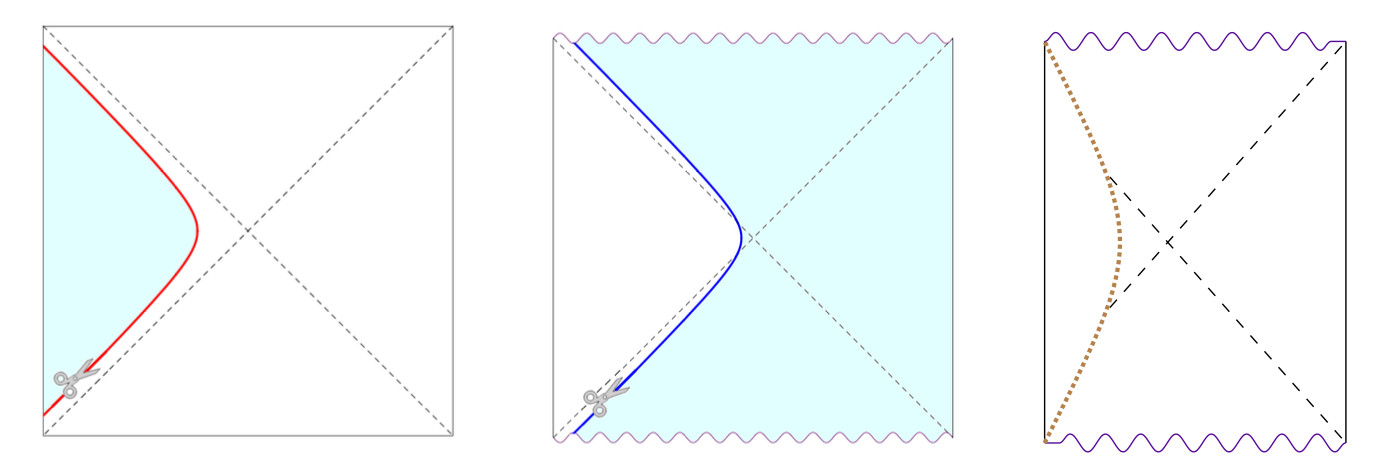}
\caption{Penrose diagrams for a not so small bubble $\mu_s < \mu < \mu_0$.
}
\label{penrose-collapsing-bubble-2}
\end{figure}

\begin{figure}
\includegraphics[scale=0.28]{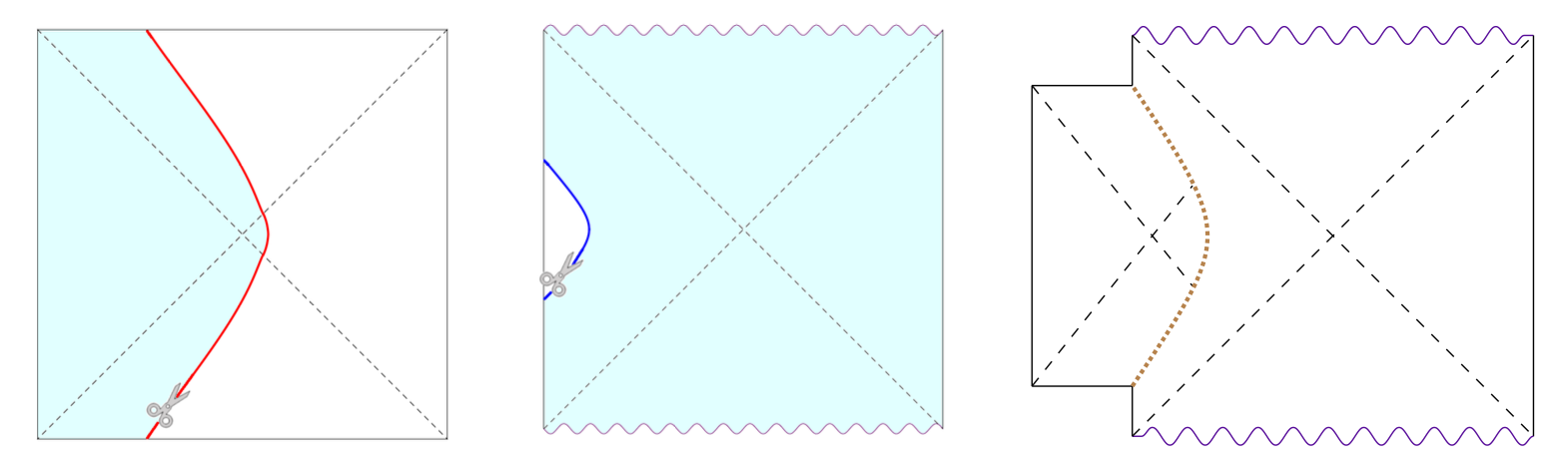}
\caption{Penrose diagrams for a very large bubble $0 < \mu < \mu_h$ (top panel)
and for a not so large bubble  $\mu_h < \mu < \mu_0$ (lower panel).}
\label{penrose-expanding-bubble-2}
\end{figure}

\begin{figure}[h]
\includegraphics[scale=0.29]{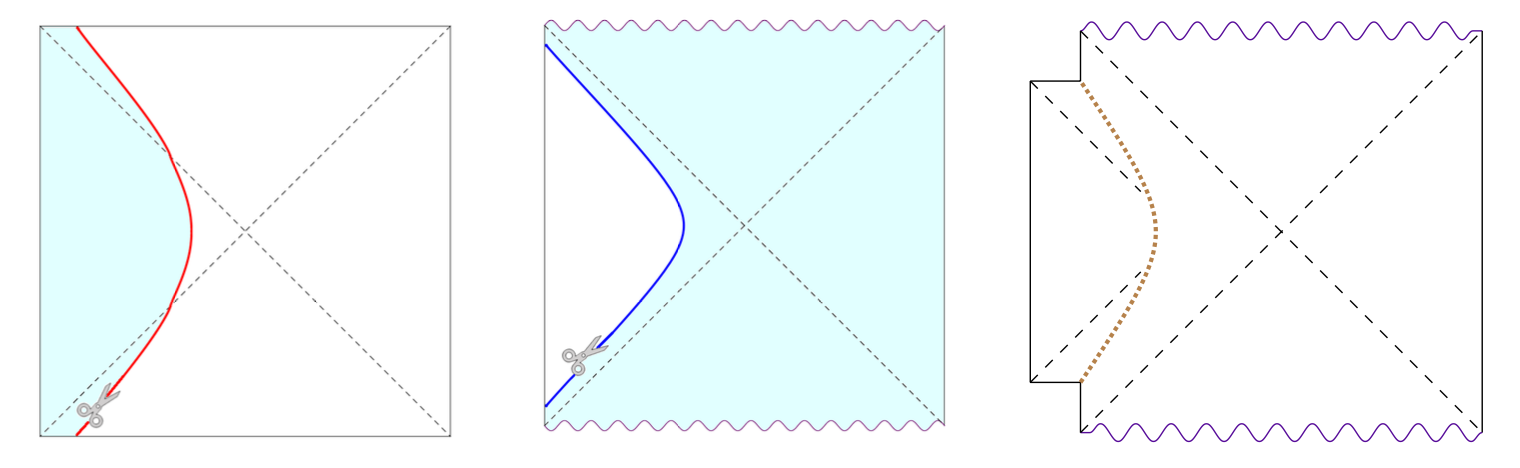}
\caption{Penrose diagrams for a not so large bubble  $\mu_h < \mu < \mu_0$.
}
\label{penrose-expanding-bubble-1}
\end{figure}

For a fixed $\mu <\mu_0$, there are two time-reversal invariant solutions:
\begin{itemize}
\item A bubble collapsing in a finite proper time,
for which the interior portion of dS has finite spacetime volume.
We refer to this case as "small bubble" solution.
We have to further distinguish between two subcases.
For $0 < \mu < {\mu}_s $, where
\beq
{\mu}_s=\frac{1}{\kappa^2+\l} < \mu_0 \, ,
\label{mu-s}
\eeq
 the dS bubble is initially on the 
 same side of the Penrose 
 diagram as the AdS boundary.
We refer to this configuration as a "very small bubble", see figure \ref{penrose-collapsing-bubble-1}.
For ${\mu}_s<\mu< \mu_0$ the bubble is initially on 
the opposite side of the Penrose diagram with respect to the AdS boundary. 
We call this situation a "not so small bubble", see figure \ref{penrose-collapsing-bubble-2}.
\item A bubble expanding for an infinite proper time,
which contains an infinite portion of the dS spacetime. 
We refer to this case as "large bubble" solution. 
We here introduce
\beq
{\mu}_h=\frac{1-\kappa^2}{\l} <\mu_0\, .
\label{mu-h}
\eeq
For $0 < \mu < \mu_h$,  the interior of the bubble 
contains as a subset a dS static patch. 
We call this situation a "very large bubble", see figure \ref{penrose-expanding-bubble-2}.
By contrast, we refer to the solution with ${\mu}_h<\mu< \mu_0$  as a "not so large bubble", 
see figure \ref{penrose-expanding-bubble-1}.
Note that very large bubbles can only be obtained for small enough
domain wall tension, \emph{i.e.} $\kappa<1$.
\end{itemize}
In the parameter space, the static bubble configuration 
is at the border between  the small and the large bubble regimes.

%

With the exception of the very small bubble,
in all cases an observer outside the horizon can not directly
see the dS bubble at any time, because it is 
screened by the black hole horizon.
The presence of the dS bubble is instead 
detected by the volume complexity,
because the extremal surface always penetrates
the dS region of the geometry.

The traditional way to apply the CV proposal
is to consider the volume of extremal surfaces which are anchored 
at the AdS boundary and which are smooth both in the AdS and in the 
dS portions of the spacetime. This prescription can be applied
 both for large and small  bubbles. With the exception of the static bubble configuration,
we find that at late time  holographic  complexity   grows linearly 
 with the same slope as for the BTZ black hole.
In the fine-tuned case of the  static bubble,  
we find instead that volume complexity is time-independent.

 The time dependence of complexity that we obtain
 reveals that the hyperfast growth
 is not necessarily related to the exponential growth
 of spacetime, which is a feature of large bubbles.
 If we consider the prescription in which
smooth extremal surfaces are anchored
just at the AdS boundary, we find that the dS portion of the extremal 
surface always remains in the dS static patches,
without entering the regions behind the dS horizon.
This is the reason why the exponential growth
of large bubbles fails to be detected by
 volume complexity.

As proposed in \cite{Freivogel:2005qh},
  the boundary dual of a large bubble configuration
should be a a quantum field theory in a mixed state.
This observation can be justified as follows.
For time-reversal symmetric bubbles,  eq. (\ref{mu-zero}) implies that
\beq
\l \mu \leq \l \mu_0 \leq 1 \, ,
\eeq
which is equivalent to
\beq
S_{\rm BH} \leq S_{\rm dS} \, ,
\label{entropie-dS-vs-AdS}
\eeq
where $S_{\rm BH}$
is the entropy of the external AdS black hole  
and $S_{\rm dS}$ is the entropy of
 the internal dS static patch  $S_{\rm dS}$,  \emph{i.e.}
\beq
S_{\rm BH}= 2 \pi \sqrt{\mu} \, , \qquad 
S_{\rm dS}= \frac{2 \pi}{\sqrt{\l}} \, .
\eeq
For  large bubbles, the interior region contains a portion of the
dS horizon, so the number of degrees of freedom 
accessible from the AdS boundary is less than the number of degrees of freedom
of the internal dS region.  We are  then forced to interpret
 large bubbles configurations in AdS/CFT 
as gravity duals of a density matrix,
obtained by tracing over a part of the degrees of freedom
of the dS region inside the bubble.

In the context of static patch horizon holography in dS 
  \cite{Susskind:2021esx,Shaghoulian:2021cef,Shaghoulian:2022fop}, 
we can conjecture that the purification of the dual mixed
state is a generalization of thermofield double state \cite{Maldacena:2001kr}  
in which the CFT living at the boundary of AdS and the
quantum system living on the stretched dS horizon
are entangled. This suggests another   way to apply the CV proposal
in the case of very large bubbles: we can anchor 
extremal surfaces both 
 at the AdS boundary and at the  dS static patch stretched horizon.
In this case, we find that extremal surfaces cross the dS horizon
and tend to bend towards the future infinity.
The complexity growth is hyperfast and diverges in finite time,
as in the dS case.

The paper is organized as follows. In section \ref{section-setup}
we describe the bubble setup and the thin wall approximation.
In section \ref{section-volume-functional} we present the 
equations of the extremal surfaces both in AdS and in dS and
we discuss the refraction law for the extremal surface on the bubble.
In section \ref{section-complexity-smooth-extremal surface} we study 
volume complexity, using the prescription in which the extremal
surfaces are anchored just at the AdS boundary and are smooth
both in the AdS and in the dS regions of the spacetime.
In section \ref{section-complexity-stretched-horizon}
we study volume complexity for very large bubbles, using extremal
surfaces which extend between the AdS boundary
and the dS stretched horizon.
We conclude in section \ref{section-conclusions}.
Several technical details are presented in appendices.


\section{Theoretical setup}
\label{section-setup}

We consider a spherically symmetric dS$_{3}$ bubble
inside an asymptotically AdS$_{3}$ spacetime.
The  spacetime  metric is taken as follows
\bea
ds^2_{i,o}&=&(g_{i,o})_{\mu \nu} dx_{i,o}^{\mu} dx_{i,o}^{\nu} \nl
&=&  -f_{i,o}(r) \, dt_{i,o}^2+ \frac{dr^2}{f_{i,o} (r)} + r^2 d \theta^2 \, ,
\label{metric-zero}
\eea
where the subscripts $i$ and $o$ refer to the inside and outside regions, respectively.
The outside geometry is a BTZ  black hole \cite{Banados:1992wn} with
\beq
f_o(r)=r^2 -\mu \, ,
\label{f-BTZ}
\eeq
where the mass of the black hole is proportional to $\mu$.
For simplicity we set the AdS length $L=1$.
The inside geometry is a dS$_3$ spacetime with radius $r_{dS}=1/\sqrt{\l}$, namely
\beq
f_i(r)=1-\l \, r^2 \, .
\label{f-dS}
\eeq
Introducing the tortoise coordinate
\beq
r^*_{i,o}=\int \frac{d \tilde{r}}{f_{i,o}(\tilde{r}) } \, ,
\label{tortoise}
\eeq
we can define the light-cone coordinates $v,u$ as follows
\beq
v_{i,o}= t_{i,o} +r^*_{i,o}(r) \, , \qquad
u_{i,o}= t_{i,o} -r^*_{i,o}(r) \, .
\label{u-v-def}
\eeq
An explicit evaluation of the integral (\ref{tortoise}) gives
\bea
r^*_o(r)&=&\frac{1}{4 \sqrt{\mu} } \log \le \frac{r-\sqrt{\mu}}{r+\sqrt{\mu}} \ri^2 \, ,
\nl
r^*_i(r)&=&\frac{1}{4 \sqrt{\l}} \log \left(\frac{1+ r \sqrt{\l}}{1- r \sqrt{\l}} \right)^2 \, ,
\eea
where the integration constants are chosen in such a way that $r^*_o( \infty) = 0$
and $r^*_i(0)=0$.

When considering spacetime regions nearby the black hole
and the cosmological horizons, 
it is convenient to write the metric in Eddington-Finkelstein (EF) coordinates $\le v_{i,o},r \ri$ or $\le u_{i,o},r \ri$:
\bea
ds^2_{i,o} &=&-f_{i,o}(r) \, dv_{i,o}^2+2 dr \, dv_{i,o} + r^2 d \theta^2 
\nl 
&=& -f_{i,o}(r) \, du_{i,o}^2-2 dr \, du_{i,o} + r^2 d \theta^2  \, .
\nl
\label{EF-metrica-esplicita}
\eea
 In order to describe the maximally extended versions
of the  dS$_3$ and the BTZ spacetimes, it is
necessary to introduce two copies of the EF coordinates $u$ and $v$,
which we denote by $u_{\rm L},v_{\rm L}$ and $u_{\rm R},v_{\rm R}$,
where ${\rm L}$ and ${\rm R}$ stand for left and right, respectively.
Penrose diagrams for these geometries with constant 
$u,v$ lines are shown in figure \ref{penrose0}.
Our conventions for Penrose diagrams are discussed in appendix \ref{penrose-appe}.

\begin{figure*}
\includegraphics[scale=0.4]{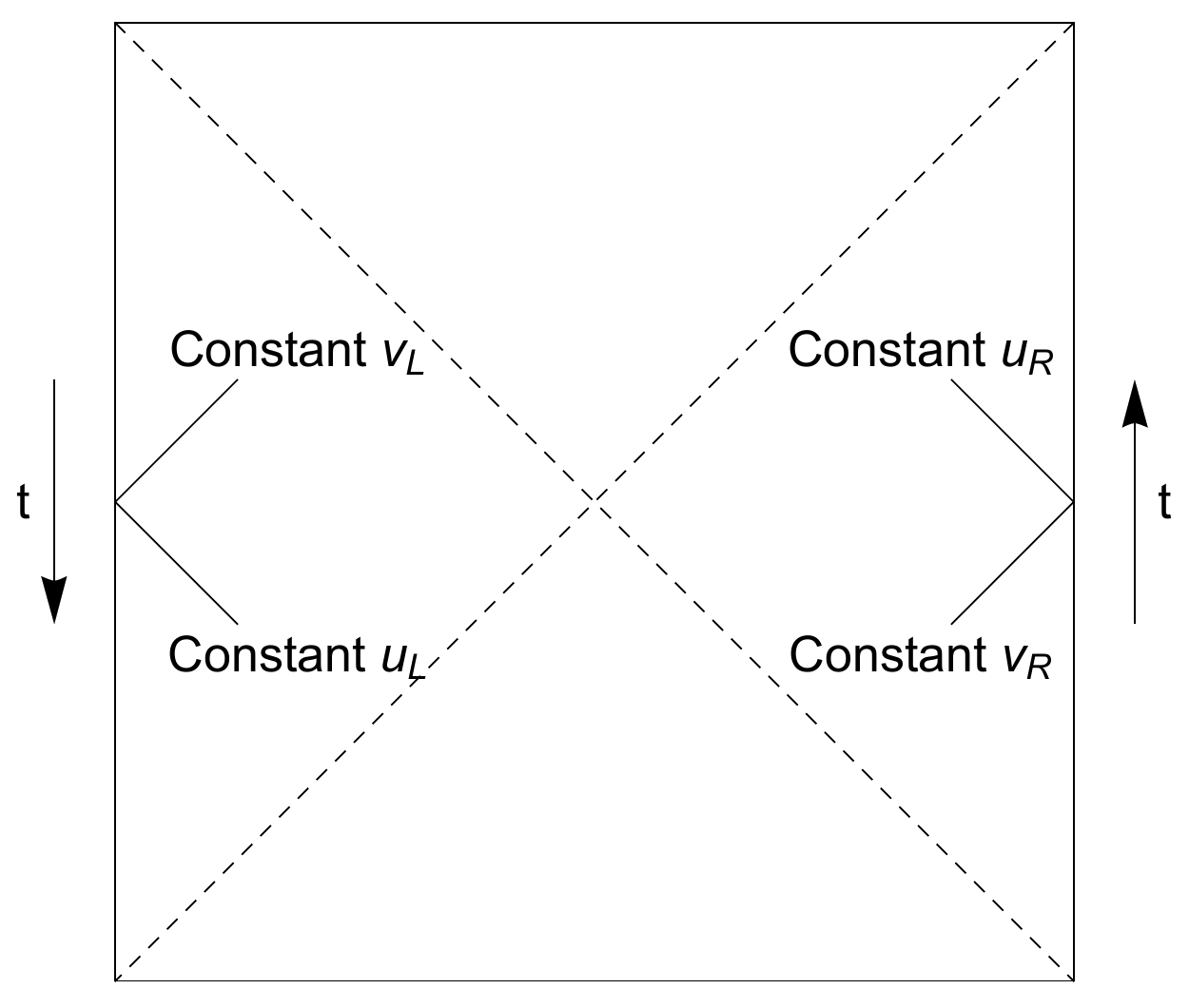}
\qquad
\includegraphics[scale=0.4]{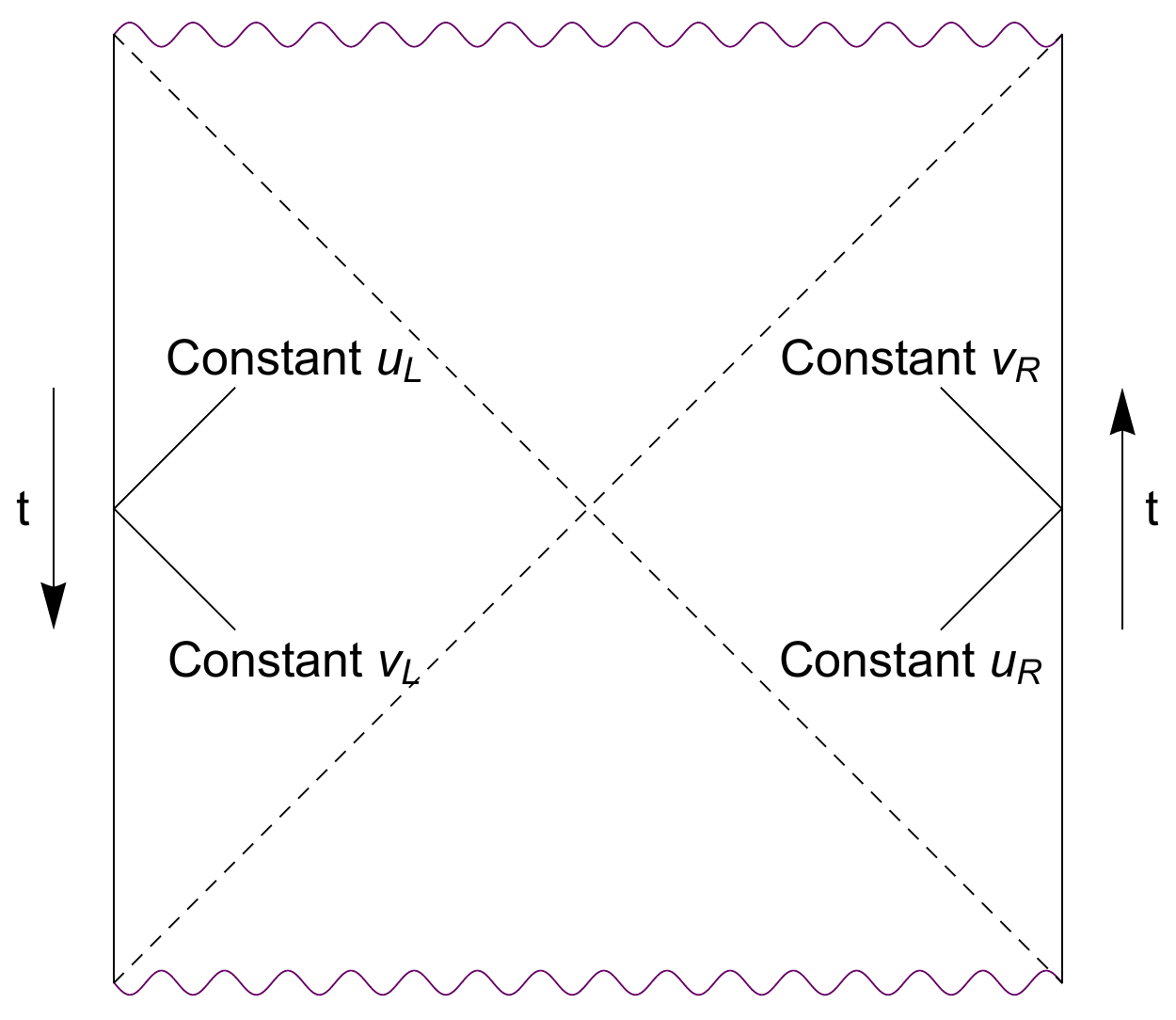}
\caption{ Penrose diagrams  for dS$_3$ 
spacetime (left) and for the  BTZ black hole (right).
The arrows show the directions of increasing coordinate $t$
on both the sides of the diagrams.
}
\label{penrose0}
\end{figure*}

\subsection{The domain wall}

The inside and the outside geometries are patched together along
a domain wall with negligible thickness,
 whose trajectory on each side of the spacetime is parametrized by
\beq
r=R(\tau) \, ,
\eeq
with $\tau$ the proper time measured on the domain wall itself.
We will denote by a dot $\dot{}$ the derivative with respect to the proper time $\tau$.

The equation of motion  for $R(\tau)$ follows from Israel's junction
conditions \cite{Israel:1966rt}
(see \cite{Blau:1986cw,Freivogel:2005qh,Poisson:2009pwt,Fu:2019oyc} for reviews),  \emph{i.e.}:
\begin{itemize}
\item The metric must be continuous across the wall.
The coordinate $r$ multiplies the metric of the transverse sphere $S^1$, so it is continuous.
Instead, the coordinate $t_{i,o}$ is in general discontinuous in passing from inside to outside.
\item The discontinuity in the extrinsic curvature $K_{a b}$ across the wall is fixed by the energy-momentum tensor.
For spherically symmetric geometries, it is customary to introduce the quantities
\beq
\b_{i,o}=(K^{\theta}_{\theta})_{i,o} \, R \, .
\eeq
The jump between $\b_i$ and $\b_o$ is 
\beq
\b_i - \b_o= \kappa \, R \, , 
\label{general-V-bubble-0}
\eeq
where in our case
\bea
\b_i &=& \pm \sqrt{\dot{R}^2+f_i(R)}  \, , 
\nl \b_o &=& \pm \sqrt{\dot{R}^2+f_o(R)} \, , 
\nl \kappa &=& 8 \pi G \, \s \, .
\label{general-V-bubble-1}
\eea
In eq. (\ref{general-V-bubble-1}), $\s$ is the domain wall tension
and $G$ is the Newton's constant.
The sign of $\b_{i,o}$ is positive if the coordinate $r$ increases as the wall is approached from the 
interior or as one moves away from the wall in the exterior, and negative in the opposite situations.
If both $\b_{i,o}$ have the same sign, $r$ is monotonic near the wall. 
If  $\b_{i,o}$ have different signs, $r$ is locally extremised at the location of the wall.
\end{itemize}

Squaring twice eq. (\ref{general-V-bubble-0}), independently of the choices
of signs in eq. (\ref{general-V-bubble-1}), the equation of motion 
can be expressed as
\beq
\dot{R}^2+V (R)=0 \, , 
\label{general-V-bubble}
\eeq
where the second term, playing the role of an effective potential, is 
\beq
V(R)=f_o(R) - \frac{(f_i(R) -f_o(R) -\kappa^2 \, R^2)^2}{4 \, \kappa^2 \, R^2}  \, .
\label{general-V-bubble-potenziale}
\eeq

\begin{figure*}
\includegraphics[scale=0.3]{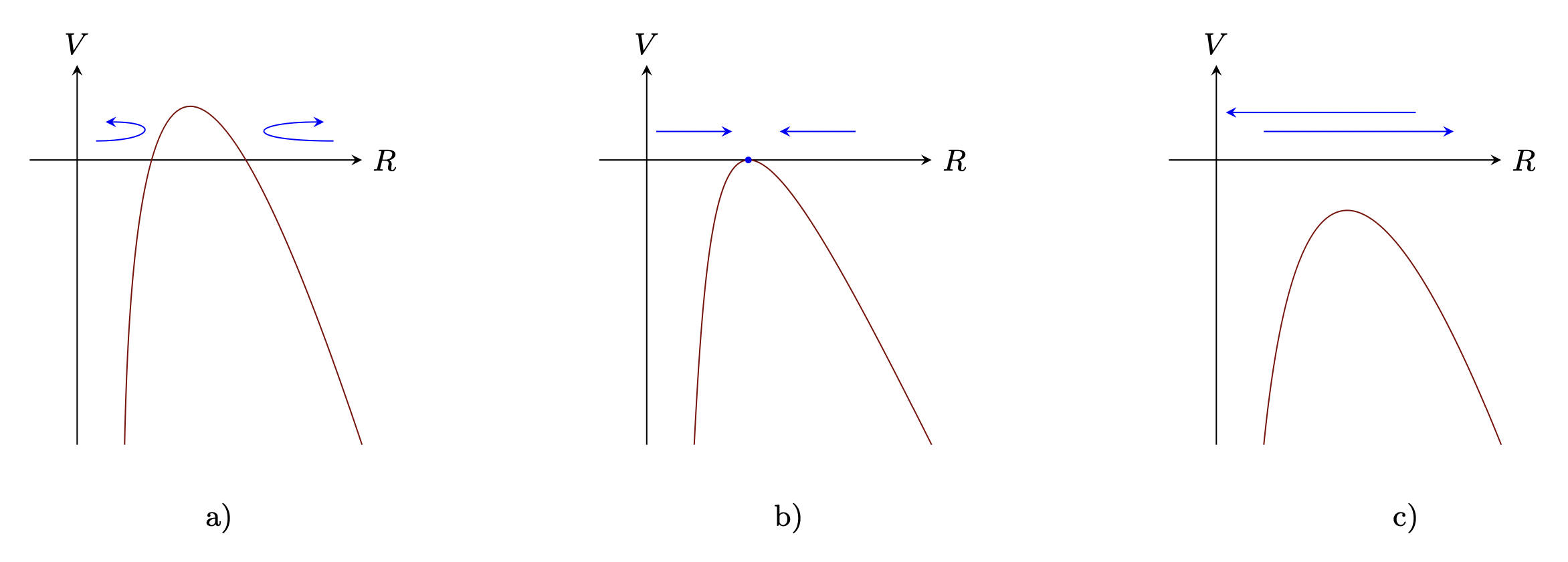}
\caption{ Qualitative plots of the effective potential $V(R)$ in eq. (\ref{V-di-erre-esplicito}) 
	with positive (plot a), vanishing (plot b) and negative (plot c) maximum. 
	Blue arrows represent the possible evolution of the bubble radius
	$R(\tau)$.}
\label{V-effettivo}
\end{figure*}

Specializing the general expression (\ref{general-V-bubble-potenziale}) to
 eqs. (\ref{f-dS}) and (\ref{f-BTZ}), we find
\beq
V(R) =-A  \, R^2 
+B
- \frac{C}{R^2} \, ,
\label{V-di-erre-esplicito}
\eeq
where
\bea
A&=&\frac{(\l+\kappa^2-1)^2 + 4 \l }{4 \kappa^2} \, , \nl
B&=&\frac{1+\kappa^2+\l +\mu-\kappa^2 \mu +\l \mu}{2 \kappa^2} \, , \nl
C&=& \frac{(1+\mu)^2}{4 \kappa^2} \, .
\eea
Note that  $A>0$ and $C>0$.
It is also convenient to introduce
\beq
\b=\frac{B}{A} \, , \qquad \gamma=\frac{C}{A} \, .
\eeq
Depending on the values of parameters, we can have three 
physically different situations:
\begin{enumerate}
\item[a)] The maximum of $V(R)$ is positive.
In this case, the radius of the bubble 
as a function of the bubble proper time $\tau$
has either a maximum or a minimum .
\item[b)]  The maximum of $V(R)$ is exactly zero. In this case, we have an unstable static bubble solution, besides other solutions which approach the maximum from both sides with asymptotically zero velocity $\dot{R}$.
\item[c)]  The maximum of $V(R)$ is negative.
In this case, the radius of the bubble monotonically contracts
or expands without any maximum or minimum.
\end{enumerate}
The qualitative behavior of $V(R)$ in these three cases is shown in figure \ref{V-effettivo}.

Time-reversal invariance selects a potential $V(R)$ 
with non-negative maximum, as in cases a) and b).
This requires 
\beq
\b^2 \geq 4 \gamma \, .
\label{condizione-simmetria-inversione}
\eeq
For time-reversal symmetric bubbles,
an explicit integration of eq. (\ref{general-V-bubble}) gives
\beq
R(\tau)=\sqrt{\frac{\b}{2} \mp \frac{\sqrt{\b^2 - 4 \gamma}}{2} \cosh ( 2 \sqrt{A}  \tau ) } \, .
\label{R-di-tau-simmetrico}
\eeq
The $-$ sign solution corresponds to a small bubble 
(the bubble has maximal radius at $\tau=0$
and collapses in a finite proper time),
whereas the $+$ sign solution describes a large bubble 
(the bubble has minimal radius at $\tau=0$
and expands forever).

The condition in eq. (\ref{condizione-simmetria-inversione}) is 
 is satisfied for $0 < \mu \leq \mu_0$, where 
the limiting value $\mu=\mu_0$ is defined in eq. (\ref{mu-zero})
and corresponds to a static bubble.

\begin{figure*}
\includegraphics[scale=0.6]{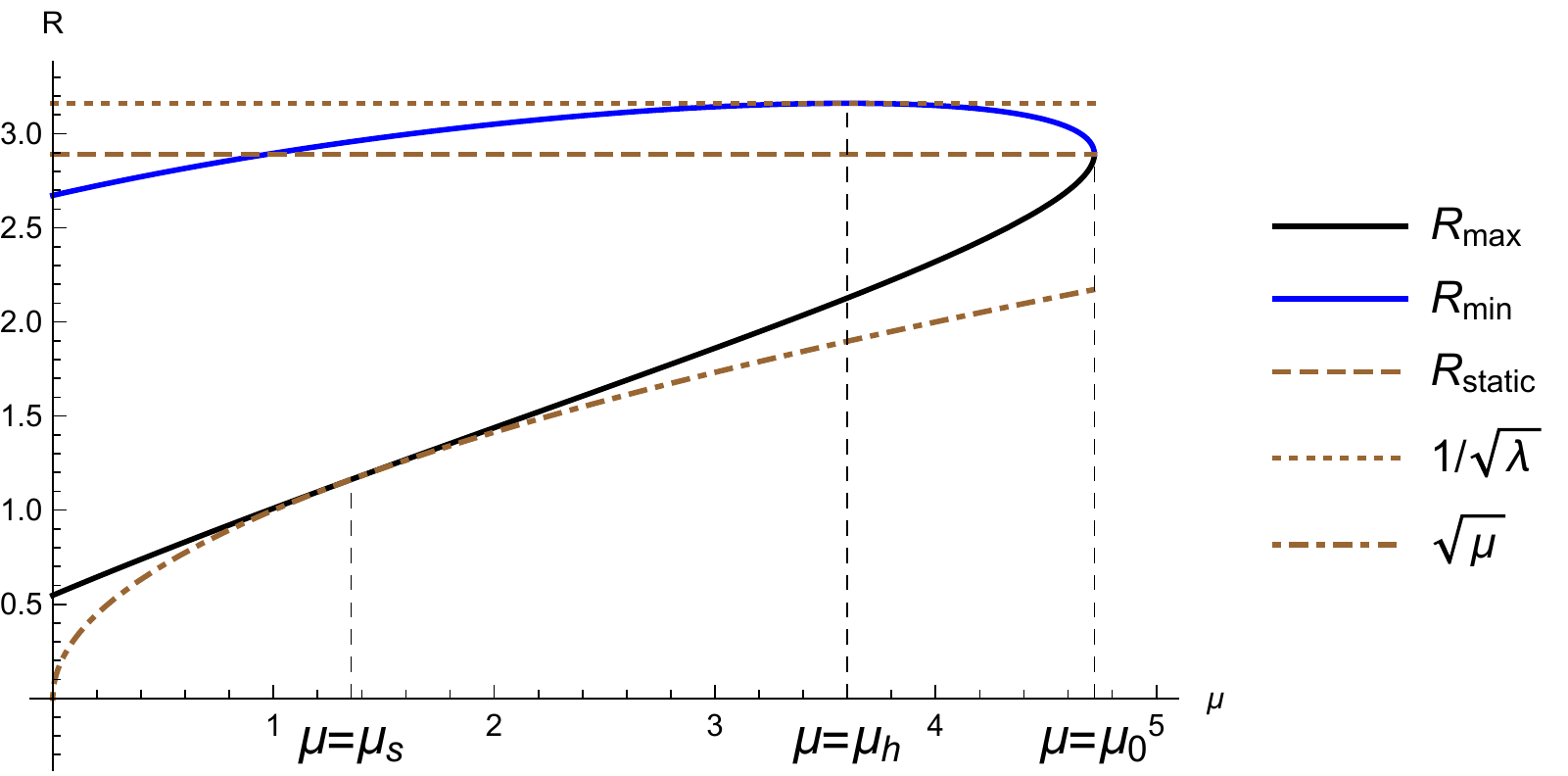}
\caption{ Illustrative plot of $R_{\max}$ and $R_{\min}$ 
as functions of $\mu$, for $\l=0.1$ and $\kappa=0.8$.}
\label{R-max-min}
\end{figure*}

For a given choice of the parameters $\l,\mu, \kappa$ with $\mu<\mu_0$
it is possible to have both a contracting and an expanding solution.
The maximal radius for the contracting bubble is
\beq
R_{\max} (\mu)=\sqrt{\frac{\b}{2} - \frac{\sqrt{\b^2 - 4 \gamma}}{2}  } \, ,
\eeq
while the minimal radius for the expanding bubble is
\beq
R_{\min}(\mu) = \sqrt{\frac{\b}{2} + \frac{\sqrt{\b^2 - 4 \gamma}}{2}  } \, .
\eeq
Both $R_{\rm max}$ and $R_{\rm min}$ are 
 real positive numbers for $0 < \mu \leq \mu_0$.
 The values of  $R_{\rm max}$ and $R_{\rm min}$ satisfy the following constraints:
 \begin{itemize}
 \item $R_{\rm max}  \leq R_{\rm min}$, with the equality saturated for $\mu=\mu_0$.
 \item $R_{\min} \leq 1/\sqrt{\l}$, with the equality saturated for the special value $\mu=\mu_h$,
see eq. (\ref{mu-h}).
 \item $R_{\max} \geq \sqrt{\mu}$
with the equality saturated for the special value $\mu=\mu_s$,
see eq. (\ref{mu-s}).
 \end{itemize}
In figure \ref{R-max-min} we show a  plot of $R_{\rm max}$ and 
$R_{\rm min}$ as functions of $\mu$, for a fixed value of $\l$ and $\kappa$.

In order to specify the solution, we should also determine the time coordinates $t_{i,o}$ on the surface of the bubble,
both in the inside and in the outside regions.
Such time coordinates, which we denote by $T_{i,o} (\tau)$,
are specified by the equation
\beq
\dot{T}^2_{i,o} =\frac{1}{f_{i,o}(R)} \le 1+\frac{\dot{R}^2}{f_{i,o}(R)} \ri \, ,
\label{T-di-tau-eq}
\eeq
following from the normalization of the bubble velocity vector $w^\a$
\beq
w^\a w_\a=-1 \, , \qquad
w^\a=(\dot{T},\dot{R},0) \, .
\eeq
Combining the equation of motion (\ref{general-V-bubble}) with eq. (\ref{T-di-tau-eq}) we find
\begin{equation}
	{dT_{i,o} \over dR} = \pm {\sqrt{f_{i,o}(R)-V(R)} \over f_{i,o}(R) \sqrt{-V(R)}}\,.
	\label{timecoreq}
\end{equation}
Plugging the explicit expressions for $f_i$ and $f_o$ in eq. (\ref{timecoreq}),
we get 
\bea
\frac{dT_i}{d R} &=& \pm \frac{1}{2 \kappa}  \frac{1+ \mu - R^2 (1+\l-\kappa^2)}{(1- \l R^2)\sqrt{A R^4 - B R^2 +C} } \, ,
\nl
\frac{dT_o}{d R} &=& \pm \frac{1}{2 \kappa}  \frac{1+ \mu - R^2 (1+\l+\kappa^2)}{(R^2-\mu)\sqrt{A R^4 - B R^2 +C} } \, .
\nl
\label{eq-diff-Ti-To}
\eea
Time-reversal solutions to these equations 
are obtained by imposing the boundary condition
\beq
T_{i,o}(R_{\max})=0 \, \qquad {\rm or} 
\qquad T_{i,o}(R_{\min})=0 
\eeq
in the collapsing and the expanding case, respectively.

\subsection{Static bubbles}

A static bubble solution is realized for $\b^2=4 \gamma$,
or equivalently $\mu=\mu_0$, see eq. (\ref{mu-zero}).
The radius of the bubble is thus
\beq
R_{\rm static}=R_{\min}(\mu_0)=R_{\max}(\mu_0) \, .
\eeq
The matching condition between the $T_{i,o}$ coordinates is
\beq
\frac{d T_i}{d T_o}
 =\pm \sqrt{\frac{f_o(R_{\rm static})}{f_i(R_{\rm static})}}=\pm \mu_0 \, .
   \label{matching-tempi-bubble-statica}
\eeq

\subsection{Small bubbles}

 In this case, eq. (\ref{eq-diff-Ti-To}) admits  a smooth solution for
  $T_i$.  On the other hand, the trajectory is not smooth in the
 coordinate $T_o$, which should then be replaced  by 
an EF coordinate, see eq. (\ref{u-v-def}).
Referring to the Penrose
diagrams for the collapsing bubble sketched 
in figures \ref{penrose-collapsing-bubble-1} and \ref{penrose-collapsing-bubble-2},
we distinguish between two cases:
\begin{itemize}
\item {\bf Very small bubble, $0 < \mu < \mu_s$.} The initial position of the bubble
$R=R_{\rm max}$ is on the same side of the AdS boundary. The smooth coordinates
that should be used are $v_{\rm R}$ for the black hole (or $u_{\rm R}$ for the white hole).
Denoting by $U_{{\rm R}, o}$ and $V_{{\rm R}, o}$
the right EF coordinates on the bubble surface in BTZ spacetime,
the equations of motion are
\bea
\frac{dV_{{\rm R},o}}{d R} &=&  -\frac{1}{2 \kappa}  \frac{1+ \mu - R^2 (1+\l+\kappa^2)}{(R^2-\mu)\sqrt{A R^4 - B R^2 +C} } 
\nl
  &+&  \frac{1}{R^2-\mu}  \, ,  \nl
\frac{dU_{{\rm R},o}}{d R} &=&  \frac{1}{2 \kappa}  \frac{1+ \mu - R^2 (1+\l+\kappa^2)}{(R^2-\mu)\sqrt{A R^4 - B R^2 +C} } 
 \nl
&-& \frac{1}{R^2-\mu} 
\, .
\label{eq-diff-VU-ssmall-lambda}
\eea
\item {\bf Not so small bubble,  $\mu_s < \mu < \mu_0$.} In this case we should use
$u_{\rm L}$ for the black hole (or $v_{\rm L}$ for the white hole).
Denoting by $U_{{\rm L}, o}$ and $V_{{\rm L}, o}$
the left EF coordinates on the bubble surface in BTZ spacetime,
the equations of motion are
\bea
\frac{dU_{{\rm L},o}}{d R}&=&-  \frac{1}{2 \kappa}  \frac{1+ \mu - R^2 (1+\l+\kappa^2)}{(R^2-\mu)\sqrt{A R^4 - B R^2 +C} } \nl
&-& \frac{1}{R^2-\mu}  \, ,
\nl
\frac{dV_{{\rm L}, o}}{d R}&=&
 \frac{1}{2 \kappa}  \frac{1+ \mu - R^2 (1+\l+\kappa^2)}{(R^2-\mu)\sqrt{A R^4 - B R^2 +C} } \nl
&+&\frac{1}{R^2-\mu}  \, .
\eea
\end{itemize}


\subsection{Large bubbles}

Large bubbles are examples of bags of gold \cite{Marolf:2008tx,Fu:2019oyc},
which are defined as spacetimes in which an eternal  black hole exterior is attached by an Einstein-Rosen bridge
 to an interior which is a portion of Friedman-Lemaitre-Robertson-Walker (FLRW) cosmology
 with an infinite spacetime volume.
In these geometries,  the entropy of the interior 
 can exceed the exterior Bekenstein-Hawking entropy, see eq. (\ref{entropie-dS-vs-AdS}),
 so the bulk states cannot be put in correspondence with
 the CFT states of the dual field theory.
 It has been suggested
that the Bekenstein-Hawking  entropy does not in general count all the states
inside the black hole, but only those which are distinguishable from the outside
\cite{Jacobson:1999mi,Marolf:2008tx}.
In particular, large bubble solutions has been proposed to
provide the holographic dual of a density matrix \cite{Freivogel:2005qh}.

Equation (\ref{eq-diff-Ti-To}) for the outside time  $T_o$  gives a smooth solution,
while $T_i$ is singular because the dS horizon is crossed 
at some time by the bubble.
We then pass to EF coordinates, see eq. (\ref{u-v-def}).
With reference to the Penrose
diagrams for the expanding bubble sketched 
in figures \ref{penrose-expanding-bubble-2}
and \ref{penrose-expanding-bubble-1}, 
we discriminate between two cases:
\begin{itemize}
\item {\bf Very large bubble, $0 < \mu < \mu_h$.} The initial position of the bubble $R=R_{\rm min}$ is in the right static patch. 
Denoting by $U_{{\rm R}, i}$ and $V_{{\rm R}, i}$
the right EF coordinates on the bubble surface in dS,
the equations of motion are
\bea
\frac{dV_{{\rm R}, i}}{d R} &=& 
  \frac{1}{2 \kappa}  \frac{1+ \mu - R^2 (1+\l-\kappa^2)}{(1- \l R^2)\sqrt{A R^4 - B R^2 +C} } \nl 
&+&\frac{1}{1-\l R^2} \, ,
\nl
\frac{dU_{{\rm R}, i}}{d R} &=&
 - \frac{1}{2 \kappa}  \frac{1+ \mu - R^2 (1+\l-\kappa^2)}{(1- \l R^2)\sqrt{A R^4 - B R^2 +C} } \nl
&-&\frac{1}{1-\l R^2} \, .
\eea
\item {\bf Not so large bubble,  $\mu_h < \mu < \mu_0$.} The initial
position of the bubble $R=R_{\rm min}$ is in the left static patch.
Denoting by $U_{{\rm L}, i}$ and $V_{{\rm L}, i}$
the left EF coordinates on the bubble surface in dS,
the equations of motion are
\bea
\frac{dU_{{\rm L}, i}}{d R}&=&
\frac{1}{2 \kappa}  \frac{1+ \mu - R^2 (1+\l-\kappa^2)}{(1- \l R^2)\sqrt{A R^4 - B R^2 +C} } \nl
&-& \frac{1}{1-\l R^2}  \, ,
\nl
\frac{dV_{{\rm L}, i}}{d R}&=&
-\frac{1}{2 \kappa}  \frac{1+ \mu - R^2 (1+\l-\kappa^2)}{(1- \l R^2)\sqrt{A R^4 - B R^2 +C} } \nl
&+&\frac{1}{1-\l R^2}  \, .
\eea
\end{itemize}


\section{Volume functional}
\label{section-volume-functional}

According to the CV conjecture \cite{Stanford:2014jda}, 
complexity of the boundary state is proportional to the volume
of a maximal codimension-one surface anchored at the given boundary time.
The volume complexity $C_V$ is usually normalized as
\beq
C_V = \frac{\mathcal{V}}{G L} \, ,
\eeq
where $\mathcal{V}$ is the volume of the maximal slice, $G$ the Newton's constant and $L$
the AdS radius.
In this section we discuss the volume of extremal surfaces in both the AdS and the dS parts of the geometry.
Then, we address  the matching condition on the domain wall.

Due to spherical symmetry, the extremal surface can be parameterized as
\beq
r= r( l)  \,  , \qquad t=t(l) \, ,
\label{parametrizzazione-superficie-estremale}
\eeq
where $l$ is a single-valued coordinate along the surface.
Since the coordinate $t$ is singular nearby the horizons,
it is useful to write the volume functional in both the versions of the EF coordinates $u$ and $v$:
\bea
\mathcal{V}_{i,o} &=& 2 \pi \int  \mathcal{L} \,  d l 
 \, , \nl
  \mathcal{L}&=& r \, \sqrt{-f_{i,o} (v'_{i,o})^2 +2 r' v'_{i,o}} \nl
&=& r \, \sqrt{-f_{i,o} (u'_{i,o})^2 -2 r' u'_{i,o}} 
   \, ,
 \label{volume-functional}
\eea
where $'$ denotes a derivative with respect to $l$.
The extremal surface is characterized by a conserved quantity $P_{i,o}$, which in the $(v,r)$ and in the $(u,r)$ coordinates reads 
\bea
P_{i,o}&=&\frac{\p \mathcal{L}}{\p v'_{i,o}}=\frac{r \, (-f_{i,o} v'_{i,o} +r') }{\sqrt{-f_{i,o} (v'_{i,o})^2 +2 r' v'_{i,o}} } \, ,
\nl
P_{i,o}&=&\frac{\p \mathcal{L}}{\p u'_{i,o}}=\frac{r \, (-f_{i,o} u'_{i,o} -r') }
{\sqrt{-f_{i,o} (u'_{i,o})^2 -2 r' u'_{i,o}} } \, , \nl
\label{conservata}
\eea
respectively.

\begin{figure*}[!]
\includegraphics[scale=0.5]{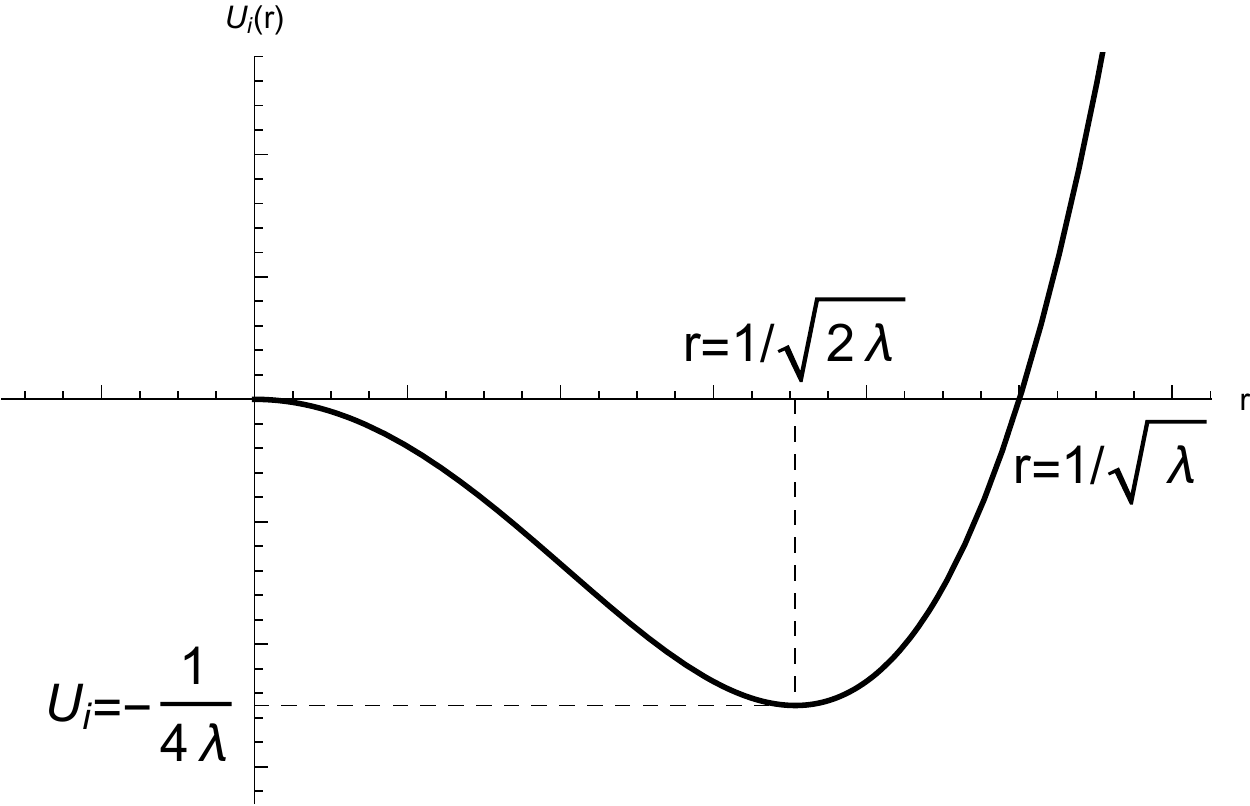}
\qquad
\includegraphics[scale=0.5]{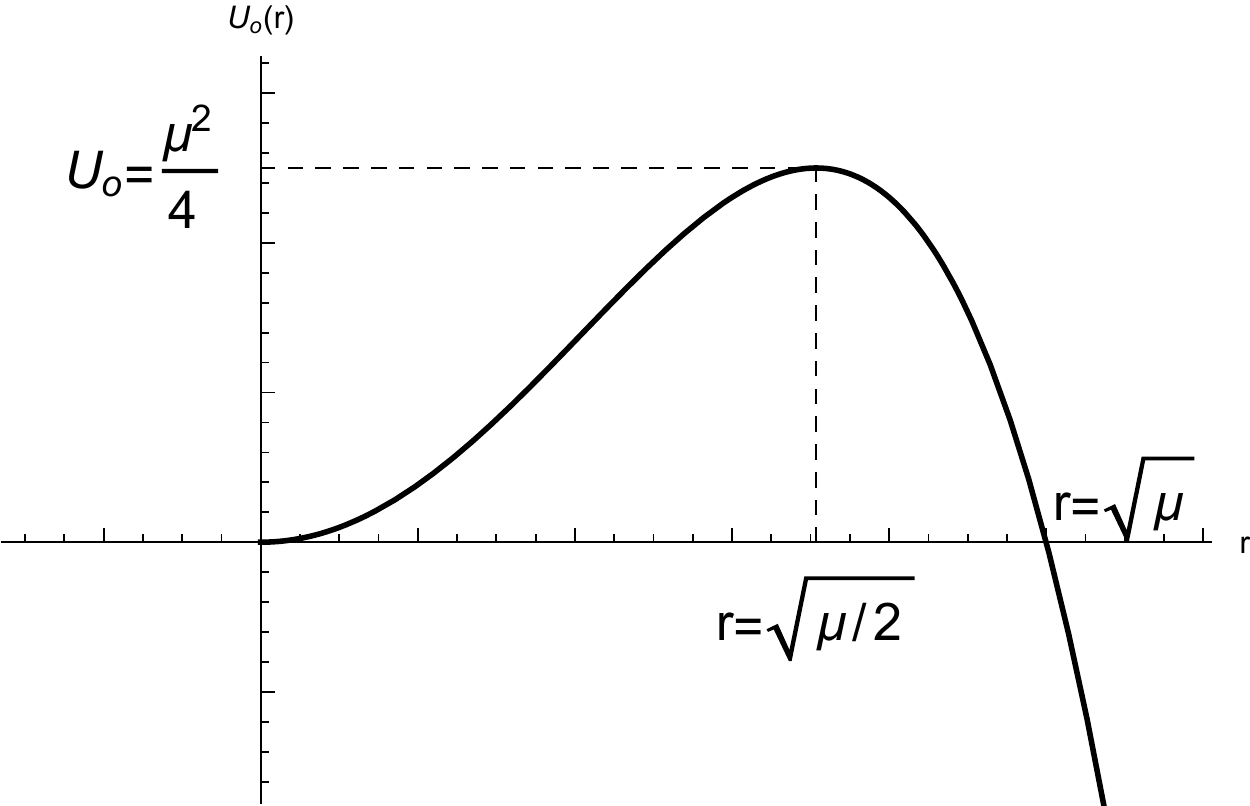}
\caption{ Effective potential for the extremal surface in the dS region (left)
and in the AdS one (right). }
\label{Uio-figure}
\end{figure*}

The volume functional in eq. (\ref{volume-functional}) is invariant
under reparameterization in $l$.  To fix this gauge freedom, 
it is convenient to impose the conditions
\bea
&& \sqrt{-f_{i,o} (v'_{i,o})^2 +2 r' v'_{i,o}} =r 
\qquad {\rm or} \nl
&& \sqrt{-f_{i,o} (u'_{i,o})^2 -2 r' u'_{i,o}} = r \, ,
\label{gauge-choice}
\eea
in such a way that the volume functional becomes 
\beq
\mathcal{V}_{i,o}
= 2 \pi \int  r^2 \,  d l \, .
 \label{gauge-fixing-generale}
\eeq
With this gauge choice, the conserved quantity $P_{i,o}$ can be expressed as follows
\bea
P_{i,o}&=& -f_{i,o} v'_{i,o} +r' \, , \qquad
v'_{i,o}=\frac{r'-P_{i,o}}{f_{i,o}} \, ,
\nl
P_{i,o}
&=&-f_{i,o} u'_{i,o} -r'  
\, , \qquad
u'_{i,o}=\frac{-r'-P_{i,o}}{f_{i,o}} \, . \nl
\label{vprimo-da-P}
\eea
By inserting this back into the gauge constraint, we get
\bea
&&(r')^2  + U_{i,o}(r) =P_{i,o}^2 \, , \qquad U_{i,o}(r)=-f_{i,o}(r) \,  r^{2} \, ,
\nl
&& r'=\pm \sqrt{P_{i,o}^2  +f_{i,o}(r) \,  r^{2}} \, ,
\label{sistema-gauge-fixed}
\eea
which is valid in both the EF coordinate systems $(v,r)$ and $(u,r)$.
The conserved quantity $P_{i,o}$ can also be written as
\beq
P_{i,o} =-f_{i,o} \, t'_{i,o}  \, .
\label{Pio-useful}
\eeq
Note that  $U_{i,o}(r)$ in eq. (\ref{sistema-gauge-fixed}) 
can be interpreted as effective potentials, see figure \ref{Uio-figure}
for qualitative plots.

The extremum of $r(l)$ represents a turning point $r_t$ 
for the extremal surface.
We can have a turning point both in the external BTZ region
and in the internal dS one:
\begin{itemize}
\item In the external asymptotically AdS region, $r_{t,{\rm AdS}}$ is a minimum of $r(l)$, given by
\beq
r_{t,{\rm AdS}}=\sqrt{\frac{\mu + \sqrt{\mu^2-4 P_o^2} }{2} }\, .
\label{turning-point}
\eeq
For a turning point to exist, it is necessary to require
\beq
P_o^2 \leq {P}_{\rm max}^2=\frac{\mu^2}{4} \, .
\label{P-critico}
\eeq
\item
In the internal dS region, $r_{t,{\rm dS}}$ is a maximum of $r(l)$, 
given by
\beq
r_{t,{\rm dS}}=\sqrt{\frac{1+\sqrt{1+4 \l \, P_i^2}}{2 \l} } \, .
\label{turning-point-dS}
\eeq
\end{itemize}

The difference in the time coordinate $t$
between two points on the extremal surface can be expressed as
\beq
t_{i,o}(r_2)-t_{i,o}(r_1)  =
 \mp \int^{r_2}_{r_1}   \frac{P_{i,o}}{f_{i,o} \, \sqrt{P_{i,o}^2  +f_{i,o}(r) \,  r^{2}}}  dr  \, ,
\label{tempo-A}
\eeq
where the $-$ sign should be chosen for 
a parameterization with $r'(l)>0$,
while the $+$ sign for a parameterization with $r'(l)<0$.
In the integral in eq. (\ref{tempo-A}), 
the Cauchy principal value prescription should be used
when crossing the horizon.

With the convention $r_2>r_1$, the volume of the extremal surface reads
\beq
\mathcal{V}_{i,o}  =  2 \pi \int_{r_1}^{r_2} \,\frac{ r^{2} }{\sqrt{P_{i,o}^2  +f_{i,o}(r) \,  r^{2}} }\, d r \, .
 \label{volume-gauge-fixed-v-var}
\eeq

\subsection{A refraction law for the extremal surface}

To determine the codimension-one extremal surfaces,
we solve
eqs. (\ref{sistema-gauge-fixed}) and (\ref{vprimo-da-P})
both in the interior and in the exterior of the bubble. 
Then,
we match the two solutions on top of the domain wall,  
imposing that the total volume is extremal.
Physically, the extremal surface is somehow "refracted" by the
domain wall.
In appendix \ref{appendice-snell}, by introducing a coordinate system
which describes both the interior and the exterior of the bubble in terms of
the same time coordinate $t_i$, we derive the refraction condition
in the thin wall approximation.

By spherical symmetry, it is not restrictive to focus just on 
the time and radial coordinates. 
We denote by $x^\mu_{i,o}(l)$ the coordinates of the extremal codimension-one
surface and by $X^\mu_{i,o}(\tau)$ the trajectory of the domain wall.
We then introduce the tangent vector to the extremal 
surface $\frac{d x^\mu_{i,o}}{d l} $ and the velocity vector of the domain wall $\frac{d X^\mu_{i,o}}{ d \tau}$, namely
\bea
\frac{d x^\mu_{i,o}}{d l} &=& (t'_{i,o}(l) , r'_{i,o}(l)) \, , \nl
\frac{d X^\mu_{i,o}}{ d \tau} &=& (\dot{T}_{i,o}(\tau), \dot{R}(\tau)) \, .
\label{derivatozze}
\eea
The matching condition on top of the domain wall is
\beq
(g_i)_{\mu \nu} \frac{d x^\mu_i}{d l} \frac{d X^\nu_i}{d \tau}= 
(g_o)_{\mu \nu} \frac{d x^\mu_o}{d l} \frac{d X^\nu_o}{d \tau} \, .
\label{snell-law-covariante}
\eeq
The details of the derivation are in appendix \ref{appendice-snell}.
A similar result was derived  in \cite{Iglesias-Zemmour} for geodesics.

Given an extremal surface intersecting the domain wall at some value
of the radial coordinate $R$, we denote by $\rho_{i,o}(R)$
the value of $r'_{i,o}(l)$ computed at the intersection,  \emph{i.e.}
\beq
\rho_{i,o}(R)=  r'_{i,o}(l_0)  \,  \qquad {\rm where} \qquad
r_{i,o}(l_0)=R \, .
\eeq
 By means of eq. (\ref{Pio-useful}),
  we can write the matching condition (\ref{snell-law-covariante}) as
\beq
P_i  \frac{ d {T}_i}{dR} +\frac{\rho_i (R) }{f_i (R)}
=P_o \frac{ d {T}_o}{dR} +\frac{\rho_o(R) }{f_o(R)} \, .
  \label{snell-law-generic-3}
\eeq
The setup with an extremal surface crossing 
a null shell of matter with negligible thickness
was studied in \cite{Balasubramanian:2011ur} for geodesics
and in \cite{Chapman:2018dem,Chapman:2018lsv} for codimension-one surfaces.
This case formally corresponds to a domain wall moving at the speed of light and the result is consistent with eq. (\ref{snell-law-generic-3}).


\section{Complexity from smooth extremal surfaces}
\label{section-complexity-smooth-extremal surface}

The conservative way to apply the CV conjecture in
asymptotically AdS geometries with an internal dS bubble is to consider
extremal codimension one surfaces which are anchored at some given time $t_b$
at the AdS boundary and which are smooth in the interior.
These surfaces lie partially in the dS and partially in the 
asymptotically AdS parts of the geometry.
By convention, we choose $l$ in eq. (\ref{parametrizzazione-superficie-estremale})
to be positive
and to vanish at the center
of the dS static patch.

In order to avoid a singularity at $r=0$ in the dS interior, we must
impose the condition $P_i=0$.
This can be checked as follows.
Using eqs. (\ref{Pio-useful}) and (\ref{sistema-gauge-fixed}),
the induced metric on the extremal surface in the gauge 
(\ref{gauge-choice}) is
\beq
d \tilde{s}_i^2= r^2(l)  \le d l^2 +   d \theta^2  \ri  \, .
\eeq
From eq. (\ref{sistema-gauge-fixed}),  $r(l)$ can have two  qualitatively
 different behaviors nearby $l \to 0$:
\begin{itemize}
\item For $P_i=0$,  $r(l) \approx r'(l)$, so the induced metric
\[
d \tilde{s}_i^2 \approx dr^2+r^2 d \theta^2 
\]
 is smooth at $l \to 0$, with $0 \leq  \theta \leq 2 \pi$.
\item For $P_i \neq 0$,   $r^2(l) \approx P_i^2 l^2 $, so the
induced metric 
\[
d \tilde{s}_i^2 \approx P_i^2( l^2 d l^2+l^2 d \theta^2) 
\]
 has a singular scalar curvature at $l \to 0$.
\end{itemize}
A similar property holds in the case of AdS Vaidya spacetime,
see  \cite{Chapman:2018dem}.

\begin{figure}
\includegraphics[scale=0.4]{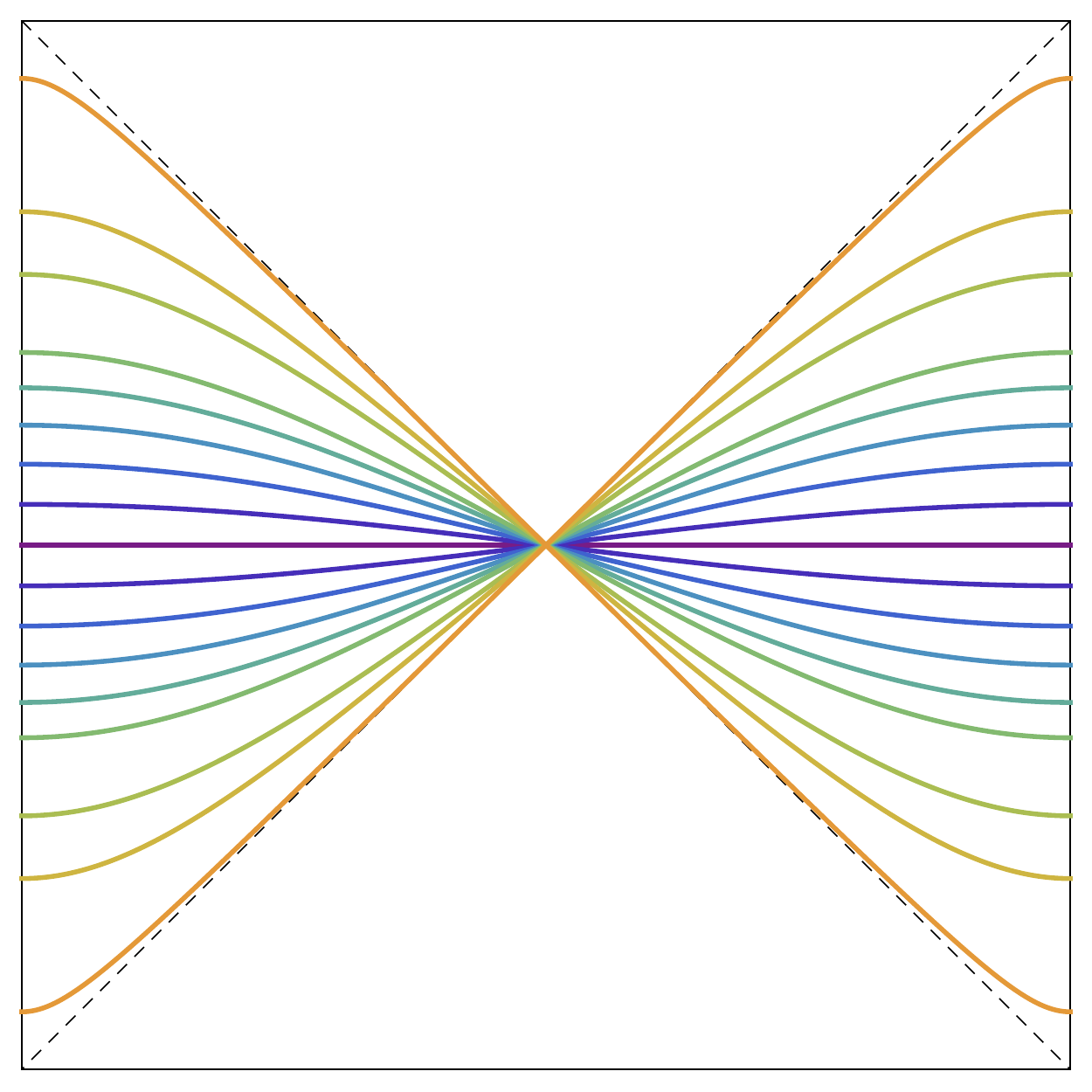}  
\caption{ 
Examples of extremal codimension one surfaces with $P_i=0$ in the dS Penrose diagram.
The interior part of the smooth extremal surfaces correspond to the portion 
of these surfaces inside the bubble.}
\label{dS-Pi-zero}
\end{figure}

Equation (\ref{Pio-useful}) implies that for $P_i=0$
 the extremal surface in the interior lies at constant $t_i$ coordinate.
Examples of these surface are shown in figure \ref{dS-Pi-zero}.

\subsection{Small bubbles}

For small bubbles, 
the inside geometry is just a region of the left static patch. 
Then, there is no turning point in the dS portion of the geometry, 
or equivalently $r_i'(l)>0$. 
As a direct consequence, we have  $\rho_i(R)>0$ 
everywhere on the domain wall.
With the condition $P_i=0$, the refraction law in eq. (\ref{snell-law-generic-3}), which must be implemented at
the bubble surface $r=R(\tau)$,  is
  \bea
&& \frac{R}{\sqrt{f_i(R)} }   
  =  \frac{ \rho_o(R) }{  f_o(R)} 
+  \frac{d T_o}{dR}  P_o \, , 
\nl
&&  P_o =\pm \sqrt{  \rho_o^2(R) -  f_o(R) \, R^{2}}
 \, .
   \label{P-matching}
  \eea
Depending on the sign of the product $\frac{d T_o}{dR}  P_o$,
two physically different solutions to the 
constraint (\ref{P-matching}) exist. 
  Let us fix for convenience the sign in  eq. (\ref{eq-diff-Ti-To}) as follows
    \bea
&& \frac{dT_o}{d R}= - \frac{1}{2 \kappa}  \frac{w_o(R)}{(R^2-\mu)\sqrt{A R^4 - B R^2 +C} } \, ,
\nl
&& w_o(R)=1+ \mu - R^2 (1+\l+\kappa^2) \, .
 \label{scelta-segno-To}
 \eea
Note that\footnote{This can be checked by the following properties:
a) the unique positive solution of the equation  $ w_o(R_{\rm max})=0$ in the variable $\mu$ is $\mu=\mu_s$; b) for $\mu=0$, $ w_o(R_{\rm max})>0$.}  
the quantity  $ w_o(R_{\rm max})$ is positive
 for $0 <\mu <\mu_s$
and negative for $\mu_s <\mu <\mu_0$.
In both cases, the sign choice in eq. (\ref{scelta-segno-To}) is such that, for $T_o \geq 0$,
the  bubble always moves
towards the upper direction of the Penrose diagram.
The configuration in which the bubble moves in the opposite direction can be recovered by a time reflection
$t \to -t$.
 
With the assumption of positive $P_o$,  eq. (\ref{P-matching}) has the following solution
\bea
\rho_o(R)=- \frac{R \left(\mu -1 +R^2 \left(\kappa ^2+\lambda -1\right)\right)}{2 \sqrt{1-\lambda  R^2}} \, .
\nl
      \label{rprimo-Po-positivo}
\eea
From a direct calculation,
we can check that $\rho_o(R_{\rm max})$ vanishes just for $\mu=\mu_s$.
Also, we have that $\rho_o(R_{\rm max})>0$ for $\mu \to 0$.
This shows that $\rho_o(R_{\rm max})$
is positive for $0 < \mu < \mu_s$
and negative for $\mu_s < \mu < \mu_0$.
In both cases, this sign is consistent with a refraction 
of the extremal surface through the domain wall.
For negative $P_o$, the solution to eq. (\ref{P-matching}) 
corresponds to a "reflection" of the extremal surface,
see appendix \ref{appe-Po-negativo}.
Thus, we discard this solution.

\begin{figure}
\includegraphics[scale=0.6]{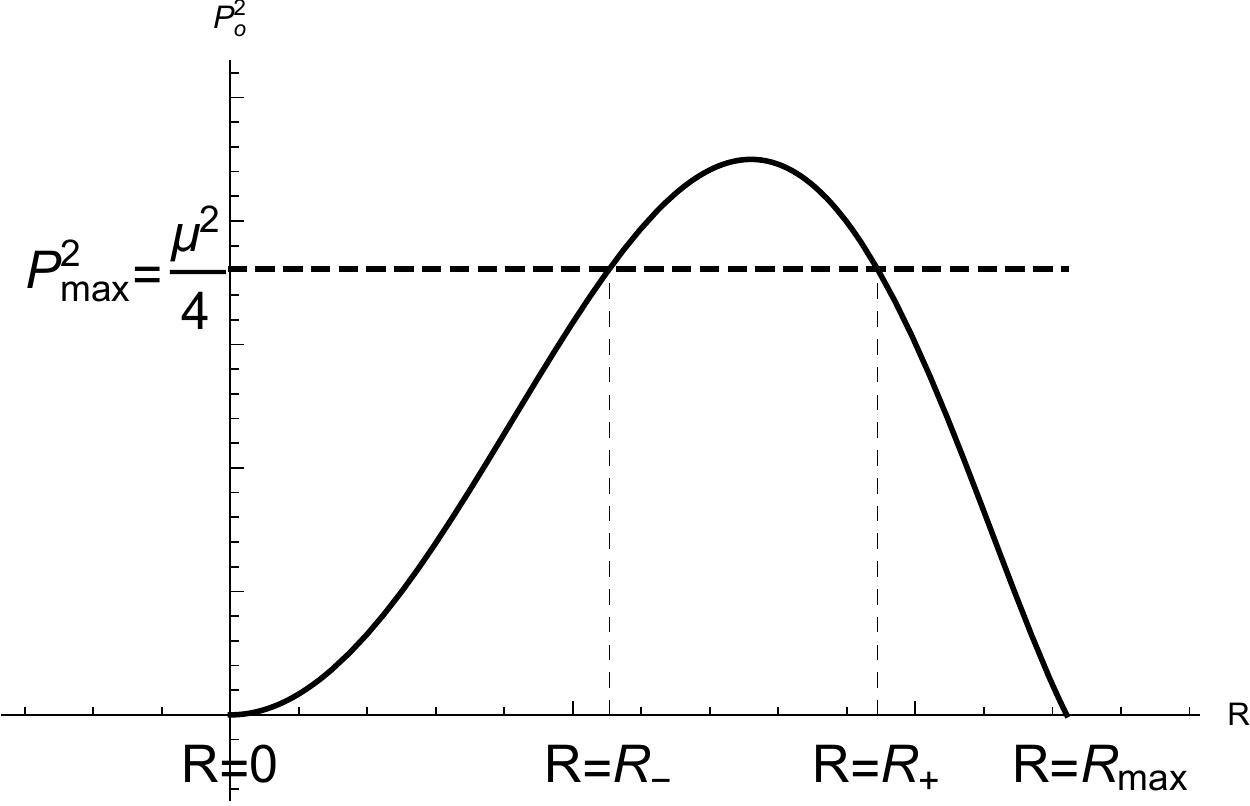} 
\caption{ Behavior of $P_o^2(R)$ in eq. (\ref{Po-positivo}) for small bubbles. }
\label{Po-quadro-di-R}
\end{figure}

Plugging eq. (\ref{rprimo-Po-positivo}) into eq. (\ref{P-matching}), we get
\bea
&& P_o^2(R) = R^2 (\mu-R^2)  \nl
&& + \frac{1}{4} R^2  \frac{\left(\mu-1 
+ R^2 \left(\kappa ^2+\lambda -1\right)\right)^2}{1-\lambda  R^2} \, ,
   \label{Po-positivo}
\eea
see figure \ref{Po-quadro-di-R} for a plot.
Note that $P_o^2(0)=P_o^2(R_{\rm max})=0$.
Using the fact that $R<1/\sqrt{\l}$ for small bubbles, 
we obtain the following inequality
\beq
P_o^2(R) 
 \geq U_o(R)=R^2 (\mu-R^2)\, ,
   \label{Po-scelta}
\eeq
where $U_o(R)$ is the effective potential in eq. (\ref{sistema-gauge-fixed}).
The maximum of $U_o(R)$ is at $R=\hat{R}=\sqrt{\mu/2}$
and its value is $U_o(\hat{R})=\mu^2/4=P_{\rm max}^2$,
defined in eq. (\ref{P-critico}). 
Then, the maximum of $P_o^2(R)$ is bigger than $P_{\rm max}^2$, as it is clear from figure \ref{Po-quadro-di-R}.
Let us define the radii $R_{\pm}$ as the  two solutions to the equation
\beq
P_o^2(R_\pm)=P_{\rm max}^2=\frac{\mu^2}{4} \, .
\label{ERRE-piu-meno}
\eeq
From the property $P_o^2(\sqrt{\mu/2})>\mu^2/4$, it follows that
\beq
R_- \leq \sqrt{\frac{\mu}{2}}\leq R_+ \, .
\label{disuguaglianza-utile}
\eeq
As we will see,
the radii $R_\pm$ determine
the behavior of the complexity rate at large time $t_b$.

\subsubsection{Complexity rate}
\label{subsec-complexity rate}

Depending on the parameter values and on $R$, the quantity $\rho_o(R)$ in eq. (\ref{rprimo-Po-positivo})
can be either positive or negative. A detailed analysis is deferred to appendix \ref{appe-segno-truccoso}.
Based on the sign of $\rho_o(R)$,
the details  of the extremal surface are 
slightly different. 
Let us distinguish between the two situations:
\begin{itemize}
\item{\bf $\rho_o(R) > 0 $.}
  The radial coordinate of the extremal surface
 is monotonic between $r=R$ and the AdS UV cutoff $r=\Lambda$.
From eq. (\ref{volume-gauge-fixed-v-var}), the volume is
\bea
\frac{\mathcal{V}}{2 \pi } &=&
    \int_{0}^R \,\frac{ r }{\sqrt{  f_{i}(r)}} \, d r
 \nl
  &+&  \int_{R}^{\Lambda}  \,\frac{ r^{2} }{\sqrt{P_{o}^2  +f_{o}(r) \,  r^{2}} }\, d r  
 \, .
\label{volume-caso-rprimo-o-positivo}
\eea
According to eq. (\ref{tempo-A}), the boundary time is
\bea
t_b &=& T_o(R) \nl
&-& \int^{\Lambda}_{R}  \frac{1}{f_{o}} \le \frac{P_{o}}{\sqrt{P_{o}^2  +f_{o}(r) \,  r^{2}}} \ri  dr \, .
\label{boundary-time-caso-rprimo-o-positivo}
\eea
In order to find the complexity rate,
we use the same strategy as in \cite{Carmi:2017jqz}.
Namely, we sum and subtract the quantity $P_o t_b$ to the
volume expressed in eq. (\ref{volume-caso-rprimo-o-positivo}): 
\bea
\frac{\mathcal{V}}{2 \pi } 
&=& P_o t_b  +    \int_{0}^R \,\frac{ r }{\sqrt{  f_{i}(r)}} \, d r  \nl 
&+& \int_{R}^{\Lambda}  \,\frac{ \sqrt{P_{o}^2+f_o r^{2}}}{f_o  }\, d r
- P_o T_o(R) \, . \nl
\label{trucco-volume}
\eea
Taking the derivative of eq. (\ref{trucco-volume}) with respect to $t_b$,
and using eq. (\ref{P-matching}), we find the complexity rate
 \beq
 W=\frac{1}{2 \pi} 
\frac{d \, \mathcal{V}}{d \, t_b }=P_o \, .
\label{rate-caso-rprimo-o-positivo}
\eeq


The asymptotic linear growth corresponds to  the
 $\mathcal{V} \to \infty$ and $t_b \to \infty$ limit, which formally comes from the divergence of the integrands in 
eqs. (\ref{volume-caso-rprimo-o-positivo}) and  (\ref{boundary-time-caso-rprimo-o-positivo})
 in correspondence of the turning point  $r_{t,\rm{AdS}}$, see eq. (\ref{turning-point}).
In particular, the turning point $r_{t,\rm{AdS}}$ satisfies
\[
P_{o}^2  +f_{o}  \,  r^{2} = 0 \, .
\]
This singularity is integrable except for
$P_o^2 \to P_{\rm max}^2=\mu^2/4 $. 
 In order to find a divergent $t_b$ in eq.
 (\ref{boundary-time-caso-rprimo-o-positivo}), we need 
 the turning point $r_{t,\rm{AdS}}=\sqrt{\mu/2}$ to lie inside the integration domain $[R,\infty]$.
  Thus, from eq. (\ref{disuguaglianza-utile}), we find that
$t_b$ is regular for $R \to R_+$ and diverges for  $R \to R_-$.
In other words,
the late time limit $t_b \to \infty$
corresponds to $R \to R_-$.

\item{\bf $\rho_o(R) < 0 $.}
For the extremal surfaces to be attached to the AdS boundary
there must be a turning point $r_{t,{\rm AdS}} < R$ inside the black hole.
From eq. (\ref{volume-gauge-fixed-v-var}), the volume is
\bea
\frac{\mathcal{V}}{2 \pi} &=&
 \int_{0}^R \,\frac{ r }{\sqrt{f_{i}(r) } }\, d r  \nl 
  &-&\int_R^{r_{t,{\rm AdS}}}  \,\frac{ r^{2} }{\sqrt{P_{o}^2  +f_{o}(r) \,  r^{2}} }\, d r  \nl
  &+&   \int_{r_{t,{\rm AdS}}}^{\Lambda}  \,\frac{ r^{2} }{\sqrt{P_{o}^2  +f_{o}(r) \,  r^{2}} }\, d r  
     \, .
\label{volume-caso-rprimo-o-negativo}
\eea
According to eq. (\ref{tempo-A}), the boundary time reads
\bea
t_b &=&  \int_{R}^{r_{t,{\rm AdS}}}   \frac{1}{f_{o}} \le  \frac{P_{o} }{ \sqrt{P_{o}^2  +f_{o}(r) \,  r^{2}}} \ri dr \nl
&-& \int^{\Lambda}_{r_{t,{\rm AdS}}}  \frac{1}{f_{o}} \le \frac{P_{o}}{\sqrt{P_{o}^2  +f_{o}(r) \,  r^{2}}} \ri  dr \nl 
&+& T_o(R) 
 \, .
\label{boundary-time-caso-rprimo-o-negativo}
\eea
The rate $W$
 can be evaluated by the same strategy as in 
 the previous case.
The result is again
given by eq. (\ref{rate-caso-rprimo-o-positivo}).
 
 As in the previous case, the asymptotic linear growth of complexity
 is in correspondence of the divergence of  the integrands in
eq. (\ref{boundary-time-caso-rprimo-o-negativo})
at $r=r_{t,\rm{AdS}}$.
The singularity is generally integrable, except
for $P_o=P_{\rm max}$, for which $r_{t,\rm{AdS}}=\sqrt{\mu/2}$.
In this case, the late $t_b$ limit corresponds to $R \to R_+$.
For $0< R < R_+$, instead, the extremal surface never reaches the AdS boundary. 
 
\end{itemize}

\begin{figure*}
\includegraphics[scale=0.4]{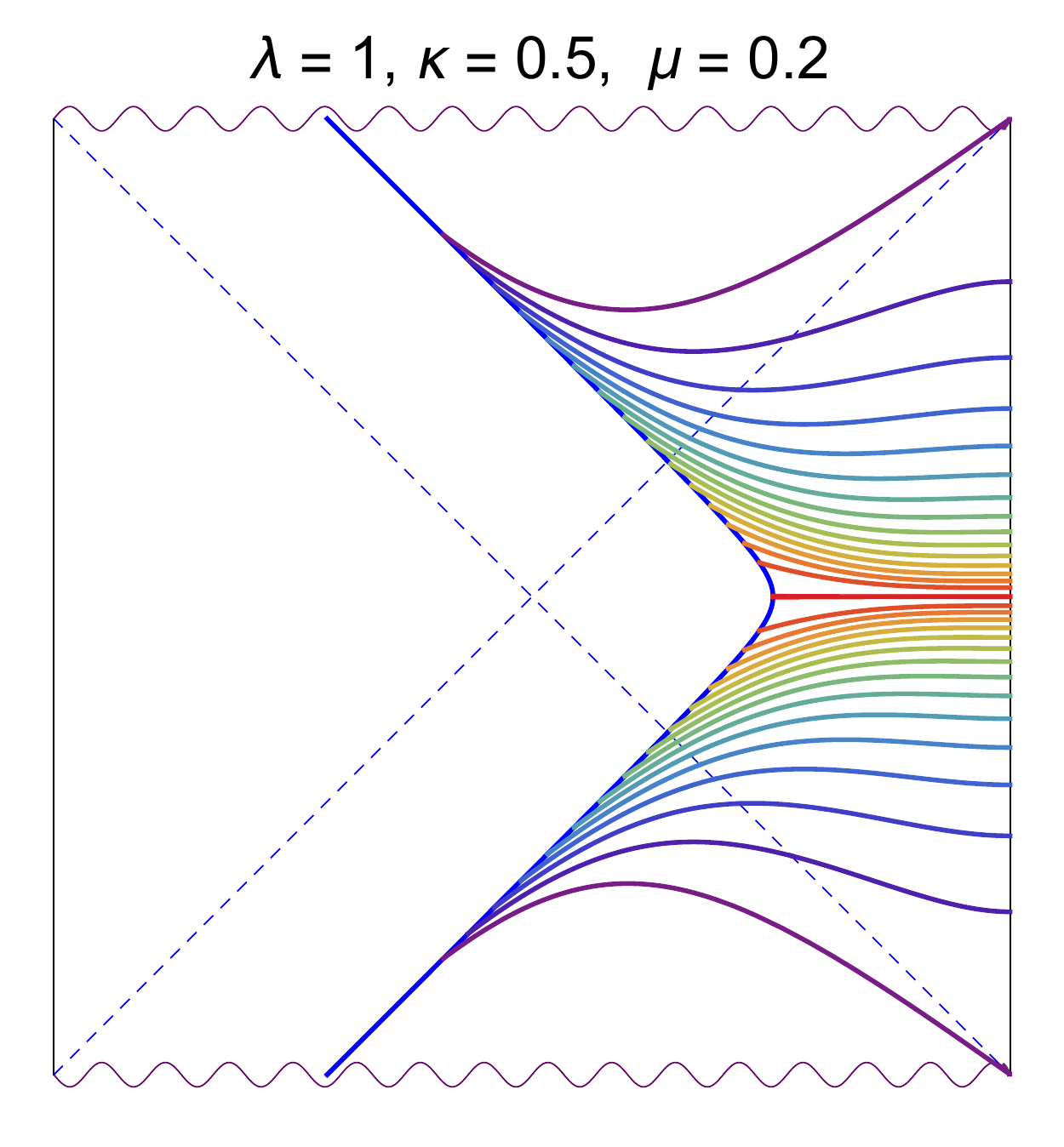}  
\qquad
\includegraphics[scale=0.6]{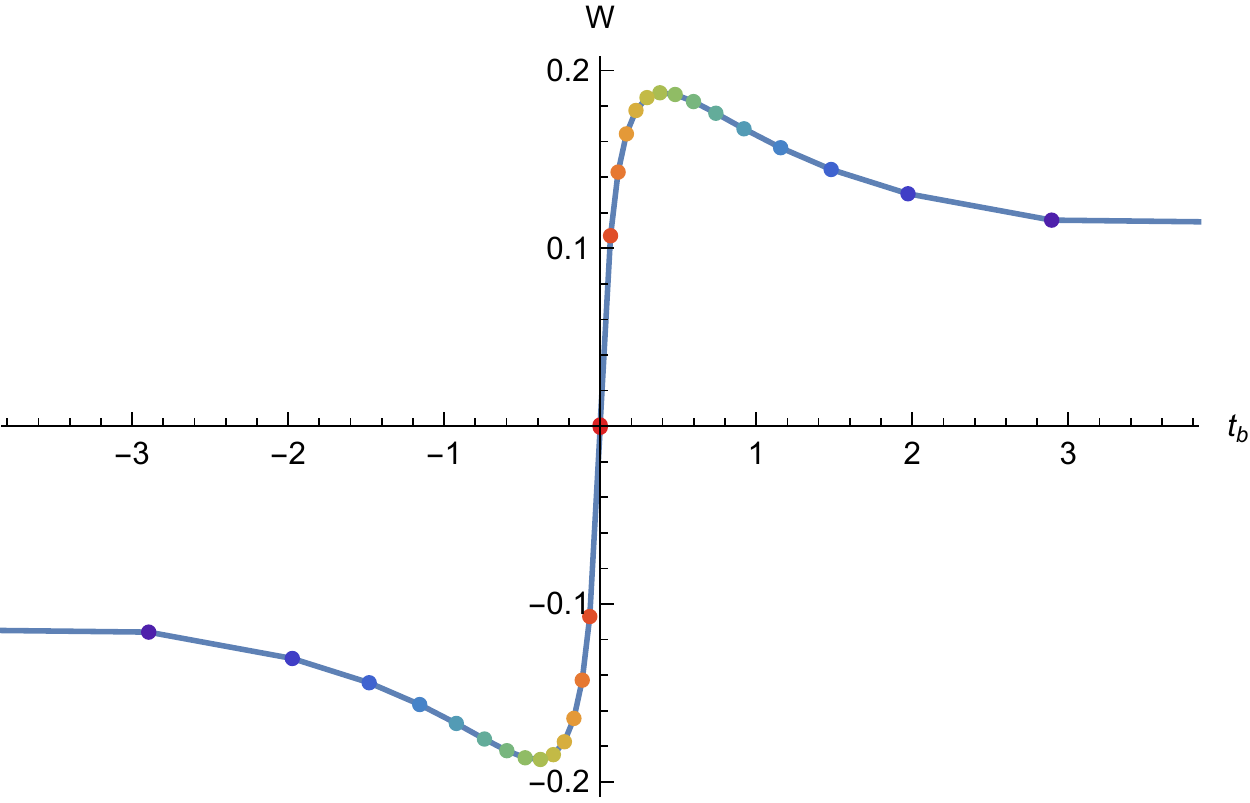} 
\caption{ 
Left panel: AdS portion of the extremal surfaces  for a very small bubble.
Right panel: plot of the complexity rate.
For every value of the boundary time $t_b$
 there exists a single extremal surface. }
\label{rainbow1a}
\end{figure*}

\begin{figure*}
\includegraphics[scale=0.4]{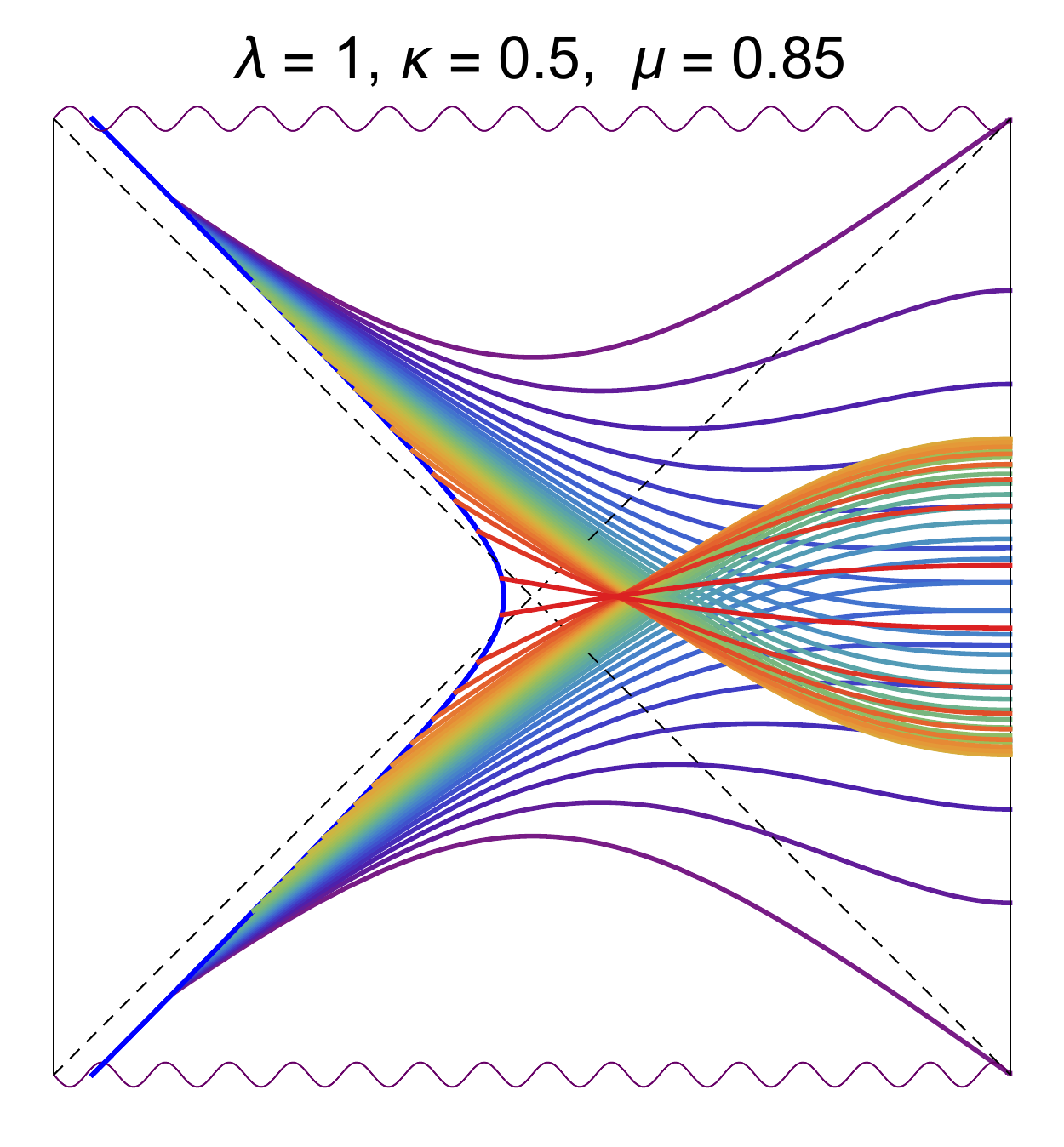}  \qquad
\includegraphics[scale=0.6]{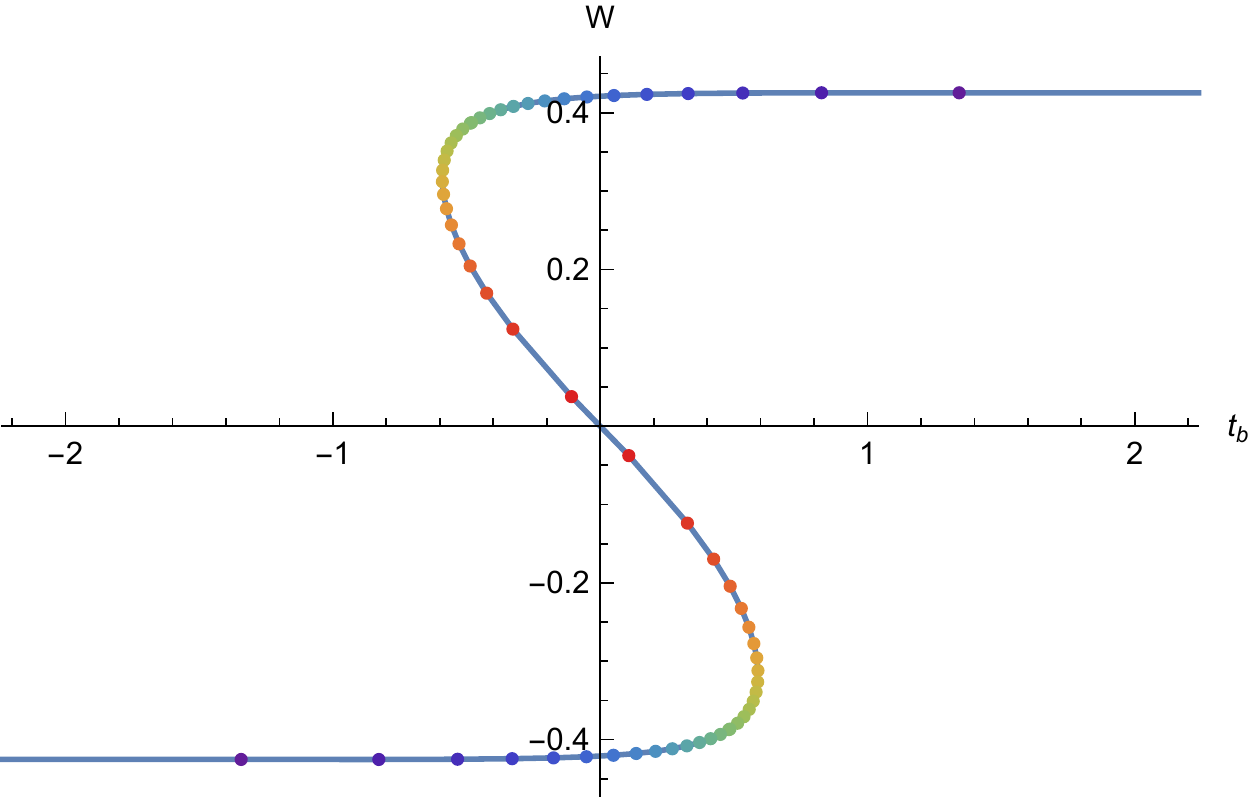}
\caption{ 
Left panel: AdS portion of the extremal surfaces
for a not so small bubble. Right panel: 
plot of the complexity rate. Note that multiple extremal surfaces exist 
for boundary times $t_b$ inside the  window
in eq. (\ref{finestra-Tb}).  }
\label{rainbow2a}
\end{figure*}

 Extremal surfaces can be found numerically by integrating the equations
 of motion  (\ref{vprimo-da-P}) and (\ref{sistema-gauge-fixed}).
 The dS portion of these extremal surfaces corresponds to the $P_i=0$
 surfaces  shown  in figure \ref{dS-Pi-zero}. The AdS portion of
 two prototypical examples of solutions is
  plotted in the left panels of figures \ref{rainbow1a} and \ref{rainbow2a}.
  The  complexity rate as a function of the boundary time $t_b$
  is shown on the right panels 
  of the same figures.

 In the example in figure \ref{rainbow1a}, for a given value
 of the boundary time $t_b$ there exists a single extremal
 surface anchored at the AdS boundary.
 Instead, in the example in figure \ref{rainbow2a}
 in a time window centered at $t_b=0$,  \emph{i.e.}
\beq
-T_b < t_b <  T_b \, ,
\label{finestra-Tb}
\eeq
 three extremal surfaces anchored at 
 the same value of the boundary time $t_b$ exist.
 The rate $W$ is then a multivalued function of the time $t_b$.
 
\begin{figure}
\includegraphics[scale=0.5]{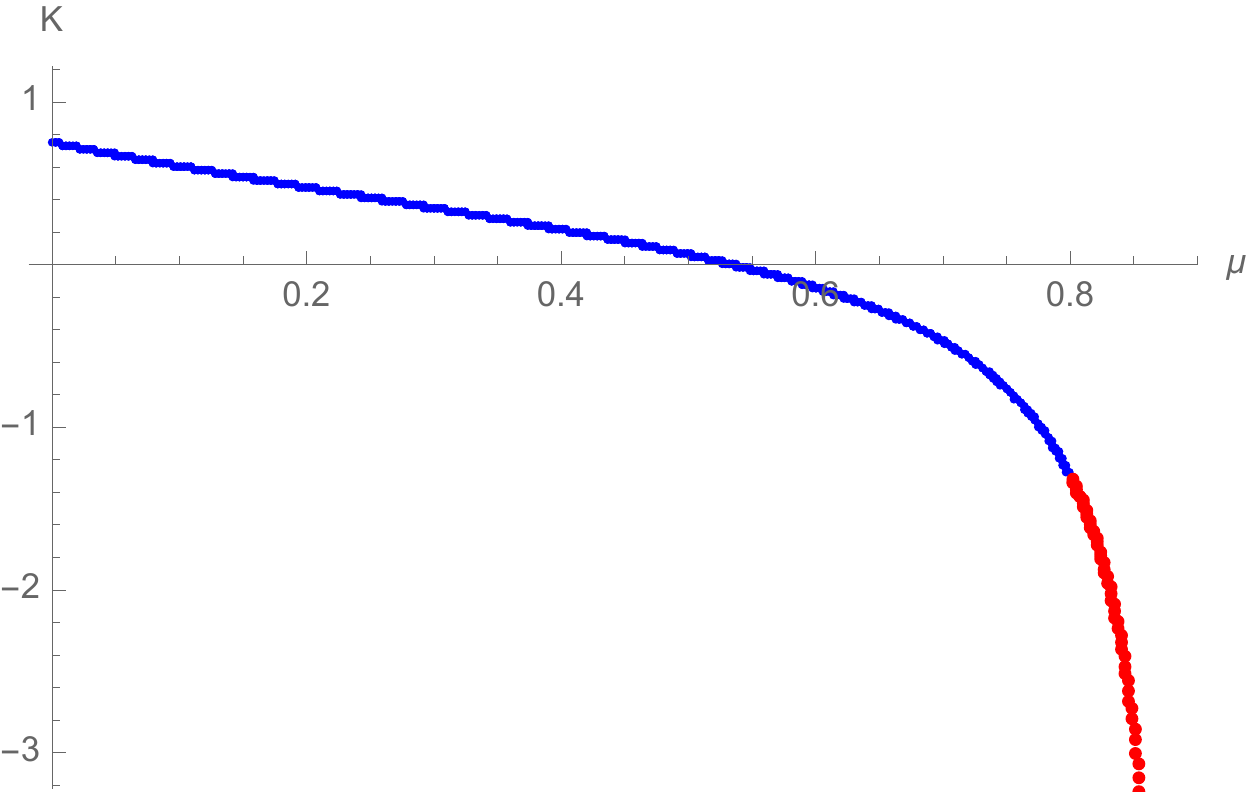}
\caption{ 
Plot of the quantity $K$ defined in eq. (\ref{definizione-K}) as a function of $\mu$ for $\l=1$, $\kappa=0.5$. 
 In this numerical example $K$ is negative for $ \mu>0.53$ 
 and approaches $-\infty$ for $\mu\to \mu_0 \approx 0.88$.
 In order to have multiple extremal surface for the same 
 value of $t_b$ as in figure \ref{rainbow2a}, we need $K<0$.
  The blue portion of the plot corresponds to very small bubbles,
  while the red part to not so small bubbles. 
 }
\label{region-S-1}
\end{figure}

\begin{figure}
\includegraphics[scale=0.35]{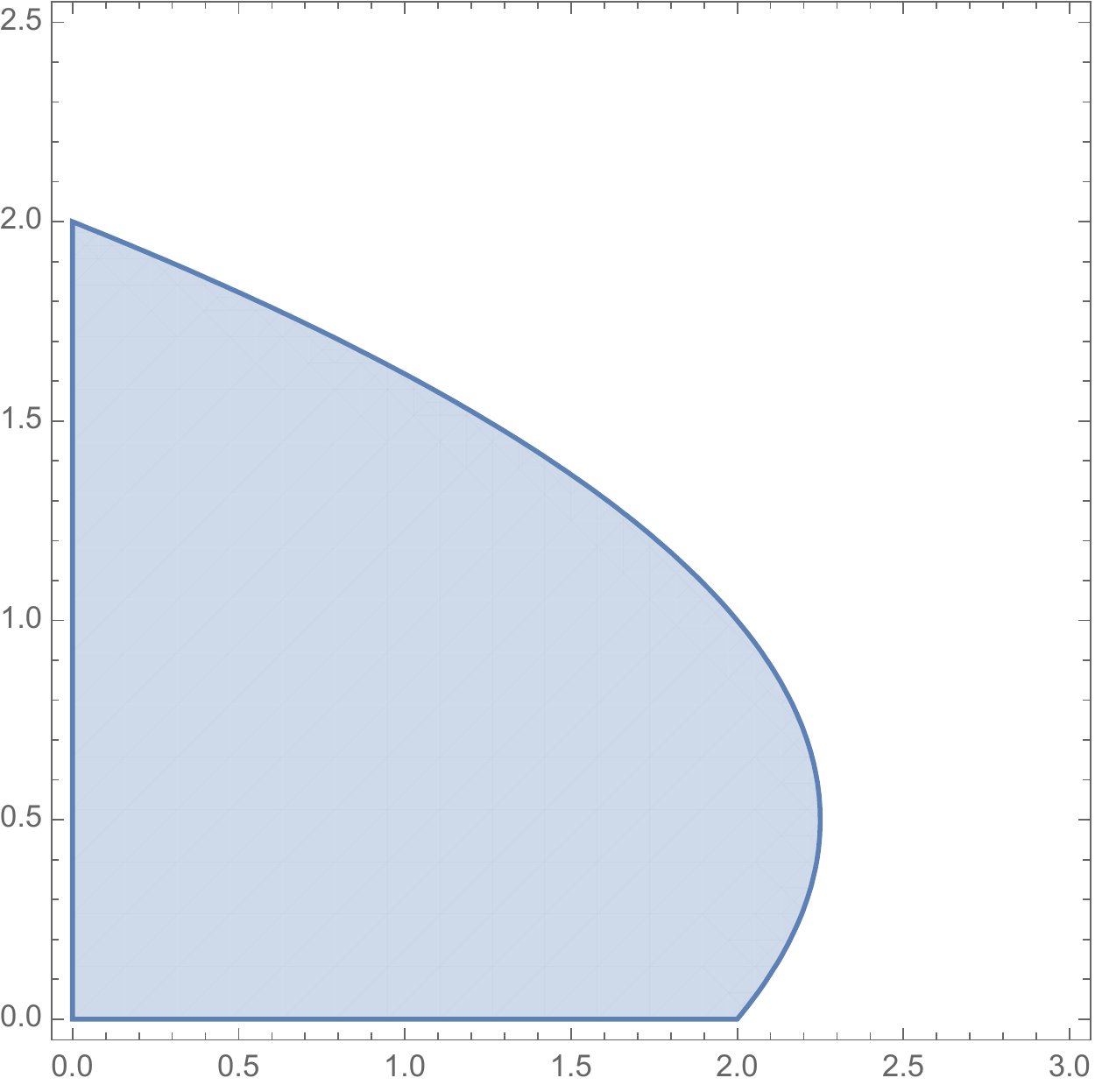}
\caption{ 
 The white region  represents the portion of the $(\l, \kappa)$ plane 
 defined in eq. (\ref{regione-tutta-S}), for which $K$ is negative for every value of $\mu$.
 }
\label{region-S-2}
\end{figure}

To discriminate between the two possible behaviors,
 it is useful to introduce the quantity
\beq
 K=\left. \frac{d  t_b}{d  P_o} \right|_{t_b=0} \, ,
 \label{definizione-K}
 \eeq
 representing the reciprocal of the slope $\frac{d W}{d t_b}$ computed at the origin
 of the plots in the right panel of figures \ref{rainbow1a} and \ref{rainbow2a}.
 The complexity rate in figure \ref{rainbow1a} is characterized by $K>0$,
contrary to the one in figure \ref{rainbow2a}, where $K<0$.
 An explicit expression for $K$ is given in appendix \ref{appe-region-S}.
  From  eqs.  (\ref{derivata-a-Po-zero-very-small-bubble})
and   (\ref{derivata-a-Po-zero-not-so-small-bubble}),
  we find that $K$ is a decreasing function
  of $\mu$ at fixed $\l$ and $\kappa$ and that  $K \to -\infty$ for $\mu \to \mu_0$.
 In  figure \ref{region-S-1}, we show
 $ K$ as a function of $\mu$ for a fixed value of $\lambda$
 and $\kappa$. 
 In the parameter region 
 \beq
 \l \geq 2+\kappa-\kappa^2 \, ,
 \label{regione-tutta-S}
 \eeq
 the quantity $K$ is always negative, so
 multiple extremal surfaces exist for every $\mu$,
 see  figure \ref{region-S-2}.


\begin{figure}[h]
	\includegraphics[scale=0.4]{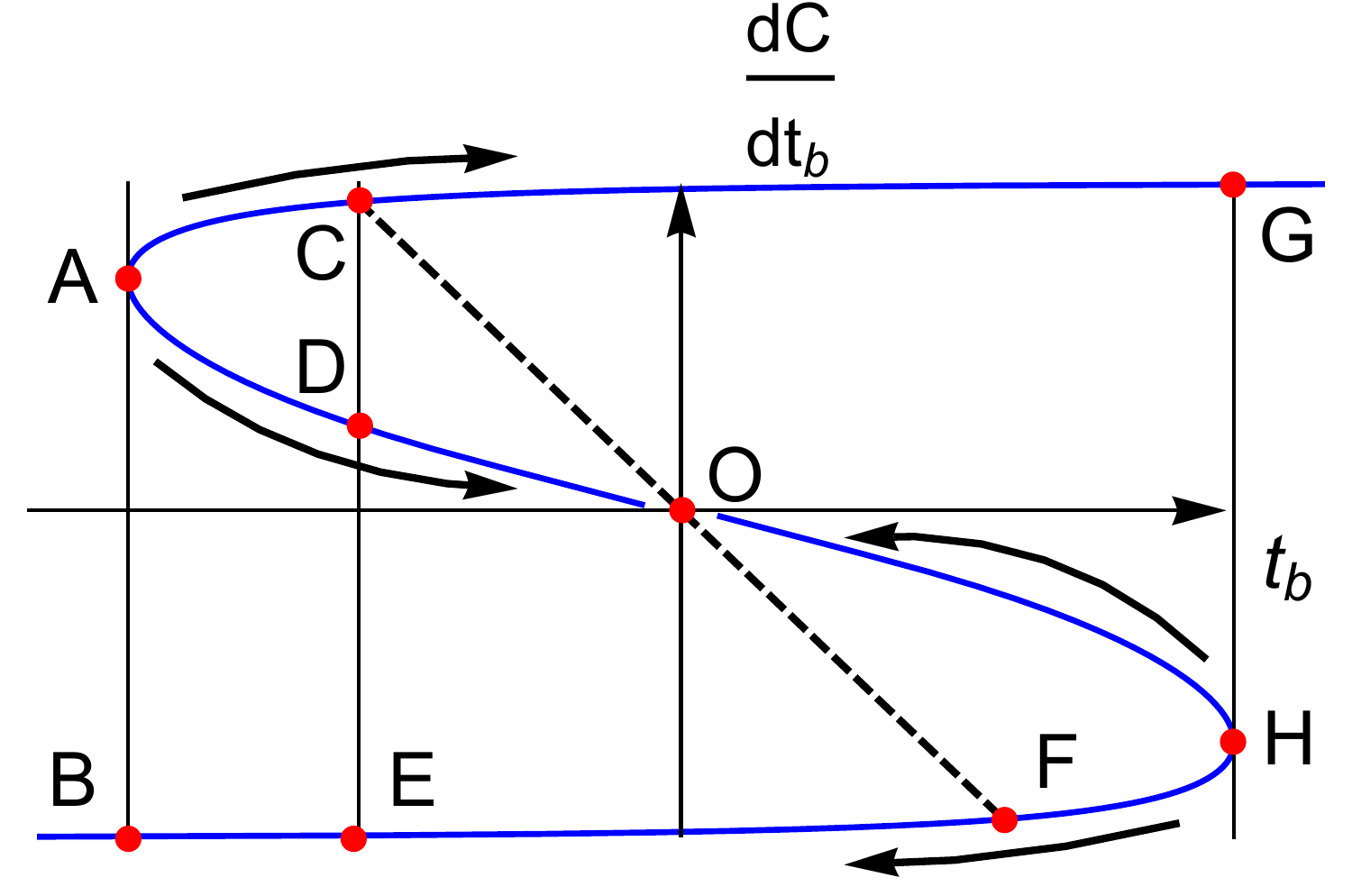}
	\caption{Schematic plot of the complexity rate in the parameter region where there are
	multiple  extremal surfaces anchored at the same boundary time.}
	\label{ExSurfChoice-1}	
\end{figure}

\begin{figure}[h]
		\includegraphics[scale=0.5]{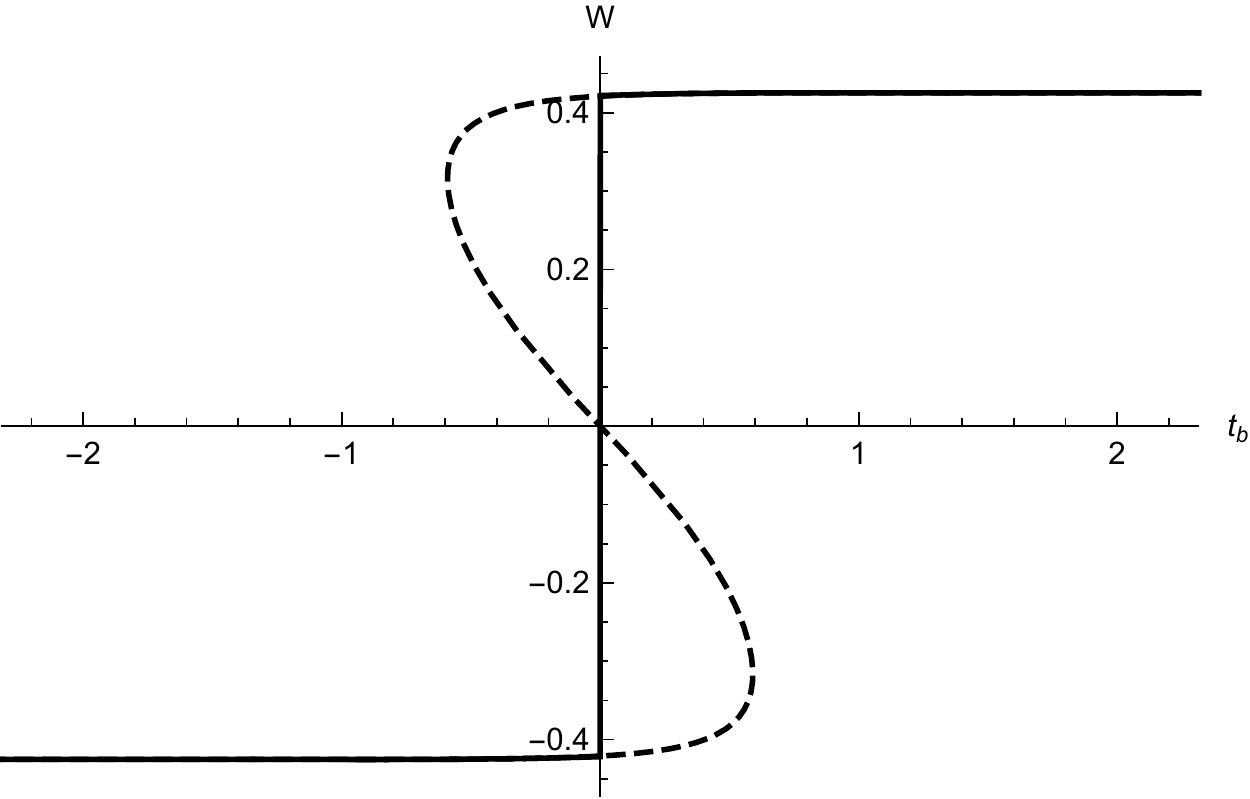}
	\caption{The requirement of maximal volume selects the step-like
	function  rate represented by the solid line.}
	\label{ExSurfChoice-2}	
\end{figure}

According to the CV conjecture, when multiple extremal  surfaces exist, 
the one with  maximal volume should be picked. 
To deal with this,
let us consider the prototypical behavior shown  in figure \ref{ExSurfChoice-1}.
The three points $C$, $D$, and $E$
are obtained by extremal surfaces
anchored at the same boundary time,
but whose volumes $\mathcal{V}$ and values of the rate differ. 
To properly choose the maximal solution,
let us first consider the two points $C$ and $D$.
   The volumes $\mathcal{V}_C$ and $\mathcal{V}_D$ of the corresponding surfaces can be inferred by starting 
   from the volume $\mathcal{V}_A$ of the surface located at the point $A$,
   and following the two arrows which bring to $C$ and $D$, respectively. 
   Since the rate is positive, 
the volume increases as we move along these arrows.   
  Hence, both $\mathcal{V}_C$ and $\mathcal{V}_D$ 
     are higher than $\mathcal{V}_A$. 
     Precisely, since the growth rate as we move from $A$ to $C$ is larger than the one going from $A$ to $D$,
      we conclude that 
       $\mathcal{V}_C > \mathcal{V}_D$.
       Moreover,
   by time-reversal symmetry, $\mathcal{V}_C = \mathcal{V}_F$.   
      Moving along the curve from 
         $E$ to $F$ the rate  is negative, which means that the volume is decreasing: $\mathcal{V}_F < \mathcal{V}_E$.      
          From this argument, we should choose the lowest value of the rate for $t_b<0$
    and the highest value of the rate for $t_b>0$, see  figure \ref{ExSurfChoice-2}.
      Consequently, the complexity rate $W$ experiences a discontinuous jump 
 at $t_b=0$. A similar analysis was performed in \cite{Jorstad:2022mls}
 for complexity in dS spacetime.

\subsubsection{Late time complexity rate}          
\label{section-late-time-small-bubbles}          
          
We now discuss the late time behavior of the rate $W$.
In terms of the bubble radius, we will show that $t_b \to \infty$
 corresponds to   $R\to R_+$ or to $R \to R_-$, depending
on the point in the  parameter space.
First note that $\rho_o(R_+)$ has the same sign as $\rho_o(R_-)$.
In fact, if the signs of $\rho_o(R_\pm)$
were different, we would have $R_-<R_0<R_+$, where $R_0$ is defined
by the condition $\rho_o(R_0)=0$,
 see eq. (\ref{ERRE-zero}). Then we should have
 $P^2_o(R_0)>\frac{\mu^2}{4}$, see figure \ref{Po-quadro-di-R}, but this is impossible because
\beq
P^2_o(R_0)-\frac{\mu^2}{4} =-\frac{\left(\kappa ^2 \mu +\lambda  \mu +\mu -2\right)^2}{4 \left(\kappa ^2+\lambda
   -1\right)^2} \leq 0 \, .
\eeq
We can then check that  the $t_b \to \infty$ limit  corresponds to $R\to R_+$ or to $R \to R_-$.
In particular, we can have
\begin{itemize}
\item {\bf Case 1}: 
If $\rho_o(R_+)<0$, the $t_b \to \infty $ limit corresponds to $R \to R_{+}$.
In this case both the integrals in eq. (\ref{boundary-time-caso-rprimo-o-negativo}) diverge,
due to a singularity in correspondence of $r_{t,{\rm AdS}}$.
For large enough $t_b>0$ the complexity rate is an increasing function of time
(see for example the right panel of figure \ref{rainbow2a}).
\item  {\bf Case 2}: 
If $\rho_o(R_+)>0$, then also $\rho_o(R_-)>0$.
From eq. (\ref{turning-point}), 
both for $R=R_+$  and $R=R_-$, we have that  $r_{t,{\rm AdS}}=\sqrt{\mu/2}$,
because $P^2_o=\mu^2/4$.
The integral which gives $t_b$,
 see eq. (\ref{boundary-time-caso-rprimo-o-positivo}), is then  divergent
for $R = R_- $ and not for $R=R_+$, because 
$R_- \leq r_{t,{\rm AdS}} \leq R_+$, see eq. (\ref{disuguaglianza-utile}).
In this case, at large enough $t_b>0$ the complexity rate is a decreasing function of time,
see for example the right panel of figure \ref{rainbow1a}.
\end{itemize}
Either way, the complexity rate at late time is  insensitive to the presence of the bubble
\beq
\frac{1}{2 \pi} \lim_{t_b \to \infty} \frac{d \, \mathcal{V}}{d \, t_b}  = \frac{ \mu}{2} \, . 
\label{late-time-rate-eq}
\eeq
However,
the sign of $\rho_o(R_+)$ discriminates between a growing or
a decreasing complexity rate at late time.
It is a complicated problem to discuss
the sign of $\rho_o(R_+)$ in a generic point of the parameter space.
Still, in some particular regions of parameters
we can determine the sign of $\rho_o(R_+)$ with some  simple arguments.
For example, for $\mu \to 0$ we are always in case 2, because
in this limit
\beq
R_+ \to R_{\rm max} \, , \qquad \rho(R_+) \to \frac{1}{(\kappa+1)^2+\l}>0 \, .
\eeq 
Instead, depending
on the value of $\l$ and $\kappa$, 
for $\mu \to \mu_0$ we can have both case 1
(for example $\l=0.5$, $\kappa=0.5$)
and case 2 (for example $\l=1$, $\kappa=1$).
See appendix \ref{appe-segno-truccoso} for further details.

According to the Lloyd's bound \cite{Lloyd}, the maximum allowed 
growth rate of quantum computational complexity
 should be proportional to the total energy. 
 In quantum systems with holographic duals, 
 it was proposed \cite{Brown:2015lvg}  that the Lloyd's bound is
 saturated at late time
  by the uncharged planar black hole solutions in AdS.
In the parameter space portion of case 2
this version of the proposal is violated,
because the asymptotic value is approached from above.
Violations of the Lloyd's bound have been previously found for the CA
conjecture, see \cite{Couch:2017yil,Swingle:2017zcd,Carmi:2017jqz,Yang:2017czx,Mahapatra:2018gig,Bernamonti:2021jyu}. 
In holographic models including just AdS boundaries,
 we do not know about any other  violation of the bound
for the CV proposal. See \cite{Yang:2019gce,Auzzi:2022bfd} for examples
of AdS hairy black holes in which the bound is instead satisfied.


\subsection{Large bubbles } 

Contrary to small bubbles,
for large bubbles
the dS part of the geometry contains 
the region beyond the cosmological horizon $r = 1/\sqrt{\l}$.
However,
recalling that $P_i=0$,
we point out that extremal codimension-one surfaces 
in the dS part cannot enter the region
with $r>1/\sqrt{\l}$, where the potential  $U_i$ is positive,
see figure \ref{Uio-figure}.
Therefore, the extremal  surfaces are confined into the static patches.
In particular,
for $0 < \mu < \mu_h$ the extremal surfaces extend into the right static patch.
Since in this case the domain wall never enter the left static patch,
we necessarily have $\rho_i(R)<0$.
On the other hand, 
for $\mu > \mu_h$ no portion of the right static patch is present in the geometry, so the extremal surfaces remain into the left static patch. 
From this argument we conclude that $\rho_i(R)>0$.

Some technical details in the calculations slightly differ in the two cases:
\begin{itemize}
\item {\bf Very large bubble, $0 < \mu < \mu_h$.} 
Equation (\ref{P-matching}), which determines $P_o$,  is replaced by
  \bea
&& -\frac{R}{\sqrt{f_i} }   
  =  \frac{1}{  f_o}  \rho_o 
+  \frac{d T_o}{dR}  P_o \, , \nl
&& P_o  =\pm \sqrt{  \rho_o^2(R) -  f_o(R) \, R^{2}}
 \, .
   \label{P-matching-very-large}
  \eea
We fix the sign of eq. (\ref{eq-diff-Ti-To}) as in eq. (\ref{scelta-segno-To}).
The solution for positive $P_o$ is
\bea
\rho_o= \frac{R \left(\mu -1 +R^2 \left(\kappa ^2+\lambda -1\right)\right)}{2 \sqrt{1-\lambda  R^2}} \, .
      \label{rprimo-Po-positivo-very-large}
\eea
From here, we obtain the same value of $P_o$ as in eq. (\ref{Po-positivo}).
It can be checked that $\rho_o(R_{\rm min}) $  in eq. (\ref{rprimo-Po-positivo-very-large}) is always negative.
Also, for fixed $\l$, $\kappa$, and $\mu$, 
 $\rho_o(R)$ is a decreasing function of $R$.
 Therefore, $\rho_o(R)$ is always negative,
 which means that there is a turning point $r_{t, {\rm AdS}}<R$ in AdS.
Note that  $\rho_o(R) \to -\infty$ for $R \to 1/\sqrt{\l}$.

The extremal surface always crosses the dS horizon on the bifurcation sphere, because $P_i=0$. Consequently, the volume is given by
\bea
 {\mathcal{V} \over 2 \pi } &=&  
 \int_{0}^{\frac{1}{\sqrt{\lambda}}} {r \over \sqrt{f_i(r)} }dr 
 -\int^{R}_{\frac{1}{\sqrt{\lambda}}} {r \over \sqrt{f_i(r)} }dr 
		\nl
	&& - \int^{r_{t,{\rm AdS}}}_{R} 	{r^{2} \over \sqrt{P_o^2 + f_o(r) r^{2}} } dr \nl 
	 && +\int_{r_{t,{\rm AdS}}}^{\Lambda} 	{r^{2} \over \sqrt{P_o^2 + f_o(r) r^{2}} } dr 		
	\, .
	 \label{volume-caso-very-large-bubble}
\eea
The boundary time $t_b$ can be obtained by eq. (\ref{boundary-time-caso-rprimo-o-negativo}).
We find that the complexity rate is again given
by eq. (\ref{rate-caso-rprimo-o-positivo}).

\item {\bf Not so large bubble, $\mu_h < \mu < \mu_0$.}
Equation (\ref{P-matching}) still holds.
We fix the sign of eq. (\ref{eq-diff-Ti-To}) as in eq. (\ref{scelta-segno-To}),
in which $w_o(R)$
 is always negative.\footnote{ 
 As $w_o(R)$ is a decreasing function of $R$, it is enough to check this property for $R=R_{\rm min}$, see eq. (\ref{wo-tecnicismo}).}
This choice corresponds to a bubble which, for $T_o \geq 0$, moves in the lower direction of the Penrose diagram.
Assuming $P_o$ to be positive,\footnote{If $P_o$ is negative, $\rho_o$ is given by eq. (\ref{Po-negativo}).
This solution corresponds to a reflection, so we discard it.}
  the $\rho_o$ solution is given by eq. (\ref{rprimo-Po-positivo}).

The quantity $\rho_o(R)$ changes sign at $R=R_0$, given in eq. (\ref{ERRE-zero}). However, 
for $\mu>\mu_h$ we have $R_0<R_{\rm min}$.
In addition, in the present case the equation $\rho_o(R_{\rm min})=0$ has no solution, so $\rho_o(R_{\rm min})$ is always negative,
in accordance with the refraction interpretation. 
Putting all together, we conclude that 
$\rho_o(R)$ remains negative in the range $R_{\rm min}<R<1/\sqrt{\l}$.
In particular, for $R \to 1/\sqrt{\l}$ we get $\rho_o \to - \infty$.

Due to the negativity of $\rho_o(R)$,
 there is a turning point at $r_{t,{\rm AdS}} < R$ in the AdS portion.
 We can then use eq. (\ref{volume-caso-rprimo-o-negativo})
 for the volume, eq. (\ref{boundary-time-caso-rprimo-o-negativo}) 
 for the boundary time, and eq.
  (\ref{rate-caso-rprimo-o-positivo})
 for the complexity rate.
 The value of $P_o^2$ is still given by eq. (\ref{Po-positivo}).
From a direct calculation
 \bea
&& P_o(R_{\rm min})=0 \, , \nl
 && P_o(1/\sqrt{\l})=\infty \, .
 \eea
 Moreover, for $R_{\rm min}<R<1/\sqrt{\l}$, $P_o$ is a monotonic function, 
 a plot of which is displayed in figure \ref{Po-quadro-di-R-large-bubble}.
 The unique solution $R^*$ to the equation 
 \beq
 P_o^2(R^*)=P_{\rm max}^2=\mu^2/4 \, ,
 \eeq
 with $R_{\rm min}<R^*<1/\sqrt{\l}$,
 corresponds to the extremal surface at $t_b \to \infty$.
 \begin{figure}
\includegraphics[scale=0.5]{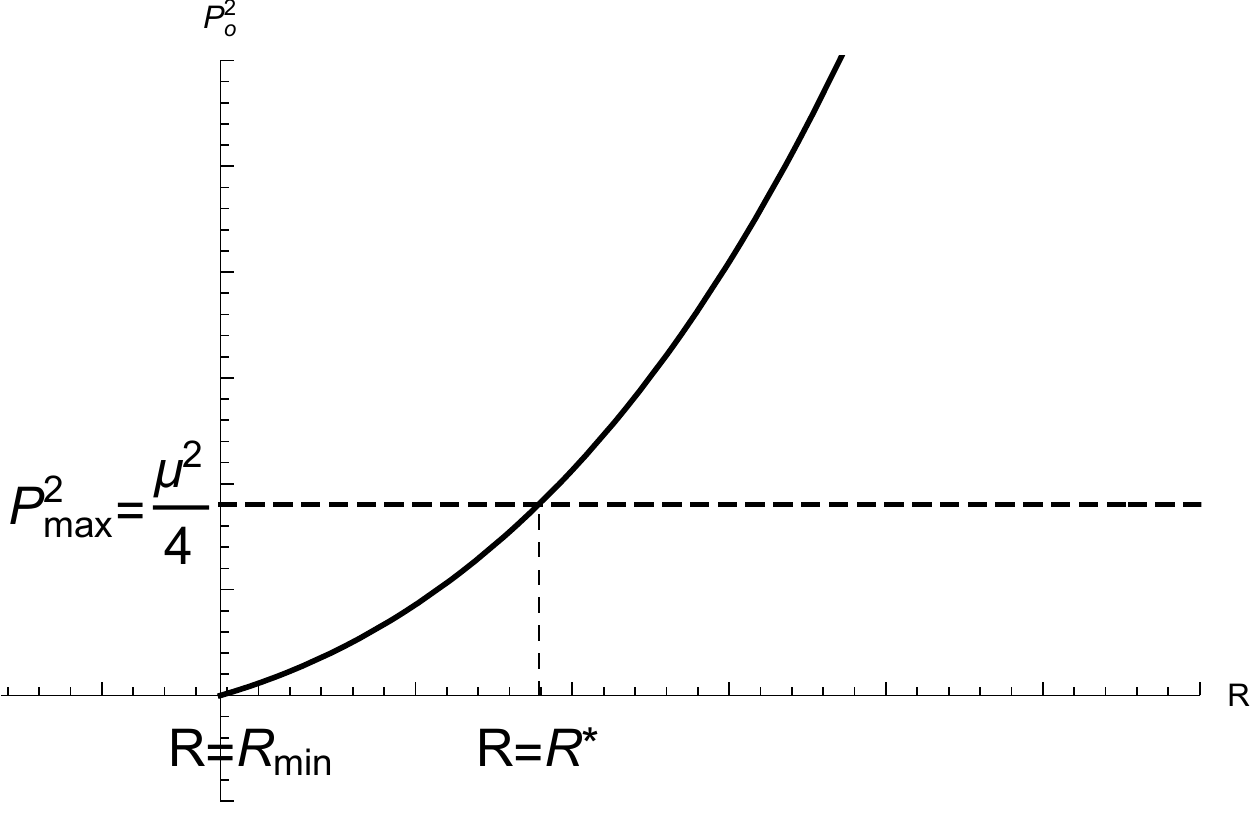} 
\caption{ Behavior of $P_o^2(R)$ in eq. (\ref{Po-positivo}) for large bubbles. }
\label{Po-quadro-di-R-large-bubble}
\end{figure}  
\end{itemize}

\begin{figure*}
\includegraphics[scale=0.4]{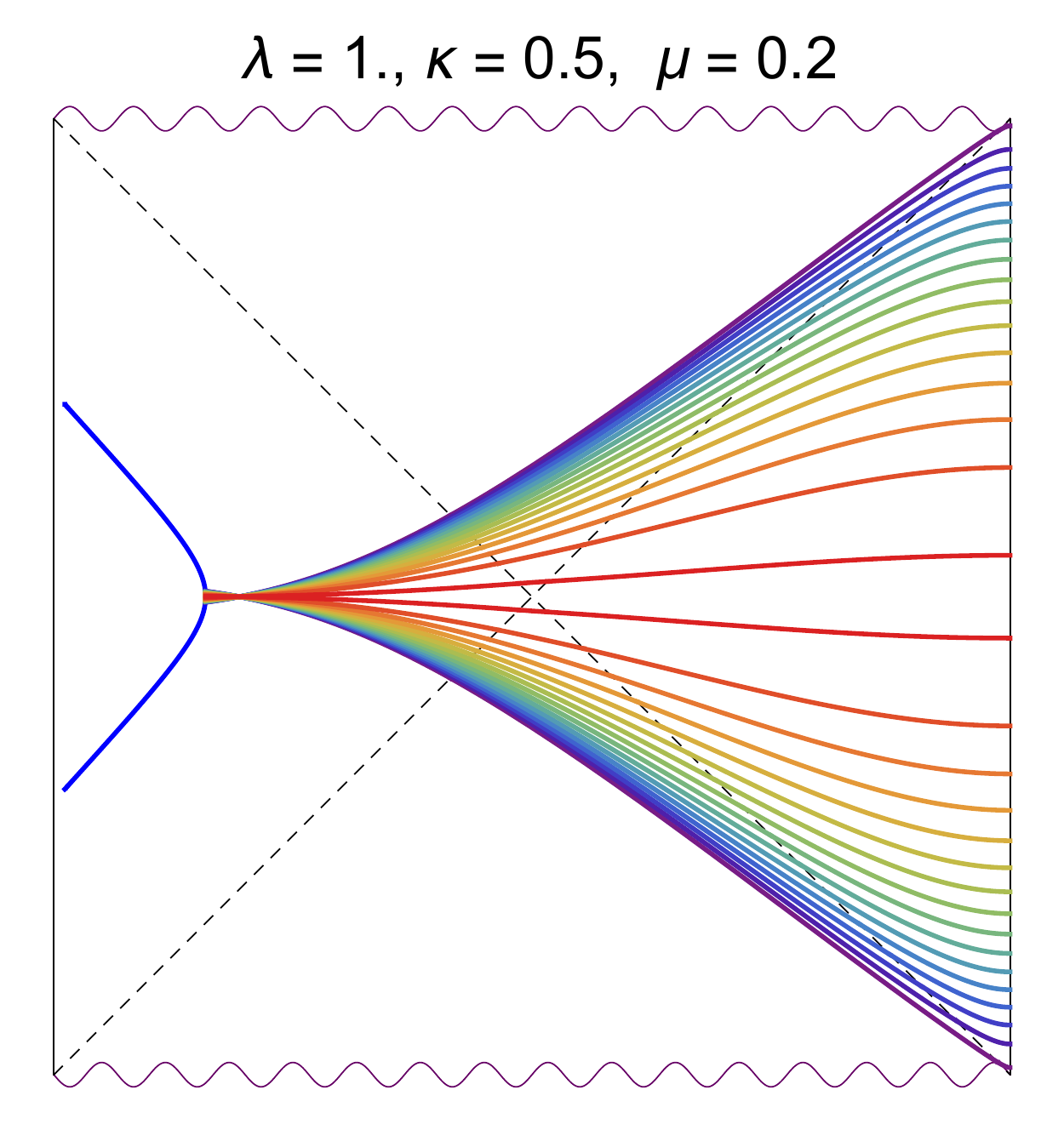}  \qquad
\includegraphics[scale=0.6]{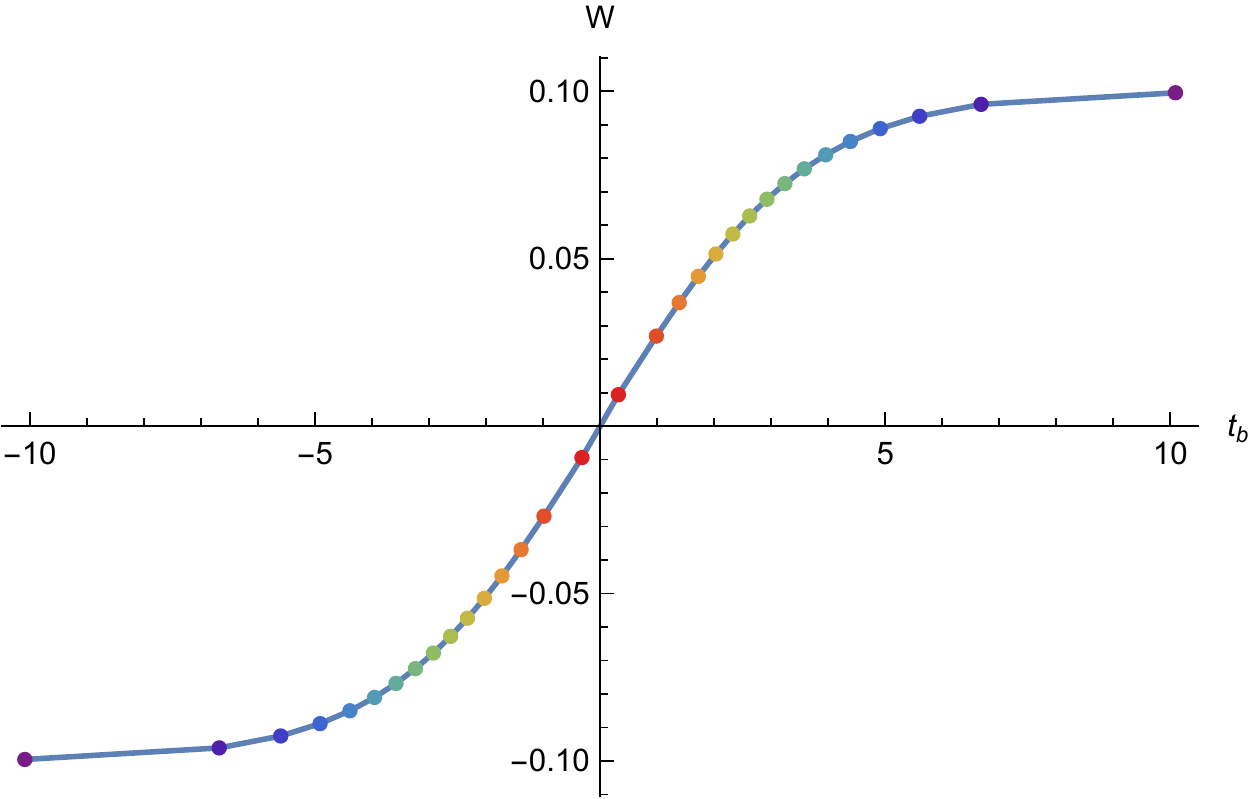}
\caption{ 
Left panel: AdS portion of the extremal surfaces
for an example of very large bubble. Right panel: plot of 
the complexity rate. 
}
\label{rainbow3}
\end{figure*}

\begin{figure*}
\includegraphics[scale=0.4]{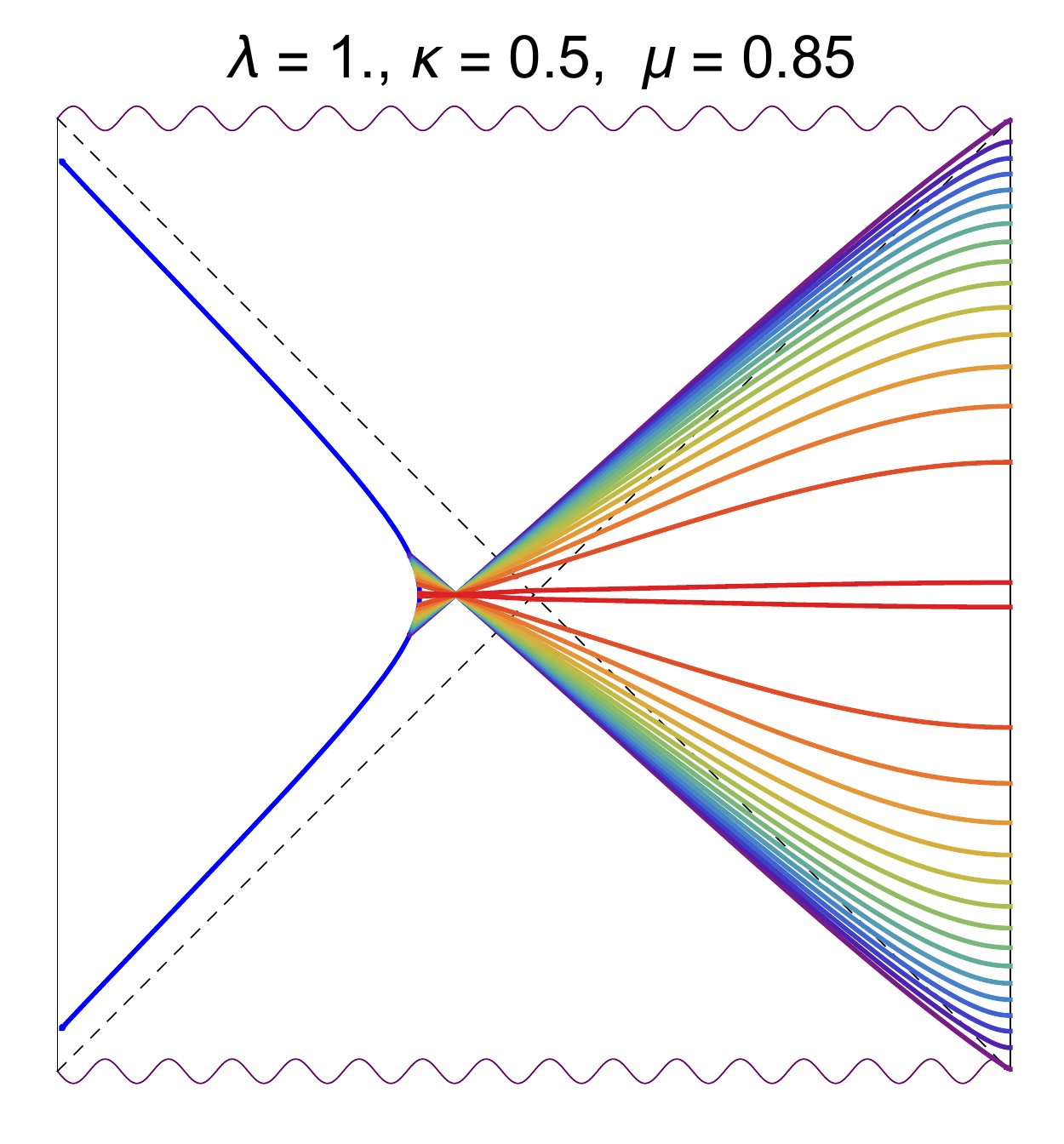}  \qquad
\includegraphics[scale=0.6]{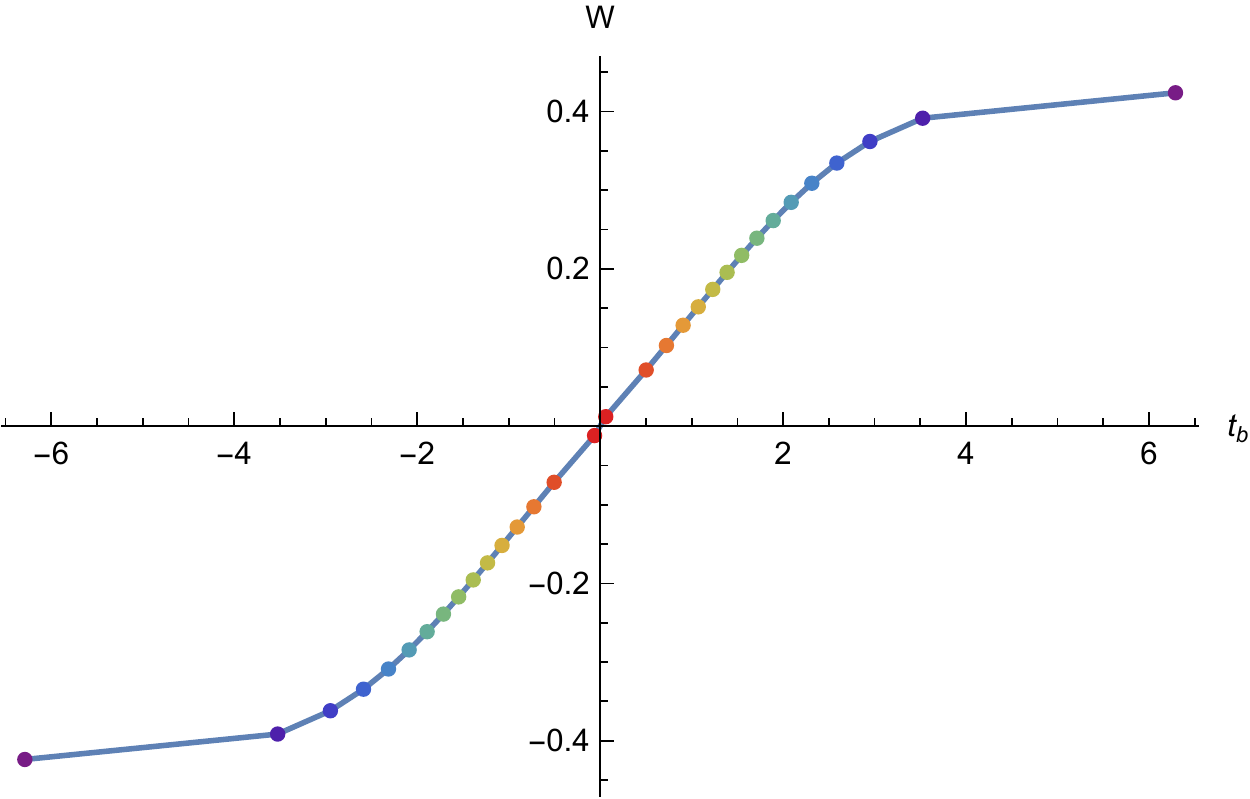}
\caption{ 
Left panel: AdS portion of the extremal surfaces
for an example of not so large bubble. Right panel: plot of 
the complexity rate. 
}
\label{rainbow4}
\end{figure*}

In figure \ref{rainbow3} and  \ref{rainbow4} we show the extremal surfaces
and the complexity rate for a very large and a not so large bubble, respectively.
In these examples, 
a single extremal surface exists for any given boundary time $t_b$ 
and the complexity rate
is an increasing function of time.

\begin{figure}
\includegraphics[scale=0.5]{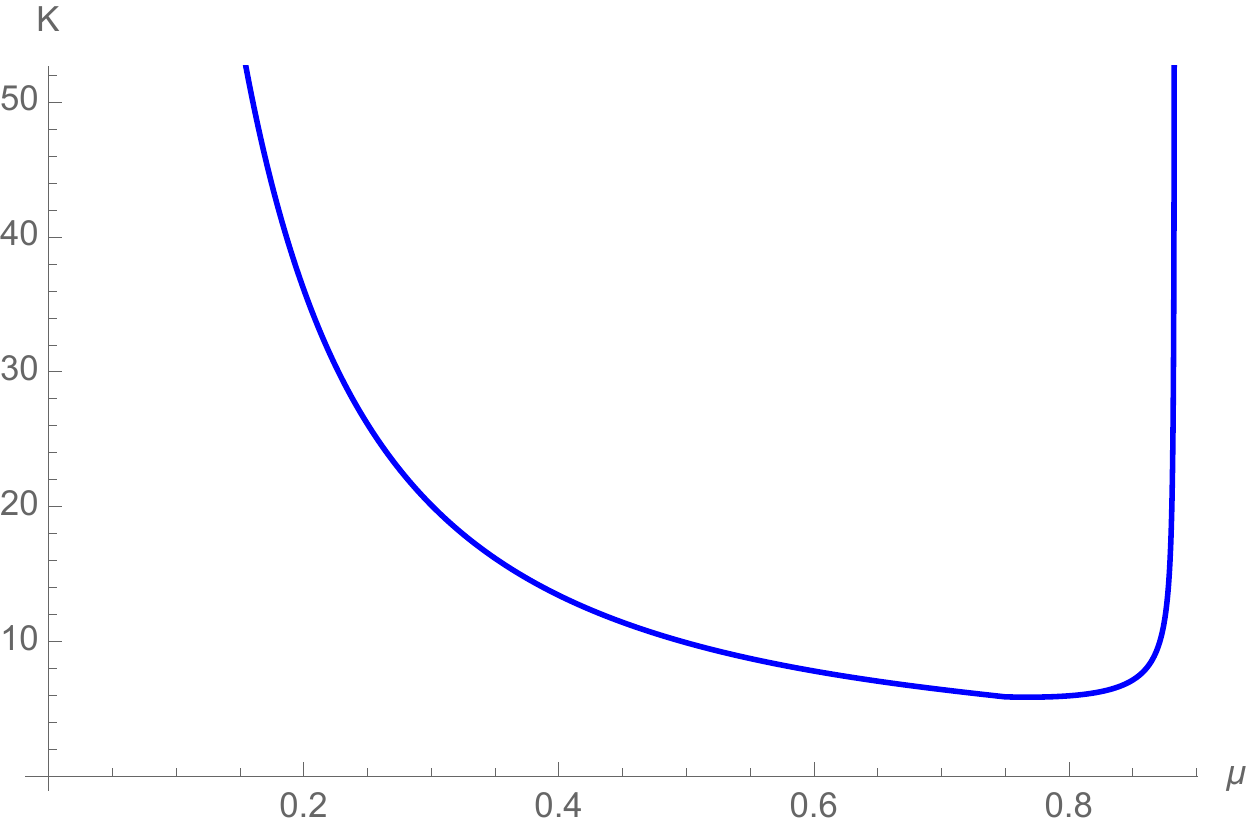}
\caption{ 
Plot of $K$, defined in eq. (\ref{definizione-K}), as 
a function of $\mu$ for $\l=1$ and $\kappa=0.5$ in the case of large bubbles.
 }
\label{no-region-S-large-bubbles}
\end{figure}

In order to check if multiple extremal surfaces 
can emerge for a given $t_b$,
we look at the behavior of $K$, expressed in eq.  (\ref{derivata-a-Po-zero-large-bubble}).
A plot of $K$ as a function of $\mu$
for fixed $\l$ and $\kappa$
is shown in figure \ref{no-region-S-large-bubbles}.
From the discussion below eq. (\ref{Htilde}),
we find that for large bubbles $K$ is always positive,
thus there are no multiple extremal surfaces 
attached at the same boundary time.

The complexity rate at large $t_b$ is still given by eq. (\ref{late-time-rate-eq}).
The rate is always an increasing function of $t_b$ at late time, 
because $P_o^2(R)$ is always an increasing function of $R$
in the range $R_{\rm min} \leq R \leq R^*$,
see figure \ref{Po-quadro-di-R-large-bubble}.
Referring to subsection \ref{section-late-time-small-bubbles},
the late time rate of complexity behaves 
as in case 1 of small bubbles.

\subsection{Static bubble limit}

\begin{figure}
\includegraphics[scale=0.4]{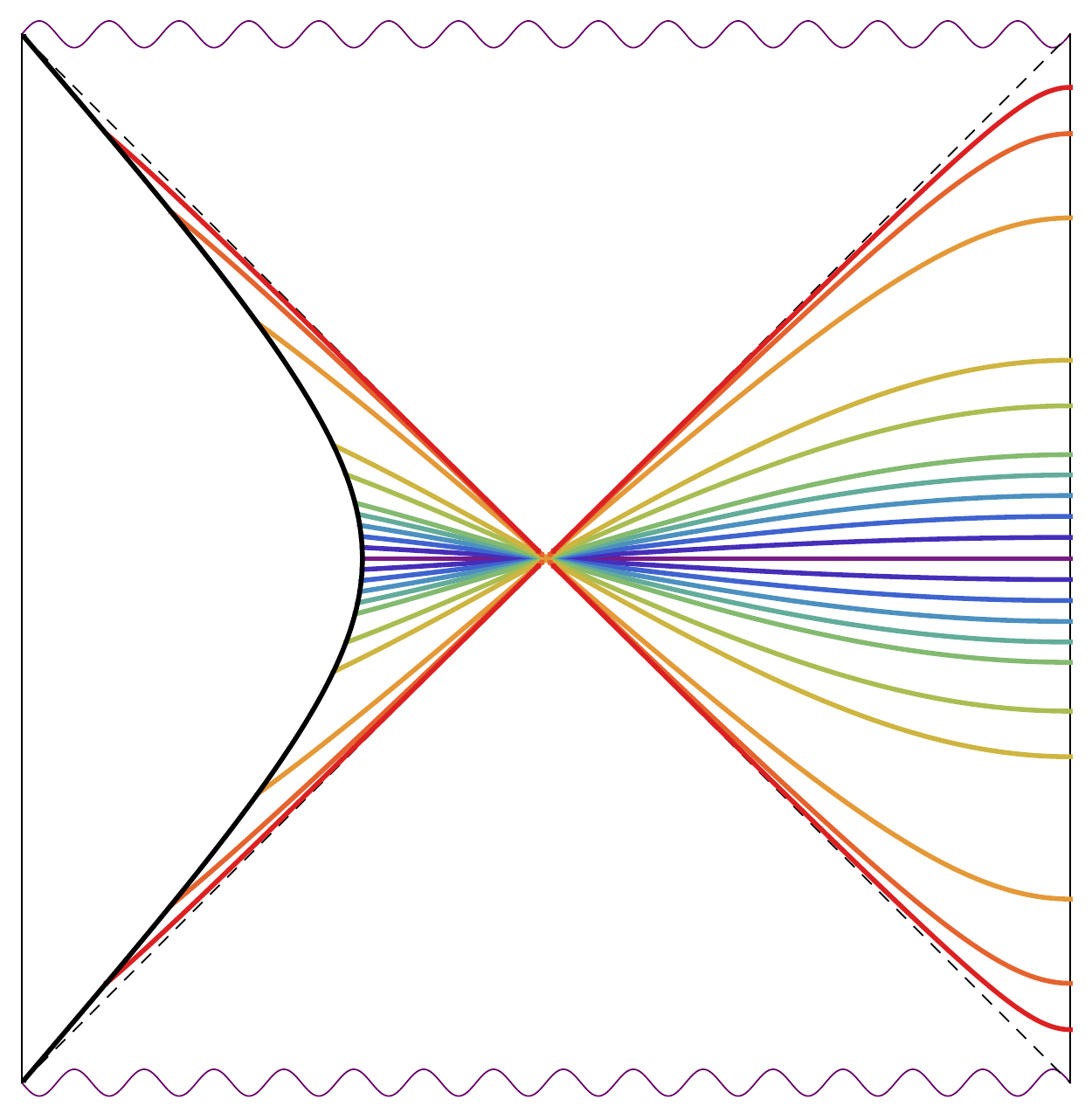}  
\caption{ 
AdS portion of the extremal solution for the static bubble
(which is shown in the black solid line).
These surfaces have all $P_o=0$ and never enter
the black and white hole 
regions of the geometry.  }
\label{static-bubble-extremal-surfs}
\end{figure}

For the strictly static bubble configuration $\mu=\mu_0$, 
eq. (\ref{static-relation-between-Pi-Po}) tells us that $P_o=0$,
so the complexity rate identically vanishes.
In this setup, the time-translation symmetry is not broken 
by the presence of the bubble and $\p/\p t$ is an everywhere well-defined Killing vector.
Also, the extremal surfaces never enter the black and white hole 
regions of the AdS geometry (see figure \ref{static-bubble-extremal-surfs}).

 \begin{figure*}
\includegraphics[scale=0.4]{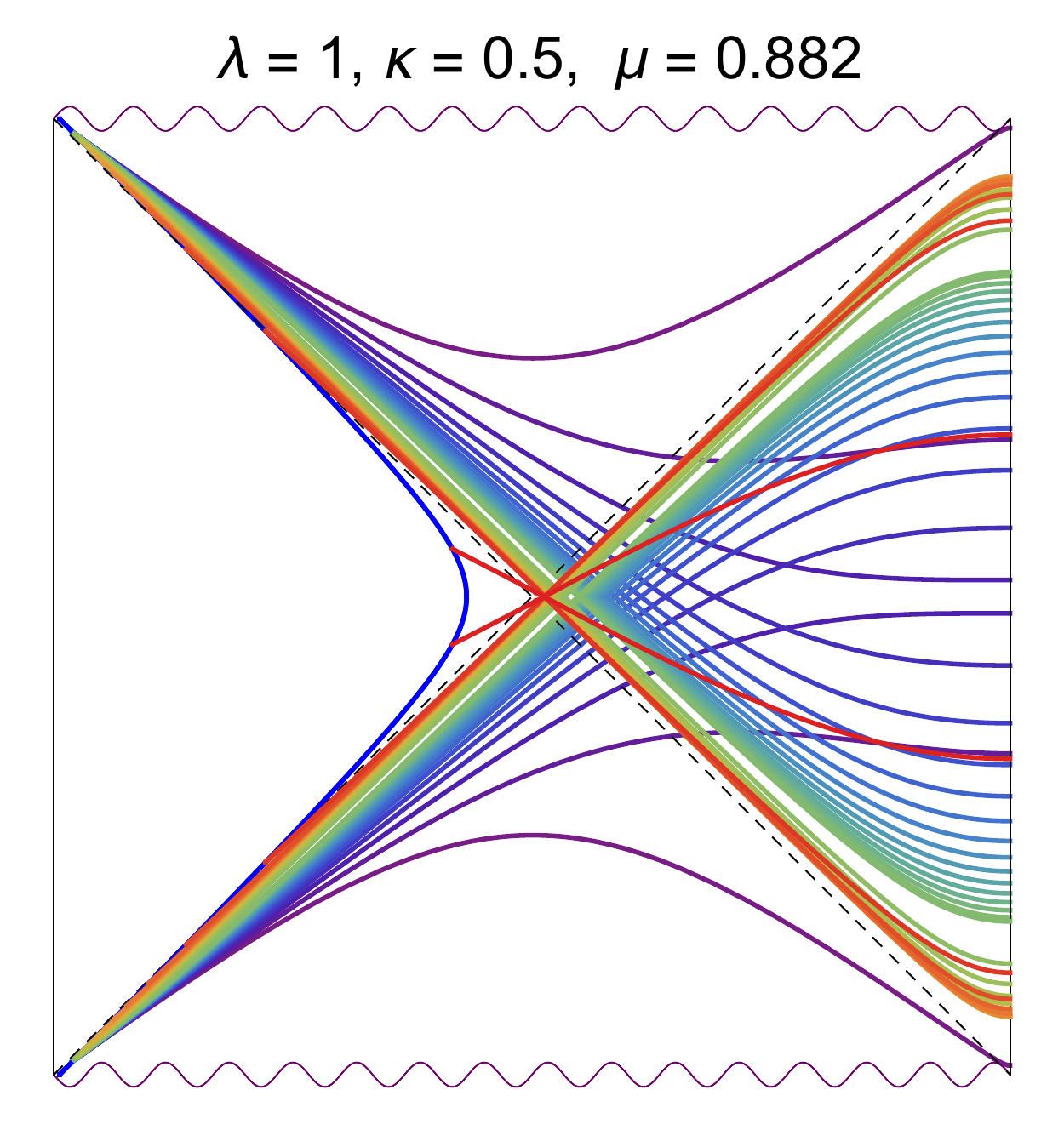}  \qquad
\includegraphics[scale=0.6]{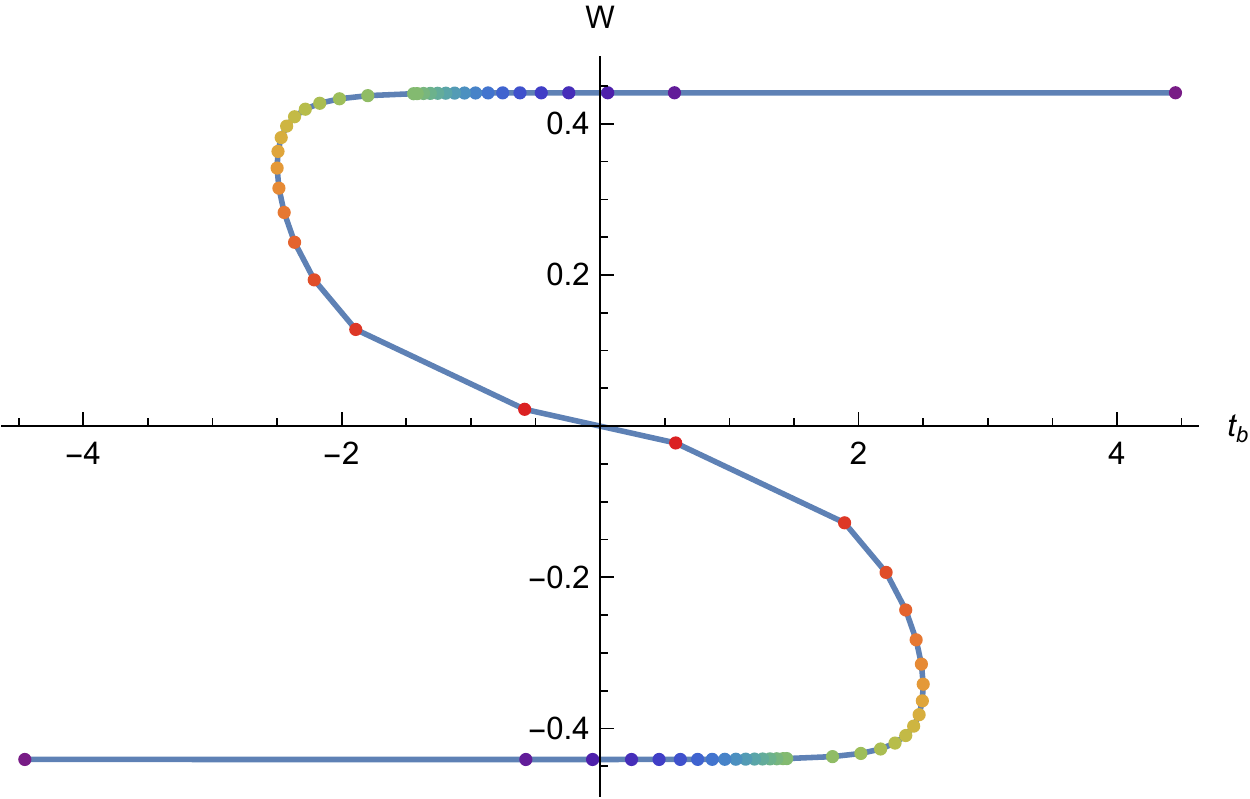}
\caption{ 
Left panel: AdS portion of the extremal surfaces
 for a not so small bubble.  
The value of $\mu$ is rather close to the static bubble configuration, which, with the chosen values of $\l$ and $\kappa$,
 is realized for $\mu_0 \approx 0.88278$.
 Right panel: plot of the
the complexity rate. 
 }
\label{rainbow-almost-static}
\end{figure*}     

\begin{figure*}
\includegraphics[scale=0.4]{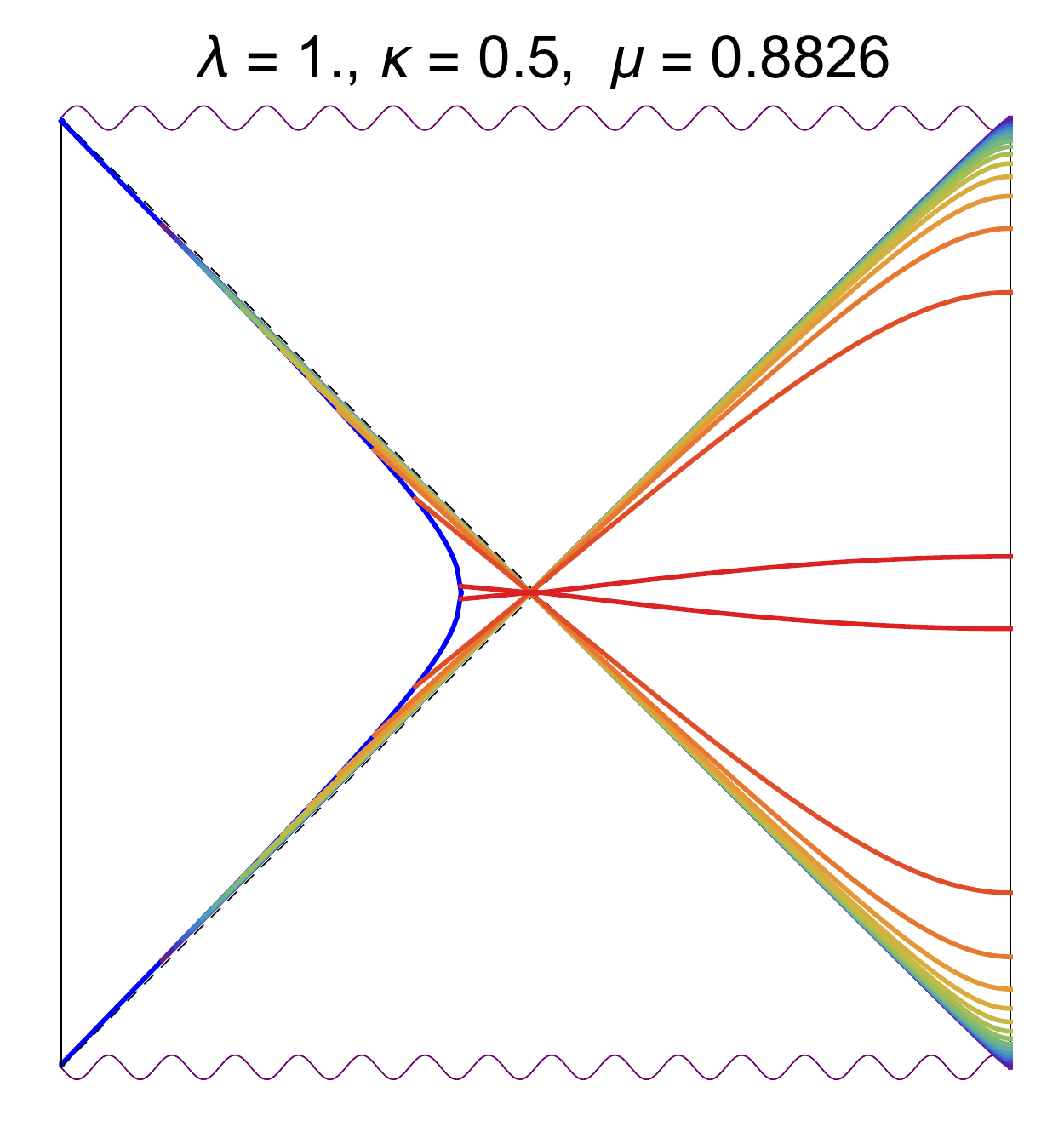}  \qquad
\includegraphics[scale=0.6]{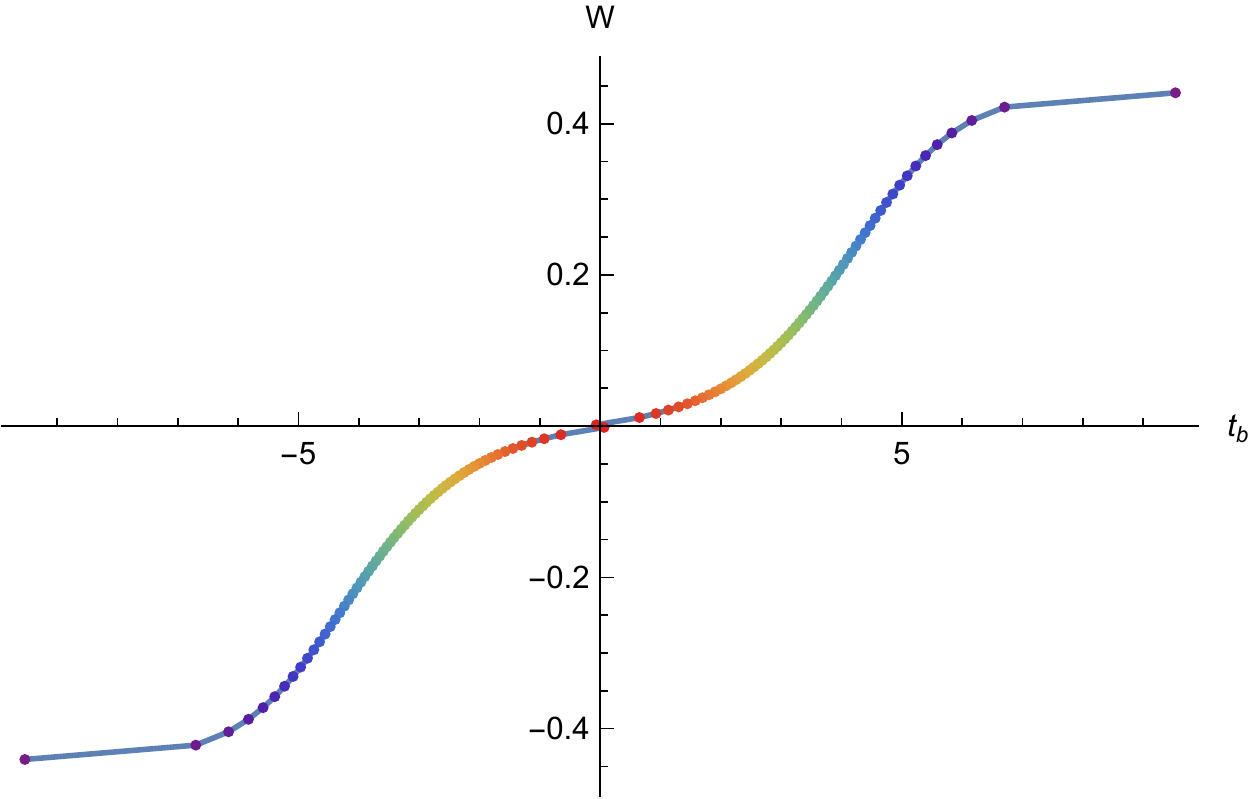}
\caption{ 
Left panel: AdS portion of the extremal surfaces
for an example of not so large bubbles with a $\mu$ rather close to
the static bubble value $\mu_0 \approx 0.88278$.
Right panel: plot of the 
the complexity rate. 
}
\label{rainbow3-almost-static}
\end{figure*}     

Let us now comment on how the static bubble is realized as a limit of the small and large bubbles:
\begin{itemize}
\item
As $\mu$ approaches the critical value $\mu_0$ from the small bubble direction,
the time $T_b$ in eq. (\ref{finestra-Tb}) tends to $+\infty$,
and the slope of the complexity rate in the central unstable branch tends to zero,
see figure \ref{rainbow-almost-static}.
This is consistent with $K \to -\infty$ for $\mu \to \mu_0$,
as shown figure \ref{region-S-1}.
In the static bubble limit $\mu \to \mu_0$, just the central branch solution
survives and the complexity rate is zero.
Interestingly, this limit is discontinuous, 
since the central branch is discarded
by the maximum volume prescription, as indicated by
the dashed line in  figure \ref{ExSurfChoice-1}.	
\item
As $\mu$ approaches the static bubble value
$\mu_0$ from the large bubble direction,
at small $t_b$ the complexity rate grows very slowly, see figure \ref{rainbow3-almost-static}.
When $\mu \to \mu_0$, the complexity rate remains frozen at zero for a large time.
This is consistent with $K$ diverging to $+\infty$
in this limit, as shown in figure \ref{no-region-S-large-bubbles}.
\end{itemize}


\subsection{Complexity of formation}

While it is possible to look at  the collapse of a very small bubble from the outside of the black hole, 
for all the other kinds of bubble an external observer just see a black hole horizon.
Thus, we can ask whether complexity can help us to discriminate,
for a given value of the boundary time $t_b$,
 between a large and small bubble state with the same $\mu$.
 To look for an answer, we consider the complexity of formation \cite{Chapman:2016hwi} at $t_b=0$, at which time $P_i=P_o=0$.
 To get a finite quantity, 
 we subtract from the volume of the bubble $\mathcal{V}_{\rm large}$ (or $  \mathcal{V}_{\rm small}$) at $t_b=0$ 
the outside volume of the BTZ black hole at the same boundary time:
 \bea
 \mathcal{V}_{\rm BTZ}&=& 2 \pi \int_{\sqrt{\mu}}^\Lambda \frac{r}{\sqrt{f_o}} dr 
 = 2 \pi \, \sqrt{\Lambda^2-\mu} \nl
 &=& 2 \pi \, \Lambda +O\le \frac{1}{\Lambda}\ri\, ,
 \eea
where $\Lambda$ is the AdS UV cutoff.
The complexity of formation is thus proportional to
  \bea
   \Delta \mathcal{V}_{\rm large}&=& \mathcal{V}_{\rm large}- \mathcal{V}_{\rm BTZ} \, ,
   \nl
      \Delta \mathcal{V}_{\rm small}&=& \mathcal{V}_{\rm small }- \mathcal{V}_{\rm BTZ} \, .
   \eea
From a direct evaluation from eqs. (\ref{volume-caso-rprimo-o-positivo}),
(\ref{volume-caso-rprimo-o-negativo}),
and (\ref{volume-caso-very-large-bubble}),
we find
\beq
\frac{ \Delta \mathcal{V}_{\rm small}}{2 \pi}=
 \frac{1-\sqrt{1-R_{\rm max}^2 \l} }{\l} \mp \sqrt{R_{\rm max}^2-\mu } 
 \label{delta-volume-1}
\eeq
where the $\mp$ sign refers to the cases
$ 0< \mu < \mu_s$ and  $\mu_s < \mu < \mu_0 $, respectively.
The regularized volume of large bubbles instead is
 \beq
 \frac{\Delta \mathcal{V}_{\rm large}}{2 \pi}=
 \sqrt{R_{\rm min}^2-\mu } +\frac{1 \pm \sqrt{1-R_{\rm min}^2 \l} }{\l}
 \label{delta-volume-2}
 \eeq
 where the $\pm$ sign refers to the cases
$ 0< \mu < \mu_h$ and $\mu_h < \mu < \mu_0$, respectively.

 \begin{figure}
\includegraphics[scale=0.7]{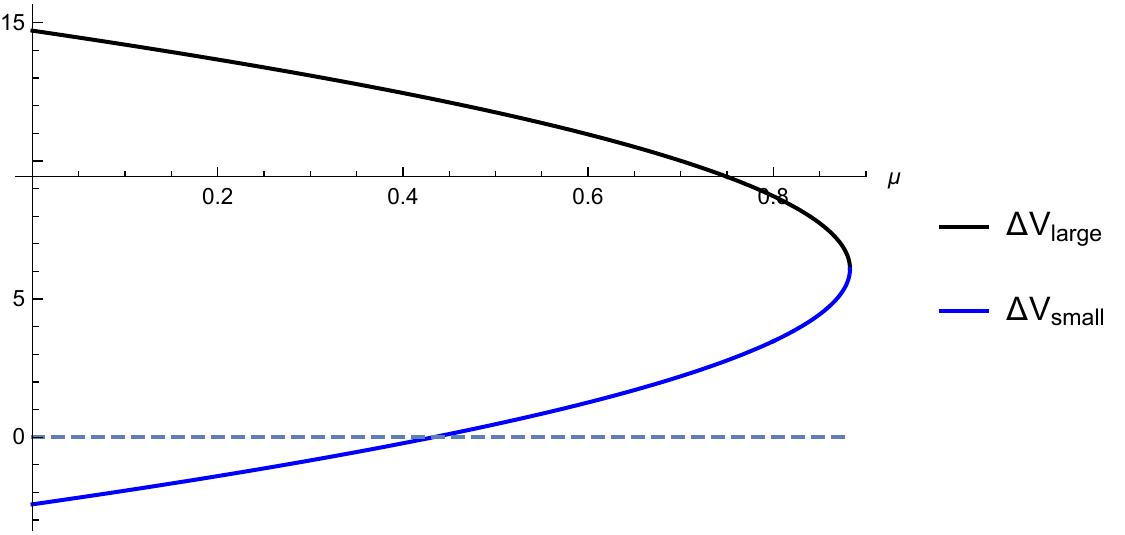} 
\caption{ Plot of  $ \Delta \mathcal{V}_{\rm small}$
 and  $\Delta \mathcal{V}_{\rm large}$ in eqs. (\ref{delta-volume-1}) and  (\ref{delta-volume-2})
 as functions of $\mu$, for $\l=1$ and $\kappa=0.5$. 
 }
\label{comple-di-formazione}
\end{figure}    

A plot of $ \Delta \mathcal{V}_{\rm large}$ and $ \Delta \mathcal{V}_{\rm small}$
as function of $\mu$ is shown in figure \ref{comple-di-formazione}.
The  large bubble has always a larger complexity than the small one.
This confirms the intuition that a large bubble state is more complex
compared to a small bubble one with the same mass.
Also,   $\Delta \mathcal{V}_{\rm small}$ is negative for $\mu \to 0$.
Very small bubbles in the limit $\mu \to 0$
then correspond to less complex states compared to the BTZ eternal black hole.


\section{Complexity with a dS stretched horizon}
\label{section-complexity-stretched-horizon}

For a very large bubble with $0 < \mu < \mu_h$, 
the geometry includes a complete dS stretched horizon.
Consequently, we can consider a  different prescription to compute 
volume complexity, in which the extremal codimension one
surface is attached both to the AdS boundary and to the dS stretched horizon.
This is an intermediate situation between 
the configuration used in the CV proposal in AdS \cite{Stanford:2014jda}
and in dS \cite{Susskind:2021esx}.
The stretched dS horizon is located
at constant $r={r_{\rm sh}}$, where
\beq
r_{\rm sh}=\frac{1}{\sqrt{\l}} \, \le 1 - \epsilon  \ri \, ,
\eeq
and $\epsilon \to 0$ is the horizon cutoff.
An $\epsilon>0$ is necessary to define a notion of time on the dS boundary,
as  the  horizon is a null hypersurface for $\epsilon=0$.

 We choose  the parameter $l$ in eq. (\ref{parametrizzazione-superficie-estremale})
   in such a way that it is positive and that 
$r(l=0)=r_{\rm sh}$.
The dS boundary removes the conical singularity
presented by extremal surfaces at the center of the static patch for $P_i \neq 0$.
For this reason, we are allowed to consider 
 an arbitrary value of $P_i$.
 
Extremal surfaces are stretched between
the dS  horizon
and the  AdS boundary,
so we can define the left and the right boundary times 
$t_L$, $t_R$
as the boundary
conditions
\bea
t_L= t_i(l=0)  \, , \qquad {\rm{where}} \qquad r(l=0)=r_{\rm sh} 
\nl
t_R= t_o(l=l_\Lambda) \, , \qquad {\rm{where}} \qquad r(l=l_\Lambda)=\Lambda 
\nl
\label{tempo-sinistro-destro}
\eea
and $\Lambda$ is the AdS UV cutoff.
In the definition of the boundary time,
we consider an arbitrary linear relation between $t_L$ and $t_R$
\beq
  t_L= \a_{t} \, t_R \, ,
 \label{legame-tempi-boundaries}
\eeq	
where $ \a_{t}$ is some opportune numerical constant.
The boundary time $t_b$ then is
\beq
t_b=-t_L =- \a_{t} \, t_R\, .
\label{boundary-time-stretched-horizon-case}
\eeq

As a technical difference compared to the usual Kruskal extension of the AdS black hole,
 there is no time-translation Killing vector $\p_t$  which is globally defined in all the geometry, because 
time invariance is broken by the trajectory of the bubble.
If such a symmetry existed, it would provide an appropriate 
value of $\a_t$ which would give a zero complexity rate.
We then expect a non-trivial time dependence of complexity 
for every value of $ \a_{t}$.

\subsection{Extremal surfaces}

In the AdS part of the geometry,
the domain wall is located into the left exterior of the black hole.
Thus, for the extremal surface to cross the black or white hole interior, 
we must require $\rho_o(R)<0$.
Instead, the sign of $\rho_i(R)$  might be positive or negative.
Explicitly,
\bea
&& \rho_o(R)=-\sqrt{P_o^2+f_o(R) \, R^{2}} \, , \nl
&& \rho_i(R)=\pm \sqrt{P_i^2+f_i(R) \, R^{2}} \, .
\eea
The condition for the extremal surface to
reach the AdS boundary without falling into the black hole singularity
is  $-\mu/2 \leq P_o \leq  \mu/2$
(see the plot of the potential in the right panel of figure \ref{Uio-figure}).
By time-reflection invariance, we take $P_o=P_i=0$ 
at $t_b=0$, when the extremal surface meets the domain wall at 
$R=R_{\rm min}$.

Under these assumptions, 
the matching condition in eq. (\ref{snell-law-generic-3}) reads
\bea
&& P_i  \frac{ d {T}_i}{dR} \pm \frac{\sqrt{P_i^2+f_i(R) \, R^{2}}}{f_i(R)} = \nl
&& =P_o \frac{ d {T}_o}{dR} - \frac{\sqrt{P_o^2+f_o(R) \, R^{2}} }{f_o(R)} \, ,
\label{snell-law-stretched-horizon}
\eea
where the $\pm$ sign corresponds  to
 $\rho_i(R) >0$ or $\rho_i(R) <0$, respectively.
 We choose the sign of  $dT_o/dR$ as in eq. (\ref{scelta-segno-To}),
and the sign of  $dT_i/dR$ in eq. (\ref{eq-diff-Ti-To}) as follows
\bea
&& \frac{dT_i}{d R}=- \frac{1}{2 \kappa}  \frac{w_i(R)}{(1- \l R^2)\sqrt{A R^4 - B R^2 +C} } \, ,
\nl
&& w_i(R)=1+ \mu - R^2 (1+\l-\kappa^2) \, .
\label{scelta-segno-Ti}
\eea
As a function of $\mu$,
the quantity $w_i(R_{\rm min})$
  is positive for $\mu>\mu_h$ and negative for $\mu<\mu_h$.
  For $\mu = \mu_h$, $w_i(R_{\rm min})$ vanishes and
 the bubble initially sits exactly at the dS bifurcation point.

Equation (\ref{snell-law-stretched-horizon}) is solved by
\bea
&& P_i=\frac{P_o (w_i \, w_o \pm\xi)+
\sqrt{\xi } (w_i \pm w_o) \sqrt{P_o^2+f_o R^2 } }
   {4 \kappa ^2 R^2 f_o} \, , 
   \nl
&& \xi =-4 \kappa^2 \, V(R) \, R^2 \, ,  
   \label{soluzione-Pi-Po}
\eea
where $V(R)$ is given in eq. (\ref{V-di-erre-esplicito}).
We point out that the solution with $P_i=0$ reproduces the smooth extremal surfaces studied 
in section \ref{section-complexity-smooth-extremal surface}.
The $-$ sign
describes an extremal surface experiencing a refraction, see 
eq. (\ref{Po-positivo}), while the $+$ sign
denotes an extremal surfaces undergoing a reflection, see eq. (\ref{Po-negativo}).
Specializing to the physical refracted solution, 
eq. (\ref{soluzione-Pi-Po}) can be further simplified to
 \bea
 P_i  &=& \frac{1}{2 f_o} \left( 
P_o \, (1-\mu +R^2(1-\l-\kappa^2) ) \right.
\nl
&& \left. + { \sqrt{\xi } \sqrt{P_o^2+f_o R^2 } }
\right)
   \, .
   \label{risultato-Pi}
\eea
An extremal surface is specified by the conserved quantities $P_i$ and $P_o$,
which are related by eq. (\ref{risultato-Pi}).
Their values for given boundary times are fixed by the boundary condition in eq. (\ref{legame-tempi-boundaries}),
 as we will explain below.

For the extremal surface to reach the AdS boundary,
the turning point
defined in eq. (\ref{turning-point-dS})
must be at  $r_{t,{\rm dS}} \geq R$,
which holds true for
\beq
P_i^2 \geq \tilde{P}_i^2 \, , \qquad  \tilde{P}_i^2 =R^2 (\l \, R^2-1) \, .
\label{rprimo-interno-zero}
\eeq

For $R_{\rm min}<R<1/\sqrt{\l}$, 
the domain wall  is located on the right side of the dS static patch,
so $\rho_i(R)<0$.
We may expect that
 at some point  $R=\tilde{R} > 1/\sqrt{\l}$, whose location depends on the bubble parameters and on the choice of $\a_t$, 
the function $\rho_i(R)$ vanishes.
Then,  $\rho_i(R) >0$ for $R>\tilde{R}$.

The condition $\rho_i(\tilde{R})=0$ is equivalent
to  $P_i^2= \tilde{P}_i^2$,
see eq. (\ref{rprimo-interno-zero}).
In terms of $P_o$, this gives 
\beq
P_o=\tilde{P}_o = \frac{R(\mu-1+R^2(\l +\kappa^2-1) )}{2 \sqrt{\l R^2-1}} \, ,
\label{Poo-tildato}
\eeq
which is well defined just for $R>1/\sqrt{\l}$.
Note that $\tilde{P}_o \to -\infty$ for $R \to 1/\sqrt{\l}$.
  For $R>1/\sqrt{\l}$ and $P_o>\tilde{P}_o$ we have
$\rho_i (R)<0$,
while for $P_o<\tilde{P}_o$ we have  $\rho_i (R)>0$.

\subsection{Complexity}

For the calculation of volume and boundary time,
let us distinguish between two cases: 
\begin{itemize}
\item {\bf $\rho_i(R) <0$. }
The volume is given by
\bea
\frac{\mathcal{V}}{2 \pi } &=&
  \int_{r_{sh}}^{r_{t,\rm dS}} \,\frac{ r^{2} }{\sqrt{P_{i}^2  +f_{i}(r) \,  r^{2}} }\, d r 
  \nl
    && -   \int^{R}_{r_{t,\rm dS}} \,\frac{ r^{2} }{\sqrt{P_{i}^2  +f_{i}(r) \,  r^{2}} }\, d r 
  \nl
&&  -\int_R^{r_{t,\rm{AdS}}}  \,\frac{ r^{2} }{\sqrt{P_{o}^2  +f_{o}(r) \,  r^{2}} }\, d r
  \nl
&&+  \int_{r_{t,\rm{AdS}}}^{\Lambda}  \,\frac{ r^{2} }{\sqrt{P_{o}^2  +f_{o}(r) \,  r^{2}} }\, d r  
 \, .
    \label{volume-streched-horizon-caso1}
\eea
From eq. (\ref{tempo-A}), we can write the following expression for the AdS boundary time
\bea
&& t_R=T_o(R)
+ \int_{R}^{r_{t,\rm{AdS}}}    \frac{P_{o} }{ f_o \sqrt{P_{o}^2  +f_{o}(r) \,  r^{2}}}  dr \nl 
&& -\int^{\Lambda}_{r_{t,\rm{AdS}}}   \frac{P_{o}}{f_o \sqrt{P_{o}^2  +f_{o}(r) \,  r^{2}}}  dr
 \, ,
\label{tempo-ads-destra}
\eea
and the following expression for the dS boundary time
\bea
&& t_L=  T_i(R)+
   \int_{{r_{\rm sh}}}^{r_{t,{\rm dS}}}    \frac{P_{i} }{ f_i \sqrt{P_{i}^2  +f_{i}(r) \,  r^{2}}}  dr \nl
&& -  \int^{R}_{r_{t,{\rm dS}}}    \frac{P_{i} }{ f_i \sqrt{P_{i}^2  +f_{i}(r) \,  r^{2}}}  dr 
 \, .
 \label{tempo-ds-sinistra-rprimo-o-neg}
\eea

\item {\bf $\rho_i(R) >0$. }
The volume can be written as
\bea
&& \frac{\mathcal{V}}{2 \pi } =
   \int_{r_{\rm sh}}^R \,\frac{ r^{2} }{\sqrt{P_{i}^2  +f_{i}(r) \,  r^{2}} }\, d r 
  \nl
&&     -\int_R^{r_{t,\rm{AdS}}}  \,\frac{ r^{2} }{\sqrt{P_{o}^2  +f_{o}(r) \,  r^{2}} }\, d r  \nl
&& +   \int_{r_{t,\rm{AdS}}}^{\Lambda}  \,\frac{ r^{2} }{\sqrt{P_{o}^2  +f_{o}(r) \,  r^{2}} }\, d r  
  \, .
    \label{volume-streched-horizon-caso2}
\eea
Equation (\ref{tempo-ads-destra}) is still valid.
Instead, from eq. (\ref{tempo-A}) we find that the time on the dS boundary is
\beq
 t_L=
T_i(R) +  \int_{{r_{sh}}}^{R}    \frac{P_{i} }{ f_i \sqrt{P_{i}^2  +f_{i}(r) \,  r^{2}}}  dr 
 \, .
 \label{tempo-ds-sinistra}
\eeq
\end{itemize}

In order to find values of  $P_i$ and $P_o$ consistent with
 the boundary condition, 
 we need to solve  eq. (\ref{legame-tempi-boundaries}) 
where $t_R$ and $t_L$ are specified by (\ref{tempo-ads-destra})
and (\ref{tempo-ds-sinistra-rprimo-o-neg}) or (\ref{tempo-ds-sinistra}),
respectively. This integral equation can be solved numerically
by the shooting method.

\begin{figure*}
\includegraphics[scale=0.3]{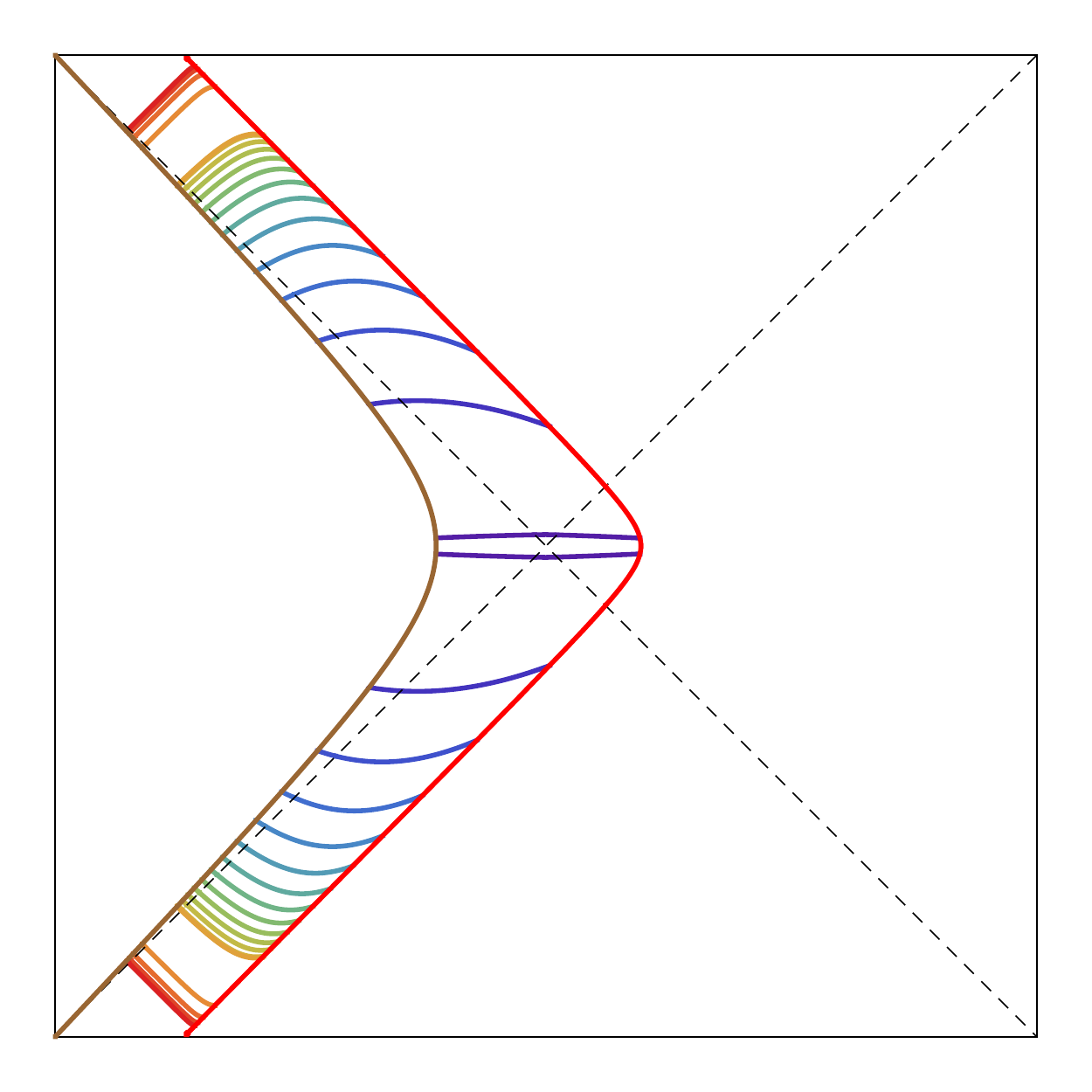}  
\qquad
\includegraphics[scale=0.3]{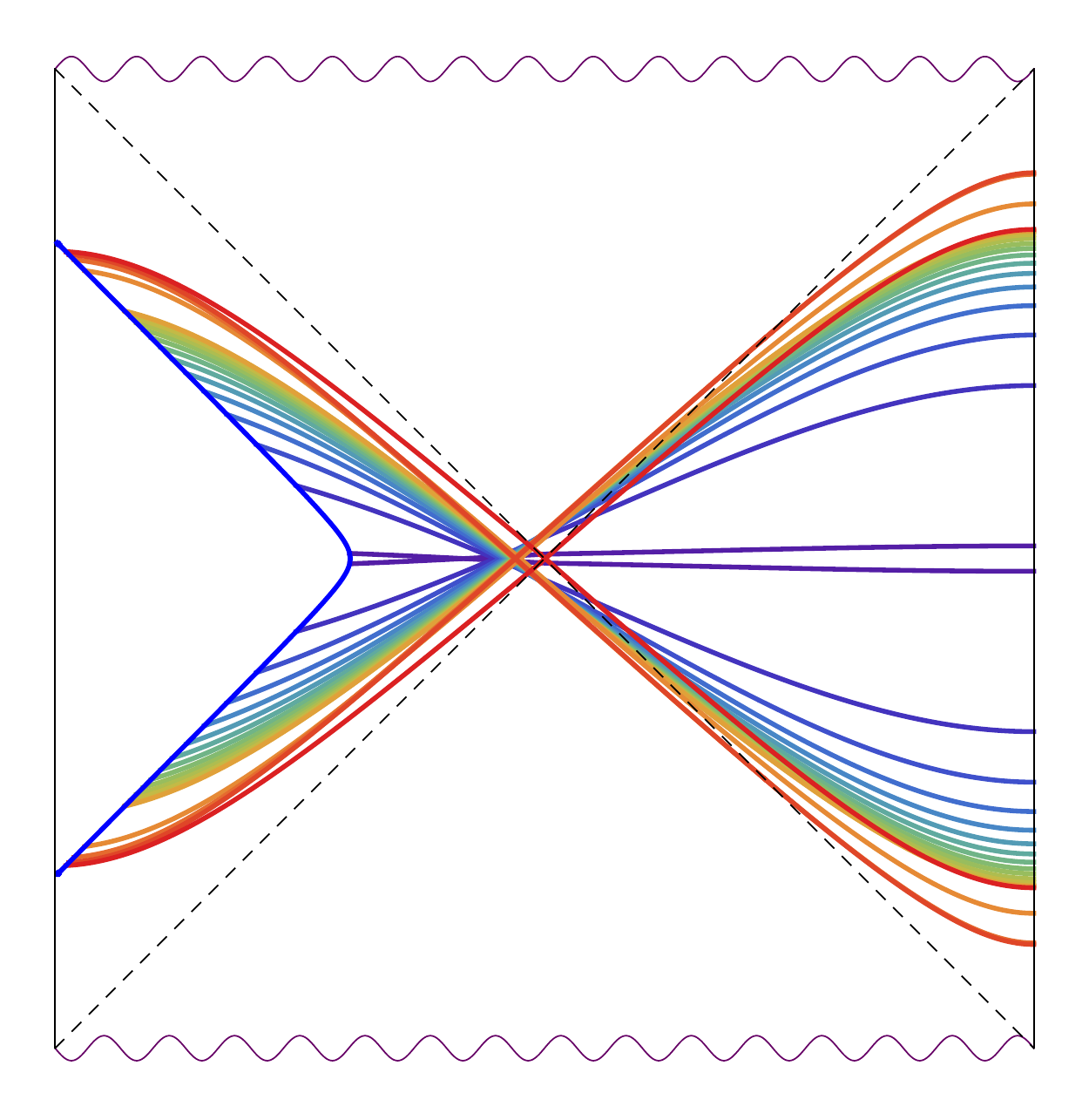} 
\qquad
\includegraphics[scale=0.4]{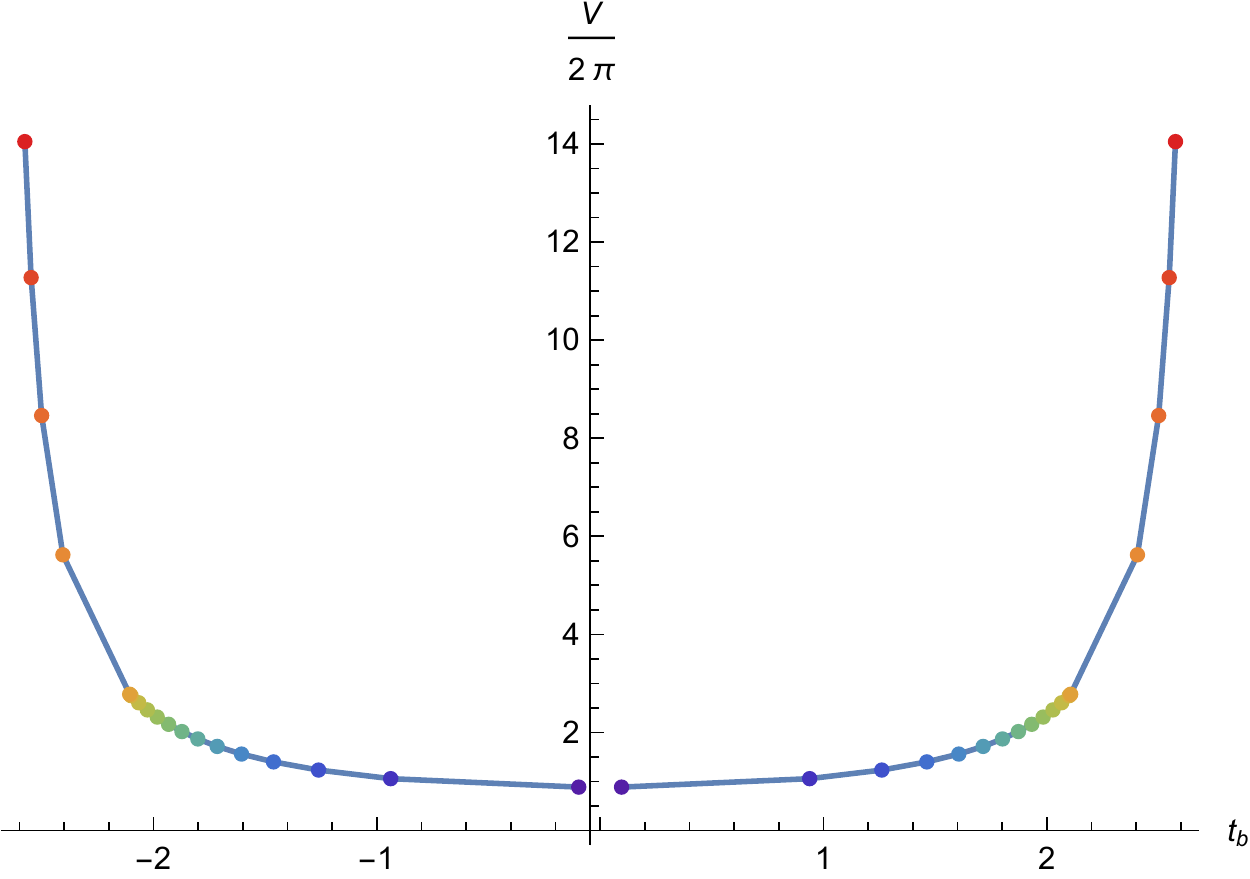} 
\caption{ Left and central panels: extremal surfaces 
in the dS and in the AdS portion of the Penrose diagram, respectively.
Right panel: plot of the volume as a function of boundary time $t_b$.
We set  $\a_t=1$, see eq. (\ref{legame-tempi-boundaries}),
 and we choose $\kappa=0.2$, $\l=1.5$, $\mu=0.4$, and $\epsilon=0.06$.
}
\label{SH-rainbow}
\end{figure*}

\begin{figure*}
\includegraphics[scale=0.3]{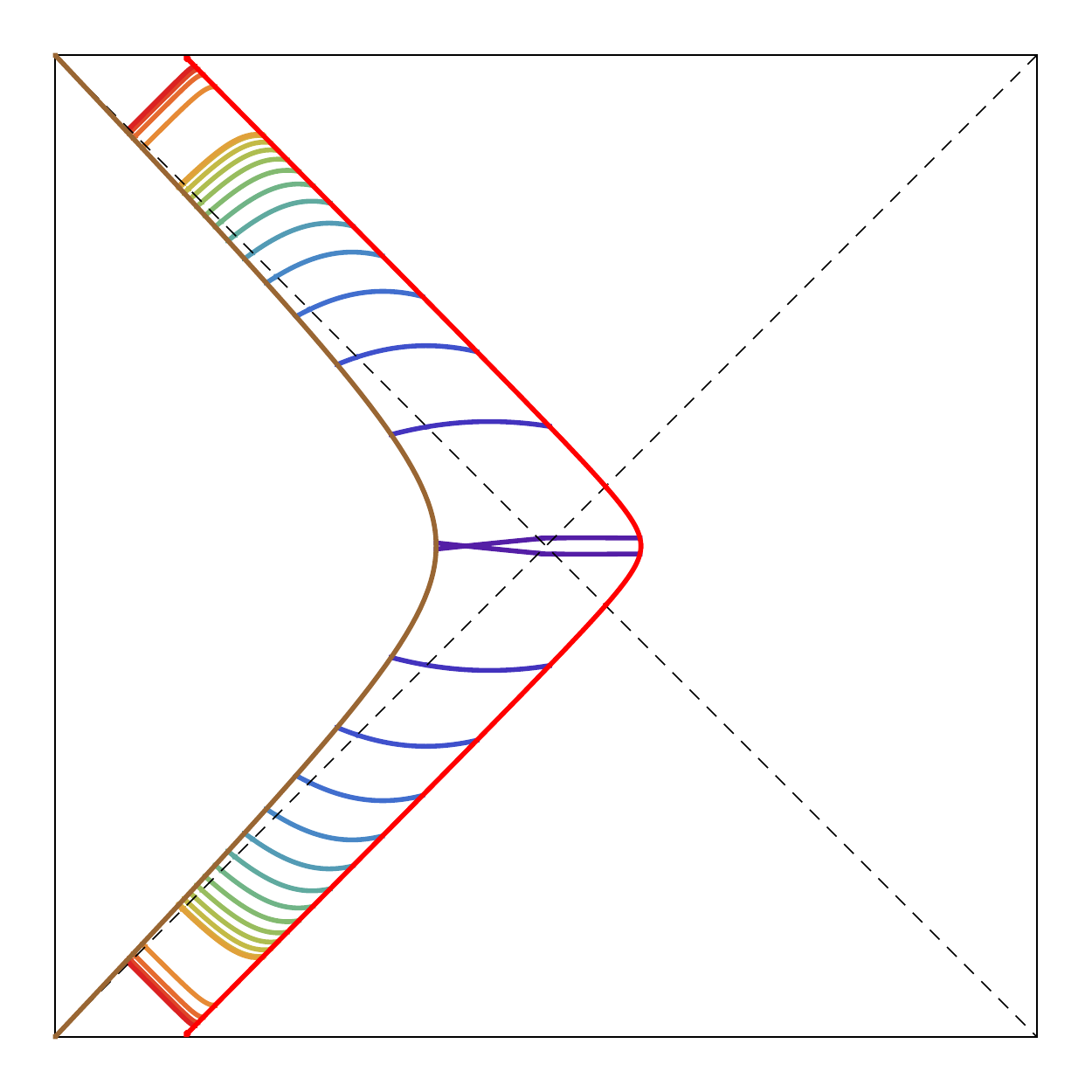}  
\qquad
\includegraphics[scale=0.3]{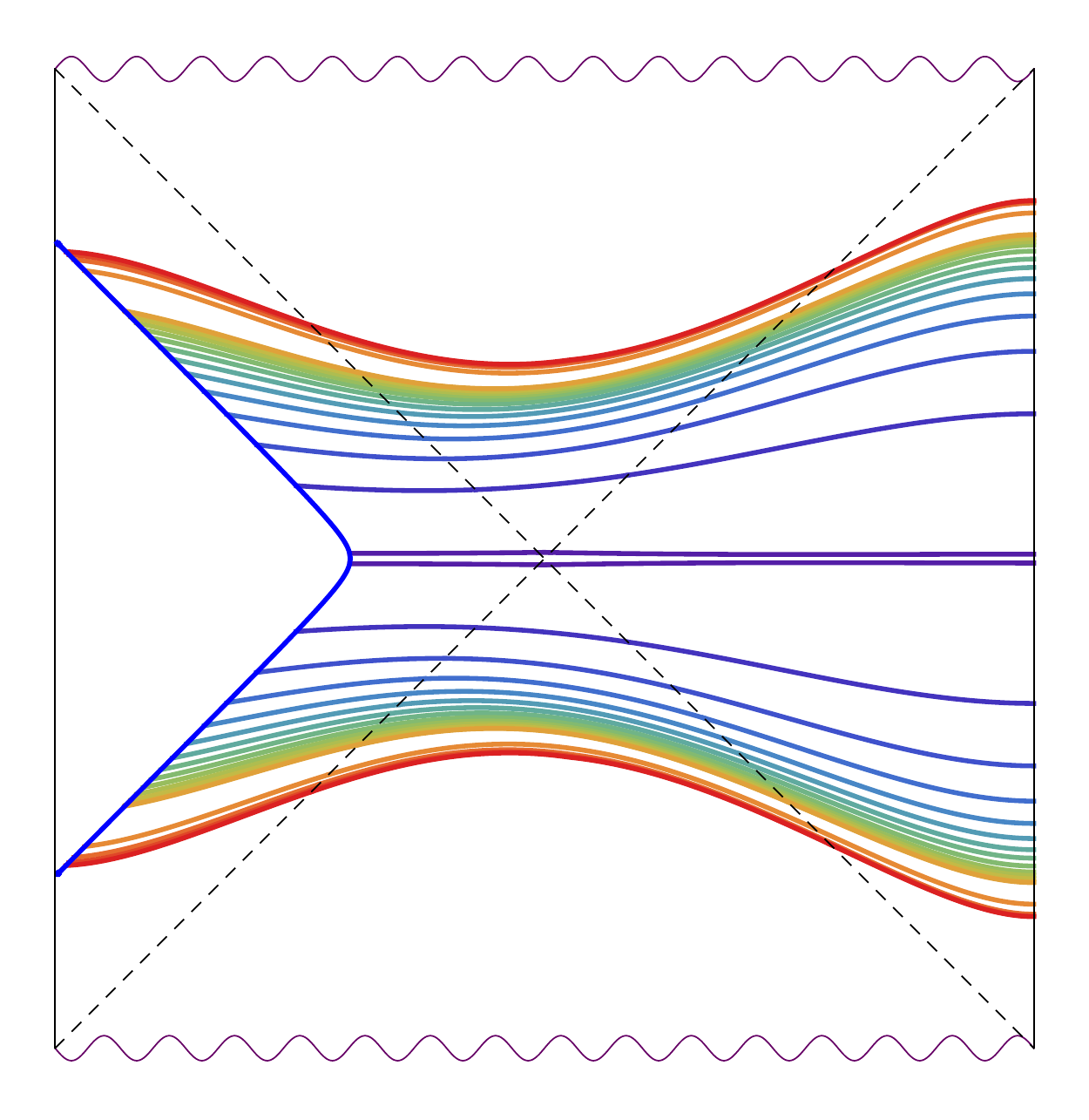} 
\qquad
\includegraphics[scale=0.4]{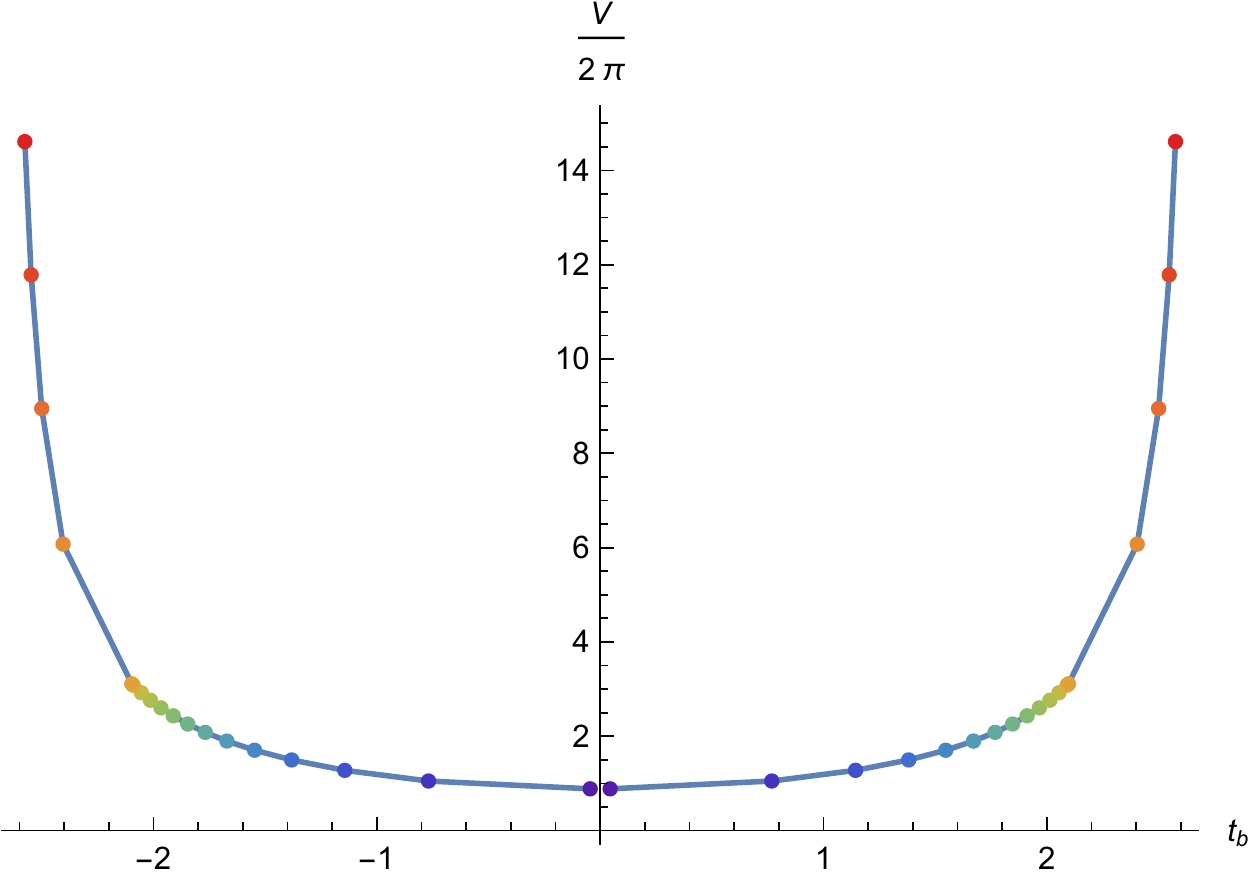} 
\caption{ Plots of the same quantities as in figure \ref{SH-rainbow},
with the choice $\a_t=-1$.
}
\label{SH-rainbow-2}
\end{figure*}

We can then compute the extremal surfaces and their volume numerically.
In figure \ref{SH-rainbow} we show the result for $\a_{t}=1$,
while in figure  \ref{SH-rainbow-2} we display the result
for the choice $\a_{t}=-1$. In both cases
we find a hyperfast complexity rate, because
the volume of the extremal surface diverges for a finite value of the boundary time $t_b$.


\subsection{Critical time behaviour}

Let us now discuss the limit of hyperfast complexity growth.
Expanding eq. (\ref{risultato-Pi}) for large $R$, we find that
\beq
P_i (R) \approx \frac{\sqrt{(\l+(\kappa-1)^2 )(\l+(\kappa+1)^2 ) }}{2} R^2 + O(R^0) \, .
\label{P-i-R-grande}
\eeq
The dependence of $P_i$ on $P_o$ shows up just at order $O(R^0)$.
At large $R$, we can approximate $T_i(R)$ in eq. (\ref{eq-diff-Ti-To}) as 
\bea
\frac{d T_i}{dR}  &\approx&  \frac{1}{f_i} \frac{1+\l - \kappa^2}{\sqrt{ (\l+\kappa^2-1)^2+ 4 \l} } \, ,
\nl
T_i(R) &\approx& \frac{1+\l - \kappa^2}{\sqrt{ (\l+\kappa^2-1)^2+ 4 \l} } r^*_i(R) + Q \, ,
\nl
\label{approssimazione-bubble-grandi tempi}
\eea
where $Q$ is an integration constant. Note that $1+\l - \kappa^2>0$, because
$\kappa < 1$ in the very large bubble regime.

Let us discriminate between two cases:
\begin{itemize}
\item $\l+\kappa^2>1$. At large $R$, 
from eq. (\ref{Poo-tildato}) we have $\tilde{P}_o \to +\infty$,
then $\rho_i(R)>0$.
At large $P_i$, we can approximate eq. (\ref{tempo-ds-sinistra}) as
\bea
t_L &\approx&
T_i(R) +  \int_{{r_{\rm sh}}}^{R}    \frac{dr  }{ f_i }  
\nl &=& T_i(R) +r_i^*(R)-r_i^*(r_{sh})  \, ,
\eea
 which gives
\bea
t_L &\approx&
\omega \frac{1}{R \l}+ Q
-\frac{1}{4 \sqrt{\l}} \log \frac{4}{ \epsilon^2} \, ,
\nl
\omega &=& 1+\frac{1+\l - \kappa^2}{\sqrt{ (\l+\kappa^2-1)^2+ 4 \l} } \, .
\label{tempo-interno-stima}
\eea
In the above expression, we have $\omega>0$
because for very large bubbles $\kappa<1$.
For large $R$,
the dS part of the volume in eq. (\ref{volume-streched-horizon-caso2})
diverges linearly:
 \bea
&& \frac{\mathcal{V}_{dS}}{2 \pi }  =
    \int_{r_{\rm sh}}^R \,\frac{ r^{2} }{\sqrt{P_{i}^2  +f_{i}(r) \,  r^{2}} }\, d r \, 
    \nl
  &&      \approx R    \int_{\frac{r_{\rm sh}}{R}}^1 \,
        \frac{ y^{2} }{\sqrt{\frac{(\l+(\kappa-1)^2 )(\l+(\kappa+1)^2 ) }{4}   -\l  y^4 } }\, d y   \, .
        \nl
        \label{large-R-divergenza-volume}
\eea
The divergence of the volume for large $R$ 
 is at the  value of $ t_L $ given in eq. (\ref{tempo-interno-stima}).

\item $\l+\kappa^2<1$. At large $R$ we have $\tilde{P}_o \to -\infty$,
which gives $\rho_i(R)<0$. In this limit,
combining eq. (\ref{turning-point-dS}) with eq. (\ref{P-i-R-grande}),
we find 
\beq
r_{t,{\rm dS}} \approx \frac{R}{\sqrt{2}} \le \frac{(\l+(\kappa-1)^2 )(\l+(\kappa+1)^2 ) }{\l} \ri^{1/4} \, .
\eeq
At large $R$, corresponding to large $P_i$,
we can approximate $t_L$ in eq. (\ref{tempo-ds-sinistra-rprimo-o-neg}) as
\beq
t_L \approx
T_i(R) +  \int_{{r_{\rm sh}}}^{r_{t,{\rm dS}}}    \frac{dr  }{ f_i } +  \int_{R}^{r_{t,{\rm dS}}}    \frac{dr  }{ f_i } \, ,
\eeq
which gives
\beq
t_L \approx
\tilde{\omega} \frac{1}{\l R}
+ Q -\frac{1}{4 \sqrt{\l}} \log \frac{4}{ \epsilon^2}  \, ,
\label{tempo-interno-stima-2}
\eeq
where
\bea
\tilde{\omega} &=& \frac{2 \sqrt{2} \l^{1/4}}{[(\l+(\kappa-1)^2 )(\l+(\kappa+1)^2 )]^{1/4}} \nl
&+& \frac{1+\l - \kappa^2}{\sqrt{ (\l+\kappa^2-1)^2+ 4 \l} } -1  \, .
\eea
In the case at hand, the quantity $\tilde{\omega}$ is always positive.
At large $R$ and at the finite time in eq. (\ref{tempo-interno-stima-2}), the volume again diverges linearly as in eq. (\ref{large-R-divergenza-volume}).
\end{itemize}
Summarizing, in both cases the volume complexity diverges for $t_b= t_{\rm cr}$,
where  the critical time is
\beq
 t_{\rm cr}=\frac{1}{4 \sqrt{\l}} \log \frac{4}{ \epsilon^2}-Q \, ,
 \label{tempo-critico-divergenza}
\eeq
and $Q$ is the integration constant defined in eq. (\ref{approssimazione-bubble-grandi tempi}).
The divergence at $t_b= t_{\rm cr}$ takes place independently
of the  parameter $\a_t$ in eq. (\ref{legame-tempi-boundaries}).
For $t_b>t_{\rm cr}$, no extremal surface exists which connects the left dS and right AdS
boundaries.
As in \cite{Jorstad:2022mls}, we may regularize the divergence in 
the complexity rate by introducing a cutoff surface at large $r=r_{\rm cut}$
nearby the future dS infinity. With this regularization, the complexity rate would remain finite 
and at $t>t_{\rm cr}$ it would saturate at a value divergent in the UV
cutoff $r_{\rm cut}$. 


 \section{Conclusions}
 \label{section-conclusions}

 In this paper we investigated the time dependence of volume complexity
 in a class of asymptotically AdS$_3$ spacetimes which include a dS$_3$ bubble in their interior.
 We first focused on extremal surfaces attached just at the 
 AdS boundary and smooth everywhere into the interior spacetime.
With the exception of the static bubble configuration,
we found that complexity asymptotically grows linearly
as a function of time, with the same rate as for the BTZ black hole.
For large bubbles, the asymptotic value of the complexity rate is always reached from below. For small bubbles, 
the asymptotic limit can be instead reached either from below (case 1) or from above (case 2), depending on the parameter space
(see section \ref{section-late-time-small-bubbles}).

The static bubble configuration gives rise to a time-independent complexity, 
so it does not match the expectation,
generically satisfied by AdS black holes,
that complexity rate at late time is of the same order of magnitude as $T S$,
with $T$ the temperature and $S$ the entropy of the system \cite{Susskind:2014moa}.
This fine-tuned solution, which is realized for $\mu=\mu_0$,
interpolates between the small and the large bubble regimes. 
As soon as a small perturbation is introduced,
the static bubble limit can be achieved
in two different ways:
\begin{itemize}
\item
If the limit $\mu \to \mu_0$ in the parameter space is approached
 from the large bubble configuration,
the complexity rate remains frozen to zero
for an initial amount of time which tends to infinity for  $\mu \to \mu_0$, 
see the right panel of figure \ref{rainbow3-almost-static}.
From this side, the asymptotic complexity rate of the BTZ black hole
is  recovered in a continuous way after an arbitrarily large time.
\item
If the limit $\mu \to \mu_0$ is approached from the small
bubble region of the parameter space, 
the static behavior of complexity emerges
from a class of extremal surfaces  with non-maximal volume,	
see figure \ref{rainbow-almost-static}.
 According to the CV prescription,
the non-maximal extremal solutions
should be discarded  for $\mu \neq \mu_0$  in favor of the global maximum,
as shown by the dashed line in figure \ref{ExSurfChoice-2}.
For $\mu \to \mu_0$ the  complexity rate resembles a step function,  
thus the limit is discontinuous, as in a first order phase transition.

Interestingly, the discarded solutions would give rise to a negative complexity rate,
 as for the two-dimensional centaur geometries studied in \cite{Chapman:2021eyy}.
 We may contemplate the possibility that 
other physical configurations could exist, which resemble our geometry nearby the discarded extremal surfaces and in which the global maximum is cut away.
If such a surgery were performed, it would give rise to
a negative complexity rate, resembling the two-dimensional case studied
in  \cite{Chapman:2021eyy}, in which no black hole horizon is present in the AdS region of the geometry.
\end{itemize}
With both procedures, for $\mu \to \mu_0$ the BTZ asymptotic complexity rate 
is recovered at late time. The limit is continuous from the large
bubble direction, while it is discontinuous from the small bubble side.
In this sense, the static bubble can be seen as a fine-tuned critical configuration
separating qualitatively different behaviors in the parameter space.

From the point of view of local observables on the AdS boundary,
the static bubble configuration is not distinguishable from 
the eternal BTZ black hole. The same holds for entanglement entropy
of subregions located on the AdS boundary.
To detect the discontinuous nature of the static bubble limit,
other quantum information probes, such as holographic complexity, are required.

Contrary to the result found in 
\cite{Susskind:2021esx,Jorstad:2022mls} for dS, 
if we consider  smooth extremal surfaces attached just to the AdS boundary
there is no hyperfast growth of complexity.
In section  \ref{section-complexity-stretched-horizon},
we checked that hyperfast growth is  recovered
in the very large bubble case
if we consider 
extremal surfaces  anchored both
at the AdS boundary and at the dS stretched horizon.
This choice should correspond to complexity of a pure thermofield 
double state which involves both an AdS and a dS boundary.
Defining the boundary time $t_b$ as in eq. (\ref{boundary-time-stretched-horizon-case}),
we find that the volume complexity diverges at $t_b= t_{\rm cr}$
given in eq. (\ref{tempo-critico-divergenza}), independently of the 
 parameter $\a_t$ in eq. (\ref{legame-tempi-boundaries}).

It is tempting to suggest the following interpretations for the two different
 ways to apply the CV conjecture in AdS geometries with very large dS bubbles:
\begin{itemize}
\item
the volume of  smooth extremal surfaces
 anchored just at the boundary of AdS 
 is proportional to
the complexity of a mixed  CFT state, obtained by
tracing over the dS degrees of freedom in the thermofield double state.
This situation is reminiscent of the
subregion complexity proposal \cite{Alishahiha:2015rta,Carmi:2016wjl,Agon:2018zso}.
\item
the volume of the extremal surface
anchored both at the AdS boundary and at the dS static patch horizon
is proportional to the complexity of the pure product  state.
\end{itemize}
Why the latter choice of boundary conditions  is possible just for very large bubbles is an interesting question.
Small bubble solutions do not contain a stretched horizon
in the dS portion of the geometry, so the only implementable prescription 
is to attach the extremal surfaces just to the AdS boundary. 
In this case, we can conjecture that the dual CFT state
does not arise from a partial trace over a
pure state involving a dS boundary.



\section*{Acknowledgements}

We thank S. Baiguera for useful discussions and comments
on the draft. 
The work of N.Z. is  supported by MEXT KAKENHI
Grant-in-Aid for Transformative Research Areas A
"Extreme Universe" No. 21H05184.

 \section*{Appendix}
\addtocontents{toc}{\protect\setcounter{tocdepth}{1}}
\appendix

\section{Details about Penrose diagrams}

\label{penrose-appe}

In this appendix we describe our conventions for 
 Penrose diagrams. 
For the BTZ external region, the relation between the
coordinates $(\tilde{t},\tilde{r})$ of the Penrose diagram
and the EF coordinates  $(u,v)$ can be expressed as
\bea
\tilde{t}&=& \frac{\pm \tan^{-1} ( e^{v \sqrt{\mu } } ) -\tan^{-1} ( e^{-u \sqrt{\mu } } ) }{2} \, , \nl
\tilde{r}&=& \frac{\pm \tan^{-1} ( e^{v \sqrt{\mu } } )  +\tan^{-1} ( e^{-u \sqrt{\mu } } ) }{2} \, ,
\label{Penrose-finali-AdS}
\eea
where the first sign is for $r > \sqrt{\mu}$ and the second sign for $r < \sqrt{\mu}$.

For the dS spacetime interior,
  we use the following choice of Penrose diagram coordinates 
\bea
\tilde{t} &=& \frac{\tan^{-1} ( e^{u \sqrt{\l } } ) \mp \tan^{-1} ( e^{-v \sqrt{\l } } ) }{2} \, , \nl
\tilde{r} &=& \frac{\tan^{-1} ( e^{u \sqrt{\l } } ) \pm \tan^{-1} ( e^{-v \sqrt{\l } } ) }{2} \, ,
\label{Penrose-finali-dS}
\eea
where the first sign is for $r < 1/\sqrt{\lambda}$ and the second sign for $r > 1/\sqrt{\lambda}$.

In this paper, Penrose diagrams are obtained by 
a parametric plot of the coordinates $(\pm \tilde{r},\pm \tilde{t})$,
using eq. (\ref{Penrose-finali-AdS}) for the external BTZ region and eq.
(\ref{Penrose-finali-dS}) for the dS interior.
The direction of increasing  $t$ on each side of the diagrams
is shown in figure \ref{penrose0}.


\section{Derivation of the refraction law for extremal surfaces}

\label{appendice-snell}

Let us assume that the change of variable between 
$t_{i,o}$ is of the form
\beq
t_o=G(r) \, t_i\, .
\label{tempo-o-tempo-i}
\eeq
The function $G$ will be specified later by requiring
the proper time on the bubble to be continuous.
In terms of the interior time coordinate $t_i$,
the outside metric in eq. (\ref{metric-zero})  reads
\bea
ds^2_o &=&- f_o  \, G^2 \, dt_i^2 -  f_o   \frac{d (G^2) }{ d r}  t_i \, dr \, dt_i \nl 
&+&  \frac{1-f_o^2 \le \frac{d G}{dr} \ri^2 t_i^2 }{f_o}  \, dr^2 
+ r^2 d \theta^2 \, ,
\eea
where $f_i$ and $f_o$ are functions of $r$ given in eqs. (\ref{f-BTZ}) and (\ref{f-dS}).
Hence, 
both the interior and the exterior metrics have the following form:
\beq
ds^2=- g(r,t_i) \, dt_i ^2+ \frac{dr^2}{f (r,t_i)} + 2 h (r,t_i) \, dr \, dt_i 
 +r^2 d \theta^2 \, \, ,
 \label{metrica-tutta-con-ti}
\eeq
where for the interior metric
\beq
g(r,t_i)=f_i  \, , \qquad f(r,t_i)=f_i \, , \qquad h(r,t_i)=0 \, , 
\label{metric-functions-interior}
\eeq
while for the exterior metric
\bea
g(r,t_i) &=& G^2 \, f_o  \, , 
\nl
 f(r,t_i) &=& \frac{f_o}{1-f_o^2 \le \frac{d G}{dr} \ri^2 t_i^2} \, , 
\nl
h(r,t_i) &=& -f_o G \, \frac{d G}{dr} \, t_i \, .
\label{metric-functions-esterno}
\eea
For later purposes, it is useful to evaluate the following quantities both inside and outside the bubble:
\beq
\left. \frac{g+ f h^2}{f} \right|_i =1 \, , \qquad
\left. \frac{g+ f h^2}{f} \right|_o =G^2 \, .
\label{useful-quantities}
\eeq
For the metric in eq. (\ref{metrica-tutta-con-ti}), the volume functional is
\bea
&& \mathcal{L} = r \, \sqrt{-g(r,t_i) \, t_i'^2 + \frac{r'^2}{f(r,t_i)}
+2 h(r,t_i) \, r' t_i' }
\nl 
&& \mathcal{V}_{i,o} = 2 \pi \int  \mathcal{L} \,  d l \, ,
 \, .
 \label{volume-functional-t}
\eea
It is convenient to fix the gauge as in eq. (\ref{gauge-fixing-generale}),
which is equivalent to
\beq
\sqrt{-g(r,t_i) \, t_i'^2 + \frac{r'^2}{f(r,t_i)}
+2 h(r,t_i) \, r' t_i' }= r \, .
   \label{gauge-fixing-Phat}
 \eeq

\subsection{Static bubble}

At constant $r$, continuity of the proper time on the domain wall fixes
\beq
G(r)  = \pm \sqrt{\frac{f_i(r)}{f_o(r)}} \, .
\label{G-grande-static}
\eeq
Let us discuss the extremal codimension-one surfaces. 
 In this case, the functions $f,g,h$ do not depend on $t_i$, so  there is a conserved quantity
\beq
\hat{P}=\frac{\p \mathcal{L}}{\p t_i'} = \frac{-g t_i' + h r'}{\sqrt{-g(r,t_i) \, t_i'^2 + \frac{r'^2}{f(r,t_i)}
+2 h(r,t_i) \, r' t_i' } }  \, r \, . 
\eeq
 With our gauge choice, the conserved quantity can be written as
 \beq
  \hat{P}^2=\le  \frac{g+ f h^2}{f} \ri (r')^2 -g r^{2} \, .
 \eeq
 Using eqs. (\ref{useful-quantities}) and (\ref{G-grande-static}), together with the fact that $g$ is continuous on the bubble, we find
 \beq
 (r'_i)^2= \frac{f_i}{f_o}    (r'_o)^2 \, .
 \eeq
From eq. (\ref{matching-tempi-bubble-statica}), we finally get
 \beq
 P_o^2=\frac{f_o(R)}{f_i(R)} P^2_i =\mu_0^2 \, P_i^2 \, .
 \label{static-relation-between-Pi-Po}
 \eeq

\subsection{Dynamical bubble}

For dynamical bubbles, 
the functions $f,g,h$ depend on $t_i$.
All the derivatives
\[
\p_{t_i} f \, , \qquad \p_{t_i} h \, , \qquad \p_r f \, , \qquad \p_r h \, , 
\]
will have a Dirac delta contribution localized on the surface of the bubble
\[
t_i=T_i(\tau) \, , \qquad r=R(\tau) \, .
\]
We impose the condition that
this delta function contribution is constant on the surface of the bubble:
\bea
 && \left. \frac{d \, f (r,t_i)}{d \tau} \right|_{\rm bubble}=
\frac{d \, f (R(\tau),T_i(\tau))}{d \tau}  \nl
&& = \dot{T}_i \,  \p_{t_i} f +\dot{R} \, \p_r f= 0 \, , 
 \eea
 which implies
 \beq
   \p_{t_i} f  = -\frac{d R}{d T_i } \,  \p_r f \, .
\label{invarianza-termini-deltiformi}
\eeq
So we expect
\bea
\p_r f &=& \frac{1}{\sqrt{ 1  + \le\frac{d R}{d T_i} \ri^2}} \, \delta (r - R(\tau)) \, \Delta f  \, , \nl
\p_{t_i} f &=&  -\frac{d R}{d T_i} \frac{1 }{\sqrt{1  + \le\frac{d R}{d T_i} \ri^2}} \, \delta (r - R(\tau))   \, \Delta f \, ,
\nonumber
\eea
where $\Delta f$ is the discontinuity of $f$ on the surface of the bubble.
An analogous equation holds for $h$.

Let us introduce the variables $s$ and $w$ such that
\beq
\left( \begin{array}{c}
t_i \\ r
\end{array}
\right) =
\left(
\begin{array}{cc}
\cos \psi & -\sin \psi \\ \sin \psi  &\cos \psi \\
  \end{array}\right) 
  \left( \begin{array}{c}
s\\ w
\end{array}
\right) \, ,
\eeq
where 
\bea
\sin \psi &=& \frac{\dot{R}}{ \sqrt{\dot{R}^2+\dot{T}_i^2 }} \, , \qquad
\cos \psi = \frac{\dot{T}_i}{ \sqrt{\dot{R}^2+\dot{T}_i^2 }} \, , \nl
\tan \psi &=& \frac{d R}{d T_i} \, .
\label{trig-psi}
\eea
The derivative of $f$
with respect to $s$ is
\beq
\frac{\p f}{\p s} = \frac{\p t_i}{\p s} \p_{t_i} f+ \frac{\p r}{\p s} \p_r f=
\frac{\dot{T}_i \p_{t_i} f+\dot{R} \, \p_r f}{\sqrt{\dot{R}^2+\dot{T_i}^2 }} \, .
\eeq
A similar property is valid for $h$. 
From eq. (\ref{invarianza-termini-deltiformi}), we thus find
that the relations  $\p_s f=0$ and  $\p_s h=0$  are satisfied by the Dirac delta term.
 
 Going back to the volume  functional in eq. (\ref{volume-functional-t}),
 we can express it in the $(s,w)$ coordinates.
 In the approximation in which we consider just the
 "fast" dependence of the Lagrangian due to discontinuities 
 at the two sides of the domain wall, the Lagrangian density 
 is independent of $s$.
We have then a conserved quantity
 of the form
 \beq
 \hat{P} = \frac{\p \mathcal{L}}{\p s'} =(h r' - g t_i') \cos \psi + \left( \frac{r'}{f} +  h t_i' \right) \sin \psi \, .
 \eeq
 The quantity   $\hat{P}$ is not globally conserved on the codimension-one extremal surface,
but just  before and after the collision with the Dirac delta domain wall.
 Plugging the gauge fixing condition (\ref{gauge-fixing-Phat}) in,
  we get
\bea
  \hat{P}^2&=& \le \frac{g+f h^2}{f g}  r' \sin \psi
  \pm \frac{(g \cos \psi -h \sin \psi)}{g} \right.
  \nl
&& \left. \sqrt{ \frac{g+f h^2}{f } (r')^2- g \, r^{2}}
  \ri^2 \, .
\eea
The conserved quantities inside and outside the domain wall are
  \bea
  \hat{P}_i^2 &=& \le \frac{\sin \psi}{f_i} r_i' \pm \sqrt{(r_i')^2-f_i r^{2 }}  \cos \psi \ri^2 \, ,
\nl
  \hat{P}^2_o &=& \le \frac{1}{  f_o}  r_o' \sin \psi  \pm 
   {\left( G \cos \psi +\frac{d G}{dr} \, t_i \,   \sin \psi \right)} \right. 
   \nl
 &&  \left.  \sqrt{  (r_o')^2-  f_o \, r^{2}}
  \ri^2 \, .
  \eea
Combining the condition $  \hat{P}_i^2=  \hat{P}^2_o$
with eqs. (\ref{trig-psi}) and (\ref{tempo-o-tempo-i}), we find 
 \bea
 && \le \frac{1}{f_i} r_i' \pm \sqrt{(r_i')^2-f_i \, r^{2 }} \frac{d T_i}{dR}\ri^2
  = 
  \nl
&& = \le \frac{1}{  f_o}  r_o' 
  \pm \frac{d T_o}{d R }  \sqrt{  (r_o')^2-  f_o \, r^{2}}
  \ri^2 \, .
  \label{snell-law-generic-2}
  \eea
 We can consider the following crosschecks of eq. (\ref{snell-law-generic-2}):
 \begin{itemize}
 \item
 For $\psi=0$, it reproduces the static bubble result in eq. (\ref{static-relation-between-Pi-Po}).  
 \item
 For a bubble moving at the speed of light, we have
 \beq
  \frac{d T_{i,o}}{dR} =\pm \frac{1}{f_{i,o}} \, .
  \label{bubble-velocita-luce}
 \eeq 
 Specializing eq. (\ref{snell-law-generic-2}) 
 to eq. (\ref{bubble-velocita-luce}) 
 and combining with eq. (\ref{vprimo-da-P}),
 we find $V'_i=V'_o$ or $U'_i=U'_o$, depending on the choice of sign.  
 This is in agreement with the results in \cite{Chapman:2018dem,Chapman:2018lsv}.
  \end{itemize}
  Using the notation in eq. (\ref{derivatozze}),
  the refraction condition (\ref{snell-law-generic-2}) can be written in the covariant form
\beq
(g_i)_{\mu \nu} \frac{d x^\mu_i}{d l} \frac{d X^\nu_i}{d \tau}= \pm
(g_o)_{\mu \nu} \frac{d x^\mu_o}{d l} \frac{d X^\nu_o}{d \tau} \, .
\label{snell-law-covariante-0}
\eeq
By continuity with the case where there is no bubble into the system,
the physical solution should be the one with the $+$ sign, which we consider in eq. (\ref{snell-law-covariante}).


\section{Technical details for smooth extremal surfaces}
 
\subsection{Solution with $P_o<0$}
\label{appe-Po-negativo}

With the assumption of negative $P_o$ and with the choice of sign for $\frac{dT_o}{d R}$ in eq. (\ref{scelta-segno-To}), eq. (\ref{P-matching}) 
is solved by
\begin{widetext}
\bea
P_o^2 &=& -R^4+\mu  R^2+\frac{\left((\mu +1)^2+R^4 
\left(\kappa ^2 (\lambda -1)+(\lambda +1)^2\right)+R^2 \left(\kappa ^2 (\mu -1)-2 (\lambda +1) (\mu
   +1)\right)\right)^2}{4 \kappa ^4 R^2 \left(1-\lambda  R^2\right)} \, .
 \nl
 \rho_o &=&  - \frac{R^4 \left(\kappa ^2 (\lambda -1)+(\lambda +1)^2 \right)-R^2 \left(2 (\lambda +1) (\mu +1)-\kappa ^2 (\mu
   -1)\right) +(\mu +1)^2 }{2 \kappa ^2 R \sqrt{1-\lambda  R^2}} \, ,
   \label{Po-negativo}
\eea
\end{widetext}

A direct calculation gives that $\rho_o(R_{\rm max})$ in eq. (\ref{Po-negativo})
is the opposite of $\rho_o(R_{\rm max})$ in eq. (\ref{rprimo-Po-positivo}).
The same holds true for $\rho_o(R_{\rm min})$. 
So, the  solution in eq. (\ref{Po-negativo}) corresponds to a "reflection" of the codimension-one extremal surface,
and as such should be discarded.
Note that both the solutions in eqs. (\ref{Po-positivo})
and (\ref{Po-negativo}) vanish for $r=R_{\rm max}=R_{\rm min}$.
 
 
 \subsection{Sign of $\rho_o(R)$ for small bubbles}
\label{appe-segno-truccoso}
 
The function $\rho_o(R)$ in eq. (\ref{rprimo-Po-positivo}) vanishes for $R=R_0$, where
\beq
R_0(\mu) = \sqrt{\frac{1-\mu}{\kappa^2+\l-1}} \, .
\label{ERRE-zero}
\eeq
Therefore,
if either $R_0(\mu) $ is not a real number or 
$R_0>R_{\rm max}$, the sign of $\rho_o(R)$
does not change along the surface of the bubble.
Conversely, if $R_0(\mu) $ is real and $R_0<R_{\rm max}$,
$\rho_o(R)$ changes sign along the bubble.
With a direct calculation, it can be checked that
\beq
R_0(\mu_s)=\sqrt{\mu_s}=R_{\rm max} (\mu_s) \, .
\eeq
There are then four possible behaviors of $\rho_o (R)$ on the domain wall:
\begin{itemize}
\item[A)]  $\mu <\mu_s$ and $\l+\kappa^2>1$. The function
$\rho_o(R)$ is always positive.
\item[B)]  $\mu <\mu_s$ and $\l+\kappa^2<1$. When $R_0$ is real,
 $\rho_o(R)$ is positive for $R_0< R \leq R_{\rm max}$
and negative for $0<R<R_{\rm 0}$.
When $R_0$ is imaginary,  $\rho_o(R)$ is always positive.
\item[C)]   $\mu >\mu_s$ and $\l+\kappa^2>1$. 
When $R_0$ is real,
 $\rho_o(R)$ is negative for $R_0< R \leq R_{\rm max}$
and positive for $0<R<R_{\rm 0}$.
When $R_0$ is imaginary,  $\rho_o(R)$ is always negative.
\item[D)]  $\mu >\mu_s$ and $\l+\kappa^2<1$. The function
$\rho_o(R)$ is always negative.
\end{itemize}
With reference to subsection \ref{section-late-time-small-bubbles},
from these considerations we find that
the region of parameter space in bullet A) belongs to case 2,
while the part of parameter space in bullet D) belongs to case 1.


  
\subsection{Behavior of the complexity rate for  $t_b=0$ }
\label{appe-region-S}

In this appendix, we determine the explicit value of the quantity $K$ defined in eq. (\ref{definizione-K}). \\
Let us first consider very small bubbles.
Since for $t_b \to 0$ we have $\rho_o(R)>0$, 
we should use  eq. (\ref{boundary-time-caso-rprimo-o-positivo}) to compute $\frac{d  t_b}{d  P_o} $.
By symmetry, $t_b =0$ corresponds to $P_o=0$, so
\beq
K=\left. \frac{d  t_b}{d  P_o} \right|_{P_o=0} 
=  H - \int_{R_{\rm max}}^\Lambda 
 \frac{1}{r (f_o(r))^{3/2}} dr  \, ,
\eeq
where
\beq
H= \left. \frac{d T_o}{d R} \frac{d R}{d P_o} \right|_{R=R_{\rm max}} =\frac{a}{b}
\eeq
and
\bea
 a&=&2 \sqrt{1-\l R_{\rm max}^2}  \, (1+\mu -R_{\rm max}^2 (1+\l+\kappa^2) )  \, ,\nl
 b&=& R_{\rm max}^2 (R_{\rm min}^2-R_{\rm max}^2) (R_{\rm max}^2-\mu )
 \nl
 && (\l +(\kappa-1)^2) (\l+(\kappa+1)^2) \, .
\eea
We can use the explicit integral
\bea
 F(r) &=& \int \frac{-1}{r (r^2-\mu)^{3/2}} dr
 \nl
 &=& \frac{1 }{\mu \sqrt{r^2-\mu}}-
 \frac{2 }{\mu^{3/2} }  \tan^{-1}  \sqrt{\frac{r+\sqrt{\mu}}{r-\sqrt{\mu}}}  
 \nl
 \eea
to obtain
 \beq
K = H - \frac{\pi}{2 \mu^{3/2}}-F(R_{\rm max}) \, .
\label{derivata-a-Po-zero-very-small-bubble}
\eeq

In the case of not so small bubbles, 
 for $t_b \to 0$ we have $\rho_o(R)<0$, so $\frac{d  t_b}{d  P_o} $ 
 is given by eq.  (\ref{boundary-time-caso-rprimo-o-negativo}).
 To perform the computation,
we first send $r_{t,{\rm AdS}} \to r_{t,{\rm AdS}} +\delta$, then we  use the Leibniz integral rule:
\bea
K &=&  H
-\int^{\Lambda}_{r_{t,{\rm AdS}}+\delta}   \frac{r^2}{(P_o^2+f_o(r) r^2)^{3/2}}  dr 
\nl
&-& \int^{R}_{r_{t,{\rm AdS}}+\delta}   \frac{r^2}{(P_o^2+f_o(r) r^2)^{3/2}} dr  
\nl 
& +&   \frac{1 }{ \sqrt{P_{o}^2  +   (r_{t,{\rm AdS}}+\delta)^{2} \, f_{o}(r_{t,{\rm AdS}}+\delta) }}
 \nl
 & & 2 \frac{d r_{t,{\rm AdS}} }{ d P_o}  \frac{P_o}{f_{o}(r_{t,{\rm AdS}})}     \, .
\eea
At small $P_o$, the following approximations are valid:
\bea
&& P_{o}^2  +f_{o}(r) \,  r^{2} \approx (r-r_{t,{\rm AdS}}) (r+\sqrt{\mu})r^2 \, , \nl
&&r_{t,{\rm AdS}}  \approx\sqrt{\mu} - \frac{P_o^2}{2 \mu^{3/2}} + O(P_o^4) \, , \nl
&& \frac{d r_{t,{\rm AdS}}}{d P_o} \approx -\frac{P_o}{\mu^{3/2}}  \, , \nl
&& \frac{1}{f_o(r_{t,{\rm AdS}})} \approx -\frac{\mu}{P_o^2} +O(P_o^0) \,  . 
\eea
Hence, we get
\bea
K&=&  H
-\int^{\Lambda}_{r_{t,{\rm AdS}}+\delta}   \frac{1}{r [(r+\sqrt{\mu})(r-r_{t,{\rm AdS}})]^{3/2}}  dr
\nl
&& - \int^{R_{\rm max}}_{r_{t,{\rm AdS}}+\delta}   \frac{1}{ r [(r+\sqrt{\mu})(r-r_{t,{\rm AdS}})]^{3/2} } dr  
\nl
&& + \frac{\sqrt{2}}{{\mu}^{5/4}}
 \frac{1 }{ \sqrt{\delta }}
\nl 
& =& 
H  + F(R_{\rm max})+\frac{3\pi}{ 2 \mu^{3/2}} \, .
 \label{derivata-a-Po-zero-not-so-small-bubble}
\eea
Consistently, eqs. (\ref{derivata-a-Po-zero-very-small-bubble})
and (\ref{derivata-a-Po-zero-not-so-small-bubble}) matches for $\mu=\mu_s$.

In the case of large bubbles, similar calculations give
\beq
K= \tilde{H}  + F(R_{\rm min})+\frac{3\pi}{ 2 \mu^{3/2}} \, ,
 \label{derivata-a-Po-zero-large-bubble}
\eeq
where
\beq
\tilde{H} =
\left. \frac{d T_o}{d R} \frac{d R}{d P_o} \right|_{R=R_{\rm min}}=\frac{\tilde{a}}{\tilde{b}}
\label{Htilde}
\eeq
and
\bea
\tilde{a } &=&-{2 \sqrt{1-\l R_{\rm min}^2}  \, (1+\mu -R_{\rm min}^2 (1+\l+\kappa^2) ) } \nl
\tilde{b} &=& R_{\rm min}^2 (R_{\rm min}^2-R_{\rm max}^2) (R_{\rm min}^2-\mu ) \nl
&& (\l +(\kappa-1)^2) (\l+(\kappa+1)^2) \, .
\eea
Note that $K$ in eq. (\ref{derivata-a-Po-zero-large-bubble})
is always positive, because both $F(r)+\frac{3\pi}{ 2 \mu^{3/2}}>0$ and $\tilde{H} >0$.
The latter property follows from
\beq
w_o(R_{\rm min})<0 \, ,
\label{wo-tecnicismo}
\eeq
where $w_o$ is defined in eq. (\ref{scelta-segno-To}).
The inequality (\ref{wo-tecnicismo}) arises from the 
negativity of $w_o(R_{\rm min})$ at $\mu=0$.
Indeed,
no real solutions for $\mu$ to the equation $w_o(R_{\rm min})=0$ exist.
Consequently,  $w_o(R_{\rm min})$ never changes sign as a function of $\mu$.


\end{document}